\documentclass[12pt,dvipsnames]{article}
\usepackage{amsmath,amssymb, multirow}
\usepackage{mathrsfs}
\usepackage{graphicx}
\usepackage{color}
\usepackage[noblocks]{authblk}
\usepackage{natbib}
\usepackage{url}
\usepackage{algorithm} 
\usepackage{algpseudocode} 
\usepackage{caption}
\usepackage{subcaption}
\usepackage{bm}

\newtheorem{fact}{Fact}
\newtheorem{definition}{Definition}
\newtheorem{lemma}{Lemma}
\newtheorem{theorem}{Theorem}

\newtheorem{assumption}{Assumption}

\usepackage{xcolor}

\newcommand{\mc}[1]{\mathcal{#1}}
\def\mbf#1{\mathbf{#1}}
\def\mbi#1{\boldsymbol{#1}} 
\newcommand{\norm}[1]{\left\|{#1}\right\|} 

\newcommand{\absbigg}[1]{\left \lvert{#1}\right\rvert}
\newcommand{\firstbigg}[1]{\left ({#1}\right)}
\newcommand{\secondbigg}[1]{\left \{{#1}\right\}}
\newcommand{\thirdbigg}[1]{\left [{#1}\right]}

\def\leqsim{\stackrel{<}{\sim}}

\newcommand{\wt}[1]{\widetilde{#1}} 

\newcommand{\E}{\mathbb{E}} 
\renewcommand{\P}{\mathbb{P}} 
\newcommand{\var}{{\rm var}} 
\newcommand{\simiid}{\stackrel{\rm iid}{\sim}}

\newcommand{\floor}[1]{\left\lfloor{#1} \right\rfloor}
\newcommand{\ceil}[1]{\left\lceil{#1} \right\rceil}
\def\indic{\mathbb I} 
\newcommand{\ind}{\perp\!\!\!\!\perp} 
\DeclareMathOperator*{\argmax}{arg\,max}
\DeclareMathOperator*{\argmin}{arg\,min}
\newcommand{\reals}{\mathbb{R}} 

\def\calI{\mathcal{I}}

\newcommand{\blind}{0}

\usepackage{fullpage}
\usepackage{xcolor}
\usepackage[colorlinks=true]{hyperref}
\hypersetup{
    colorlinks=true,
    linkcolor=Maroon,   
    urlcolor=Green,
    citecolor=NavyBlue,
}

\begin{document}

\def\spacingset#1{\renewcommand{\baselinestretch}%
{#1}\small\normalsize} \spacingset{1}


\if0\blind
	{
		\title{\bf \Large  Risk-inclusive Contextual Bandits for \\ Early Phase Clinical Trials}
            \author[1]{Rohit Kanrar}
		\author[2]{Chunlin Li}
            \author[3]{Zara Ghodsi}
            \author[4]{Margaret Gamalo}
	    \affil[1]{\it Department of Statistics,
Iowa State University, Ames, Iowa, U.S.A}
       \affil[2]{\it Department of Statistics,
University of Virginia, Charlottesville, Virginia, U.S.A}
       \affil[3]{\it Pfizer R\&D Ltd., Tadworth, Surrey, U.K.}
       \affil[4]{\it Pfizer Inc., Collegeville, Pennsylvania, U.S.A}
\setcounter{Maxaffil}{0}
		
		\renewcommand\Affilfont{\itshape\small}
		\date{\vspace{-5ex}
  }
		\maketitle
	} \fi
\if1\blind
{
    \bigskip
    \bigskip
    \bigskip
    \begin{center}
        {\Large\bf Risk-inclusive Contextual Bandits for Early Phase Clinical Trials}
    \end{center}
    \medskip
} \fi

\bigskip
\begin{abstract}
Early-phase clinical trials face the challenge of selecting optimal drug doses that balance safety and efficacy due to uncertain dose-response relationships and varied participant characteristics. Traditional randomized dose allocation often exposes participants to sub-optimal doses by not considering individual covariates, necessitating larger sample sizes and prolonging drug development. This paper introduces a risk-inclusive contextual bandit algorithm that utilizes multi-arm bandit (MAB) strategies to optimize dosing through participant-specific data integration. By combining two separate Thompson samplers, one for efficacy and one for safety, the algorithm enhances the balance between efficacy and safety in dose allocation. The effect sizes are estimated with a generalized version of asymptotic confidence sequences (AsympCS), offering a uniform coverage guarantee for sequential causal inference over time. The validity of AsympCS is also established in the MAB setup with a possibly mis-specified model. The empirical results demonstrate the strengths of this method in optimizing dose allocation compared to randomized allocations and traditional contextual bandits focused solely on efficacy. Moreover, an application on real data generated from a recent Phase IIb study aligns with actual findings.
\end{abstract}

\noindent%
\vfill

\newpage
\spacingset{1.45} 

\tableofcontents

\section{Introduction}

\begin{figure}[htpb]
    \centering
    \includegraphics[width=0.75\linewidth]{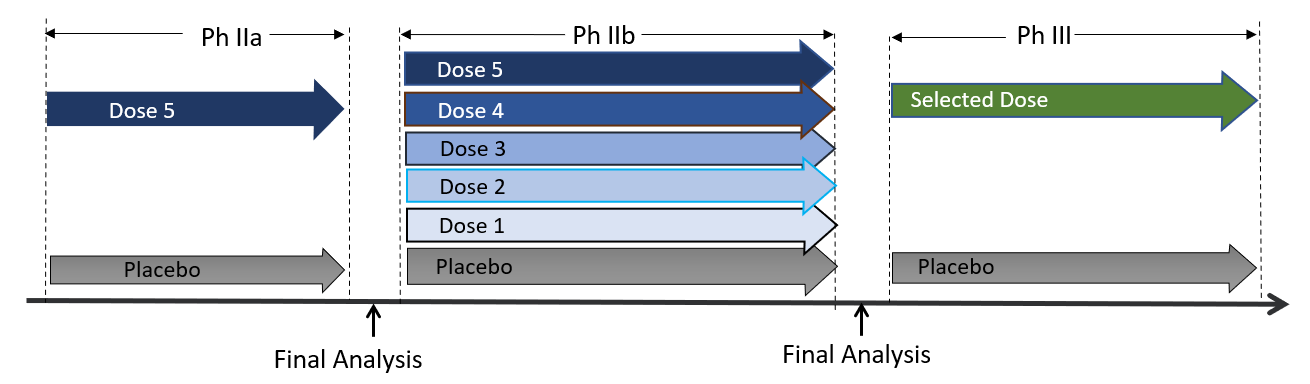}
    \caption{Study design of a Phase II/III clinical trial. 
    In this scenario, Dose 5 of a novel drug, demonstrating preliminary safety and early efficacy in Phase I and IIa, progresses to Phase IIb.
    Then, five doses and a placebo control are randomized in Phase IIb to facilitate dose-response curve characterization and optimal dose selection. Final analyses are performed at the end of Phase IIb to select the most suitable dose for the larger Phase III trial.} 
    \label{fig:study_schema}
\end{figure}

Randomized Controlled Trials (RCTs) are structured studies to evaluate the safety and effectiveness of investigational treatments. RCTs are conducted in progressive phases \citep{friedman_et_al_2020_textbook}. Phase I trials involve a small group to assess safety and tolerability. Phase II expands the sample size to evaluate preliminary efficacy and safety, establishing optimal dosage. This precedes larger Phase III trials, which confirm treatment safety and effectiveness. Of particular importance are dose-ranging studies, which are frequent in Phase IIb and aim to determine the optimal dose that balances efficacy and safety for further development. These trials administer multiple doses to participant cohorts to identify the best therapeutic benefit with minimal adverse effects, as illustrated in Figure \ref{fig:study_schema}. 

Traditional dose-ranging study designs often rely on equal randomization (ER). However, ER has several limitations \citep{hay_et_al_2014_nature}: (1) The unknown dose-response relationship in early phases exposes many participants to suboptimal doses. (2) Dose-response relationships are often heterogeneous and influenced by participant-level covariates \citep{tournaue_et_al_2009_jnci}. Adaptive methods such as response-adaptive randomization (RAR) \citep{robertson_lee_et_al_2023_stat_science} and covariate-adjusted response-adaptive (CARA) designs \citep{villar2018covariate, yang_diao_rosenberger_2024_sbs} aim to address this; nevertheless, existing adaptive designs fall short in combining efficacy, safety, and covariates to inform treatment allocation. (3) Traditional approaches only allow formal inference at the end, whereas continuous monitoring is needed to enable early stopping for efficacy or futility, saving resources and time. (4) Finally, a valid inference on effect sizes often depends on correctly specified parametric models, but practitioners often do not have substantial knowledge about the efficacy or safety in early phases, incurring model mis-specification. As early-phase trials involve small sample sizes and investigational drugs that are not yet approved, practitioners prefer inference methods that do not rely heavily on model assumptions, especially given the critical role such results play in informing Phase III planning and decision-making. Overcoming these limitations is crucial for accelerating drug development while ensuring participant safety.

Motivated by these challenges, we propose an RAR framework extending recent developments in {online reinforcement learning } \citep[RL,][]{agrawal_et_al_2013_icml} and {sequential} causal inference \citep{waudby-smith_et_al_2024+} to optimize {dose-ranging} trials. Our framework has two components: trial optimization and trial evaluation. For optimization, we frame the trial as a contextual multi-arm bandit (CMAB) problem, incorporating covariate information. To balance efficacy and safety, we combine two separate CMAB models, one for each, with a pre-specified weight parameter. The weight parameter enables informed decision-making based on both trial data and insights from preclinical and earlier-phase studies. Trial evaluation focuses on estimating effect sizes for all active doses. Moving beyond standard post-trial inference, we perform anytime-valid inference by extending Asymptotic Confidence Sequences \citep[AsympCS,][]{waudby-smith_et_al_2024+} to the CMAB setup, offering coverage guarantees during continuous data monitoring and allowing statistically principled early trial stopping. While parametric models are used for efficient trial optimization, inference during trial evaluation is robust to model mis-specification, making our approach model-assisted.

Extensive simulations and analysis based on a real-life Phase IIb study demonstrate the utility and effectiveness of the proposed approach. Notably, in our experiments, we also consider some logistical challenges to operate the proposed method in a multi-center, multi-regional setup. We further discuss unaddressed issues in Section \ref{sec:discussion}.

The rest of this article is organized as follows: Section \ref{subsec:related_work} reviews relevant work; Section \ref{sec:methods} defines notations, introduces the setup, and details our proposed method; Section \ref{sec:simulation} examines its operating characteristics on simulated data; Section \ref{sec:real_data} presents its application to a real-world dataset; and Section \ref{sec:discussion} concludes by discussing future directions.

\subsection{Related Work} \label{subsec:related_work} 

\paragraph{{Adaptive Designs in Clinical Trials}}

Adaptive clinical trials offer advantages over traditional fixed designs by enabling modifications based on accumulating data \citep{pallmann2018adaptive, guidance2018adaptive, burnett2020adding}. However, adaptive designs, including Response Adaptive Randomization (RAR), may perform sub-optimally in long-duration trials where treatment effects evolve over time \citep{proschan_evans_2020}. Classical RAR methods are often criticized for reacting too quickly to certain trends in the data and for ignoring the covariates of participants during arm allocation. Most existing methods fall short of analyzing data from adaptive experiments \citep{thall_et_al_2015}. Despite these challenges, the advantages of adaptive designs have motivated methodological research in the past decade \cite[e.g.,][]{villar_wason_bowden_2015_biometrics, williamsom_villar_2020_biometrics, robertson_lee_et_al_2023_stat_science, norwood_davidian_laber_2024_biometrics}. In this article, we focus on short-duration trials with approximately constant treatment effect sizes, common in areas such as oncology and dermatology. Our method addresses the foregoing issues in classical adaptive designs by (1) making appropriate modifications, such as ``clipping,'' to encourage exploring sub-optimal arms, (2) using participants' covariate information during arm allocation, and (3) performing sequential causal inference that allows valid model-robust inference on effect sizes.

\paragraph{{Safety Consideration and Multi-objective Reinforcement Learning}}

RAR methods that balance efficacy and safety remain underexplored. To address this gap, we utilize Thompson Sampling \citep[TS,][]{thompson_1933_biometrika} and leverage recent developments in online RL \citep{agrawal_et_al_2013_icml}. Although multi-objective contextual bandit methods allow integration of multiple responses \citep[e.g.,][]{kim_iyengar_zeevi_2024+}, existing methods typically treat all outcomes equally. Clinical trials, however, often have unique safety tolerance limits for efficacy, such as in oncology, where greater risks might be accepted for higher therapeutic benefits, versus dermatology, where minimal adverse effects are preferred. Furthermore, by Phase IIb, preliminary safety knowledge usually exists, warranting its incorporation. Although weighted composite outcomes are used in precision medicine \citep{luckett_et_al_2021_jmlr, zhu_shen_fu_qu_2024_aoas}, existing work focuses on individualized treatment recommendations. To our knowledge, no existing work effectively combines both efficacy and safety endpoints to develop multi-objective contextual bandits for early-phase clinical trials.

\paragraph{{Statistical Inference from Reinforcement Learning}}

Research in online RL mainly focuses on regret minimization, balancing exploration and exploitation \citep{agrawal_goyal_2017_acm, dimakopoulou_et_al_2019_aaai, dimakopoulou_et_al_2021_nips, kim_et_al_2021_nips}. However, all phases of clinical trials typically involve smaller sample sizes than online RL literature often considers. In small samples, some suboptimal arm allocations are {\color{blue} necessary} due to exploration, with valid estimation and inference of estimands like effect sizes being the primary interest. Recent inference tools enable us to perform valid inference with adaptively collected data \citep{bibaut_et_al_2021_neurips, hadad_et_al_2021_pnas, zhang_et_al_2020_nips, zhang_et_al_2021_nips, shen_et_al_2023_jasa}. While these methods primarily focus on post-trial inference, continuous monitoring in early-phase trials facilitates faster progression to Phase III or early termination when the investigational drug is evidently effective or deemed unsafe, helping reduce unnecessary time and resource expenditure. Valid sequential testing offers this flexibility, finding success in other domains \citep{lindon_et_al_2022,maharaj_et_al_2023}. More recently, \cite{waudby-smith_et_al_2024+} proposed AsympCS for time-uniform inference of effect sizes in both randomized and observational data. Based on this, \cite{dalal_et_al_2024+} extended its validity for general causal parameters. \cite{ham_et_al_2022+} proposed a design-based confidence sequence for anytime-valid inference of finite-sample estimands. While non-asymptotic CSs \citep{howard_ramdas_mcauliffe_sekhon_2021_aos, waudby-smith_ramdas_2023_jrssb} are popular due to non-asymptotic inferential guarantees under less assumptions, we advocate for AsympCS because it yields narrower confidence bounds than non-asymptotic CSs for practical sample sizes common in early-phase clinical trials.

\section{Methods} \label{sec:methods}
\subsection{Framework} \label{subsec:setup}

Our setup for dose-ranging trials includes $K$ arms. Arm $1$ corresponds to placebo, and Arms $2,\ldots,K$ are active doses of a novel drug with increasing order of doses. For each participant, we consider their covariate information and two (continuous) endpoints, efficacy and safety. 
Specifically, assume the complete profiles of participants $(\mbi{X}_n, \{R_n(a)\}_{a=1}^{K}, \{S_n(a)\}_{a=1}^{K})_{n \geq 1}$ are independent and identically distributed, where $\bm X_n\in\mathbb R^d$, $\{R_n(a)\}_{a=1}^K$, and $\{S_n(a)\}_{a=1}^K$ are the $n$-th participant's covariate vector, potential outcomes of efficacy and safety, respectively.
The complete profile of each participant cannot be observed. Thus, we randomly assign Arm $A_n$ to the $n$-th participant, and observe the outcomes $R_n$ and $S_n$ instead. For equal randomization (ER), $A_n$ is independent of the covariates $\bm X_n$ and other participants' information.
Response adaptive randomization (RAR) improves arm allocation by allowing the distribution of $A_n$ to depend on the covariates $\bm X_n$ and the historical information $\mathcal H_{n-1} = \{ (\bm X_i, A_i, R_i, S_i) \}_{i=1}^{n-1}$ from the first $n-1$ participants. In this article, we focus on identifying the most efficacious arm with {uncertainty quantification}, while allocating better and safer treatments to participants.
Define the average efficacy of Arm $a$ as $\theta(a) := \E[R_n(a)]$ and the effect size of Arm $a$ relative to the placebo as $\Delta(a) := \theta(a) - \theta(1)$. 
Average safety can also be defined in a similar manner. For ease of exposition, we focus on average efficacy in this article. The following conditions suffice to identify $\{\theta(a)\}_{a=1}^{K}$ and $\{\Delta(a)\}_{a=2}^{K}$:

\begin{enumerate}
    \item[(A1)] \textbf{Consistency:} $R_n = \sum_{a=1}^{K} R_n(a) \indic\{A_n = a\}$ and $S_n = \sum_{a=1}^{K} S_n(a) \indic\{A_n = a\}$; $n\geq 1$.
    
    \item[(A2)] \textbf{Positivity:} $\P\thirdbigg{ A_n = a | \mbi{X}_n, \mc{H}_{n-1}} > 0$ for all $a \in [K]$; $n\geq 1$.
    
    \item[(A3)] \textbf{Ignorability:} $\{R_n(a), S_n(a)\}_{a=1}^{K} \ind A_n | \{\mbi{X}_n, \mc{H}_{n-1} \}$; $n\geq 1$. 
\end{enumerate} 

By (A1), we assume participants adhere to their assigned treatments. 
We will design a trial optimization algorithm in Section \ref{subsec:trial_opt}, so that (A2)--(A3) are automatically met. It further leads to trial evaluation through inferring $\{\Delta(a)\}_{a=2}^{K}$ in Section \ref{subsec:trial_eval}.

To enable a Bayesian trial optimization, we introduce a parametric \emph{working model} for the potential efficacy and safety outcomes: \begin{equation} \label{eq:lin_rwd}
    R_n(a) = \mu^{(a)}(\mbi{X}_n) + \epsilon_n, \quad S_n(a) = \nu^{(a)}(\mbi{X}_n) + \delta_n, \quad n\geq 1,
\end{equation} where $\mu^{(a)}(\mbi{x}) = \beta_{a,0} + \mbi{x}^T\mbi{\beta}_{a}$, $\nu^{(a)}(\mbi{x}) = \gamma_{a,0} + \mbi{x}^T\mbi{\gamma}_{a}$, $\epsilon_n\sim \mathcal N(0,\sigma_{\epsilon}^2)$ and $\delta_n\sim \mathcal{N}(0,\sigma_{\delta}^2)$ are normal errors. {We will show that parametric model \eqref{eq:lin_rwd} assists in efficient learning for better treatment allocation. More importantly, 
the validity of inference for $\{\Delta(a)\}_{a=2}^{K}$ will not be affected even if \eqref{eq:lin_rwd} is mis-specified.} 

\subsection{Trial Optimization using a Risk-inclusive Thompson Sampler} \label{subsec:trial_opt}
\paragraph{An Overview of TS} 
Under consistency (A1) and parametric form \eqref{eq:lin_rwd}, $R_i = \beta_{A_i,0} + \mbi{X}_i^T\mbi{\beta}_{A_i}^{*} + \epsilon_i.$ TS starts with priors $\{\pi^{(0)}(\beta_{a,0}), \pi^{(0)}(\mbi{\beta}_a)\}_{a=1}^{K}$ on the regression parameters, updating them to posteriors $\{\pi^{(n)}(\beta_{a,0}), \pi^{(n)}(\mbi{\beta}_a)\}_{a=1}^{K}$ based on historical data from the first $n-1$ participants. Before assigning arm to the $n$-th participant, TS simulates $MK$ samples, with $M$ samples drawn from each posterior in $\{\pi^{(n)}(\beta_{a,0}), \pi^{(n)}(\mbi{\beta}_a)\}_{a=1}^{K}$. Let $\{b_{0,l}^{(n)}(a)\}_{l=1}^{M}$ and $\{\mbi{b}^{(n)}_{l}(a)\}_{l=1}^{M}$ be the $M$ samples from $\pi^{(n)}(\beta_{a,0})$ and $\pi^{(n)}(\mbi{\beta}_a)$. The next arm is assigned according to $\{(p_{n}(a, \mbi{x})\}_{a=1}^{K}$, where 
$p_{n}(a, \mbi{x}) := \frac{1}{M} \sum_{l=1}^{M} \indic \{ b_{0,l}^{(n)}(a) + \mbi{x}^T \mbi{b}^{(n)}_l(a) \in \underset{a' \in [K]}{\argmax} \ b_{0,l}^{(n)}(a') + \mbi{x}^T \mbi{b}^{(n)}_l(a') \}$ is the propensity of allocating arm $a$ to the $n$-th participant, given their covariate $\bm X_n = \mbi{x}$ and the historical data $\mc{H}_{n-1} = \{(\mbi{X}_i, A_i, R_i, S_i)\}_{i=1}^{n-1}$. Safety outcomes $\{S_i\}_{i=1}^{n-1}$, although available in $\mc{H}_{n-1}$, are ignored in this approach.

\paragraph{Outline of RiTS} To utilize and balance safety, we propose using two separate TS: one for efficacy and one for safety. The priors $\{\pi^{(0)}(\beta_{a,0}), \pi^{(0)}(\mbi{\beta}_a)\}_{a=1}^{K}$ for efficacy, and $\{\pi^{(0)}(\gamma_{a,0}), \pi^{(0)}(\mbi{\gamma}_a)\}_{a=1}^{K}$ for safety are updated to posteriors $\{\pi^{(n)}(\beta_{a,0}), \pi^{(n)}(\mbi{\beta}_a)\}_{a=1}^{K}$ and $\{\pi^{(n)}(\gamma_{a,0}), \pi^{(n)}(\mbi{\gamma}_a)\}_{a=1}^{K}$ using $\mc{H}_{n-1}$. Samples $\{b_{0,l}^{(n)}(a), \mbi{b}_{l}^{(n)}(a)\}_{l=1}^{M}$ and $\{g_{0,l}^{(n)}(a), \mbi{g}_{l}^{(n)}(a)\}_{l=1}^{M}$ are simulated from $\{\pi^{(n)}(\beta_{a,0}), \pi^{(n)}(\mbi{\beta}_a)\}$ and $\{\pi^{(n)}(\gamma_{a,0}), \pi^{(n)}(\mbi{\gamma}_a)\}$. For a weight parameter $w \in (0, 1)$, define $\omega(\mbi{x}, b_0, \mbi{b}, g_0, \mbi{g}; w) := w \cdot (b_0 + \mbi{x}^T \mbi{b}) + (1 - w) \cdot (g_0 + \mbi{x}^T \mbi{g})$. The next arm is randomized according to $\{q_{n}(a, \mbi{x})\}_{a=1}^{K}$, where given $\bm X_n = \mbi{x}$ and past data $\mc{H}_{n-1}$, \begin{equation*}
    \resizebox{\linewidth}{!}{
    $\displaystyle
    q_{n}(a, \mbi{x}) := \frac{1}{M} \sum_{l=1}^{M} \indic \left\{ \omega\firstbigg{ \mbi{x}, b_{0, l}^{(n)}(a), \mbi{b}^{(n)}_l(a), g_{0, l}^{(n)}(a), \mbi{g}^{(n)}_l(a); w} \in \argmax_{a' \in [K]} \omega\firstbigg{ \mbi{x}, b_{0, l}^{(n)}(a'), \mbi{b}^{(n)}_l(a'), g_{0, l}^{(n)}(a'), \mbi{g}^{(n)}_l(a'); w} \right\}
    $
    }
\end{equation*} 
is the propensity of assigning arm $a$ to the $n$-th participant. We refer to this approach as the ``{\bf R}isk-{\bf i}nclusive {\bf T}hompson {\bf S}ampling'' ({\bf RiTS}). Additional discussions on the choice of weight parameter is included in Web Appendix \ref{subsec:choice_of_w}. Choice of prior distributions for practical implementation are discussed in Web Appendix \ref{subsec:prior}. Exact form of posteriors for multivariate normal prior on $\beta_{a,0}, \mbi{\beta}_a$ are outlined in Web Appendix \ref{subsec:posterior}.

\subsection{Robust Trial Evaluation Using AsympCS} \label{subsec:trial_eval}

We first provide a less formal definition of AsympCS in our setting. The formal definition is originally proposed in Definition 2.1 of \cite{waudby-smith_et_al_2024+}.
\begin{definition}
    Let $\hat{\Delta}_{n}^{\times}$ be an estimator of $\Delta$ using $\{Z_i: Z_i := (\mbi{X}_i, A_i, R_i, S_i)\}_{i=1}^{n}.$ A set of intervals $(\hat{\Delta}_{n}^{\times} - L_n, \hat{\Delta}_{n}^{\times} + U_n)_{n \geq 1}$ with nonzero bounds $L_n, U_n > 0$ for all $n \geq 1$ for a $(1-\alpha)$-AsympCS for $\Delta$ if there exists a (possibly unknown) non-asymptotic $(1-\alpha)$-CS $(\hat{\Delta}_{n}^{\times} - L_{n}^{*}, \hat{\Delta}_{n}^{\times} + U_{n}^{*})_{n \geq 1}$ for $\Delta$, which satisfies $\P(\forall n \geq 1, \Delta \in [\hat{\Delta}_{n}^{\times} - L_{n}^{*}, \hat{\Delta}_{n}^{\times} + U_{n}^{*}]) \geq 1-\alpha,$ such that $L_n$ and $U_n$ become arbitrarily precise (almost-sure) approximation to $L_n^*$ and $U_n^*.$
\end{definition}

\paragraph{Construction of AsympCS} 
To begin, we randomly allocate arms to the first $n_0$ participants using equal randomization. Once we observe $\{Z_i\}_{i=1}^{n_0}$, arms are randomized adaptively using RiTS to the $(n_0+1)$-th participant onward. Odd and even-indexed $Z_i$ are placed in $\mc{D}^{trn}_n$ and $\mc{D}^{eval}_n$ respectively, i.e., $\mc{D}^{trn}_{n}:=(Z_{1}, Z_{3}, \ldots, Z_{\ceil{n/2}})$ and $\mc{D}^{eval}_n=(Z_{2}, Z_{4}, \ldots, Z_{\floor{n/2}})$. Denote $N' := |\mc{D}^{trn}_{n}| = \ceil{n/2}$ and $N := |\mc{D}^{eval}_{n}| = \floor{n/2}$. Corresponding sets of indicators are defined as $\calI^{trn}_n$ and $\calI^{eval}_n$, i.e., $i \in \calI^{trn}_n$ if and only if $Z_i \in \mc{D}^{trn}_n$. With $\{Z_{i}\}_{i=1}^{n}$, we construct $(1-\alpha)$ AsympCS $\{\mc{C}_{n}^{(\alpha)}(a)\}_{n \geq 1}$ for $\Delta(a)$ using the four steps below:
\begin{enumerate}
    \item [(Step 1)]\textbf{Regression:} Obtain $\hat{\mu}_{N'}^{(a)}(\mbi{x})$ to estimate $\E(R|A=a, \mbi{X}=\mbi{x})$ using $\mc{D}^{trn}_n \ \forall 
     a \in [K]$.
    \item [(Step 2)]\textbf{Pseudo-outcomes:} For each $i \in \mc{I}^{eval}_n$, compute the Augmented Inverse Propensity Weighted (AIPW) pseudo-outcomes for efficacy endpoint: \begin{equation*}
        \hat{g}_{N'}(Z_{i}; a) := \hat{\mu}_{N'}^{(a)}(\mbi{X}_i) + \frac{\indic\{A_i = a\}}{q_i(a, \mbi{X}_i)} \secondbigg{ R_i - \hat{\mu}_{N'}^{(A_i)}(\mbi{X}_i) }, \quad \forall a \in [K],
    \end{equation*} and pseudo-contrasts $\hat{f}_{N'}(Z_{i}; a) = \hat{g}_{N'}(Z_{i}; a) - \hat{g}_{N'}(Z_{i}; 1)$ for $a \in [K]/\{1\}$. 

    \item [(Step 3)]\textbf{Cross-fitting:} Steps 1--2 are replicated by switching the roles of $\mc{D}^{trn}_n$ and $\mc{D}^{eval}_n$ to obtain a tuple of pseudo-contrasts for all observed efficacy endpoints, i.e., for each $R_i$, generate $\{ \hat{f}_{N'}(Z_{i}; a) \}_{a=2}^{K}$ or $\{ \hat{f}_{N}(Z_{i}; a) \}_{a=2}^{K}$ depending on $i \in \calI^{eval}_n$ or $\calI^{trn}_n$.
    
    \item [(Step 4)]\textbf{Construction of $\mc{C}_{n}^{(\alpha)}(a)$:} Let $\hat{\Delta}_{n}^{\times}(a) := \big( \sum_{i\in \calI^{eval}_n} \hat{f}_{N'}(Z_{i}; a) + \sum_{i\in \calI^{trn}_n} \hat{f}_{N}(Z_{i}; a) \big)/n$ and $\hat{\sigma}_{n}^{2}(a) := \big( \hat{\var}_{N}(\hat{f}_{N'}; a) + \hat{\var}_{N'}(\hat{f}_{N};a)\big)/2$, where $\hat{\var}_{N}(\hat{f}_{N'}; a)$ and $\hat{\var}_{N'}(\hat{f}_{N}; a)$ are the sample variances of $\{\hat{f}_{N'}(Z_{i}; a): i\in \calI^{eval}_n \}$ and $\{ \hat{f}_{N}(Z_{i}; a): i\in \calI^{trn}_n \}$, respectively. Obtain the $n$-th interval of the AsympCS: \begin{equation} \label{eq:asympcs_n}
        \mc{C}_{n}^{(\alpha)}(a) := \hat{\Delta}_{n}^{\times}(a) \pm \sqrt{{2(n \rho^2 \hat{\sigma}_{n}^{2}(a) +1) \log\firstbigg{\frac{\sqrt{n\rho^2 \hat{\sigma}_{n}^{2}(a) +1}}{\alpha}}}/{n^2\rho^2}}.
    \end{equation} $\rho > 0$ is a tuning parameter and the asymptotic time-uniform guarantee of $\{\mc{C}_{n}^{(\alpha)}(a)\}_{n\geq m}$ holds for any $\rho > 0$, provided the burn-in sample size $m$ is sufficiently large. We follow the recommendation of \cite{waudby-smith_et_al_2024+} and fix $\rho = \rho_m := \sqrt{\frac{-2\log \alpha + \log(-2\log \alpha)+1}{\hat{\sigma}_m(a) m \log(\max\{m, e\})}}$ where $e$ is the natural base. The choice of $\rho = \rho_m$ ensures $\mc{C}_{m}^{(\alpha)}(a)$ is the narrowest.
\end{enumerate} 
Our proposed method is summarized in Algorithm \ref{alg:prop_algo}. For implementation, we use a weighted ridge regression estimator $\hat{\mu}_{N'}^{(a)}(\mbi{x}) := \hat{\beta}_{a,0} + \mbi{x}^T \hat{\mbi{\beta}}$ in Step 1, where for tuning parameter $\lambda > 0$, \begin{equation} \label{eq:ipw_trial_eval}
    \{ \hat{\beta}_{a,0} \}, \ \hat{\mbi{\beta}} := \argmin_{ \{\beta_{a,0}\},\ \mbi{\beta}} \thirdbigg{\sum_{a\in [K]} \sum_{i\in \mathcal I^{trn}_n} \frac{\indic\{A_i = a\}}{q_i(a, \mbi{X}_i)} (R_i - \beta_{a,0} - \mbi{X}_{i}^{T} \mbi{\beta})^2 + \lambda \norm{\mbi{\beta}}_{2}^{2} }.
\end{equation} 
For simplicity, we fix $\lambda=10$ in all of our numerical experiments in Sections \ref{sec:simulation} and \ref{sec:real_data}. A more flexible model can be used, but requires a larger burn-in sample size. It is noteworthy that the inverse propensity weight adjustment is crucial for valid coverage of AsympCS when the model is possibly mis-specified and the data are adaptively collected; see Section \ref{sec:theory}. Regression estimators $\hat{\mu}_{N'}^{(a)}(\mbi{x})$ shares same slope parameter $\mbi{\beta}$ across arms. Doses with inferior utility are likely to be allocated to fewer subjects by \texttt{RiTS}. A regression estimator with same slope parameter $\mbi{\beta}$ allows information borrowing across arms, which ultimately leads to similar level of efficiency in estimating $\{\Delta(a)\}_{a=2}^{K}$, and isolating the best arm with confidence. An alternative is to fix $\mbi{\beta}=0.$ However, observed efficacy is itself heterogeneous in $\mbi{x}.$ A mis-specified estimator in $\hat{\mu}_{N'}^{(a)}(\mbi{x})$ helps to reduce variance and cumulative miscoverage for inferior arms. 
\begin{algorithm}[htpb]
\caption{Proposed Framework of RiTS }\label{alg:prop_algo}
    \begin{algorithmic}[1]
        \Require Data $\{\mbi{X}_n\}_{n\geq1}$; Priors $\pi(\mbi{\beta}_a), \ \pi(\mbi{\gamma}_a)$; Weight $w \in [0, 1]$; Number of initial participants to randomize $n_0$; 
        Minimum propensity for arms (clipping parameter) 
        $\delta\in (0,1/K)$; Number of burn-in samples $m$; Minimum clinically significant effect size $\tau>0$.
        
        \For{$n \leq n_0$} \Comment{Burn-in begins}
        \State Randomize arms; Set $q_{n}(a, \mbi{X}_n) = 1/K$ for all $a \in [K]$.
        \State Simulate $A_n \sim \text{Multinomial}\thirdbigg{\secondbigg{q_{n}(a, \mbi{X}_n)}_{a=1}^{K}}$ and assign arm $A_n$ to $n$-th participant.
        \EndFor \Comment{Burn-in ends}
        
        \For{$n \geq n_0+1$} \Comment{RiTS begins}
        \State Utilize $\mc{H}_{n-1}$ to update priors to posteriors $\{\pi^{(n)}(\mbi{\beta}_{a})\}_{a=1}^{K}$, $\{\pi^{(n)}(\mbi{\gamma}_{a})\}_{a=1}^{K}$.
        \State Obtain $\{q_{n}(a, \mbi{X}_n)\}_{a=1}^{K}$; Clip to ensure $q_n(a, \mbi{X}_n) \in [\delta, 1-\delta]$ for $a \in [K]$.
        \State Simulate $A_n \sim \text{Multinomial}\thirdbigg{\secondbigg{q_{n}(a, \mbi{X}_n)}_{a=1}^{K}}$ and assign arm $A_n$ to $n$-th participant.
        \If{$n \geq m$}
        \State Use \eqref{eq:asympcs_n} to obtain $\{\mc{C}_n^{(\alpha)}(a):\mc{C}_n^{(\alpha)}(a) := (L_n^{(\alpha)}(a), U_n^{(\alpha)}(a))\}_{a=2}^{K}$.
        \State Stop if (1) $\max_{a \in [K]/\{1\}} L_{n}^{(\alpha)}(a) > \tau,$ or (2) $\max_{a \in [K]/\{1\}} U_{n}^{(\alpha)}(a) \leq \tau.$
        \EndIf
        \EndFor \Comment{RiTS ends}

        \Return $\{(\mbi{X}_n, A_n, R_n, S_n)\}_{n \geq 1}$ and $\{\mc{C}_n^{(\alpha)}(a): a\in \{2, \ldots, K\}\}_{n\geq m}$.
    \end{algorithmic}
\end{algorithm}
All hyper-parameters in Algorithm 1 need to be chosen carefully so that AsympCS controls the cumulative miscoverage at a nominal level. Additional discussions on the choice of role of such hyperparameters are deferred to Web Appendix \ref{subsec:hyparam_asympcs}. Possibility of batched and delayed construction of AsympCS are also discussed in Web Appendix \ref{subsec:pract_cons}.

\subsection{Validity of AsympCS} \label{sec:theory}

The validity of AsympCS hinges upon Assumptions \ref{assump:unif_cons}--\ref{assump:moment}.
{For any function $f(\cdot)$ from $\mathbb R^d$ to $\mathbb R$, define $\| f\|_{L_2} := (\E_{\bm X} |f(\bm X)|^2)^{1/2}$, where $\E_X(\cdot)$ is taken with respect to $\bm X$.}

\begin{assumption} \label{assump:unif_cons}
    For $a \in [K]$, $\| \hat{\mu}_{n}^{(a)} - \Bar{\mu}^{(a)} \|_{L_2} = o(1)$ {a.s.}, for some function $\Bar{\mu}^{(a)}$.
\end{assumption}

\begin{assumption} \label{assump:moment}
    There exists $\eta > 0$ such that $\max_{a \in [K]} \|\hat{\mu}_{n}^{(a)}\|_{L_{2 +\eta}} \vee \|\Bar{\mu}^{(a)} \|_{L_{2 +\eta}} =O(1)$ a.s.
\end{assumption}

\begin{theorem} \label{thm:main_theorem}
For Algorithm \ref{alg:prop_algo}, the following statements are true:
\begin{itemize}
    \item Conditions (A2)--(A3) in Section \ref{subsec:setup} hold by design.
    \item Under (A1), if $\{\bm X_n\}_{n\geq 1}$ are sub-Gaussian and the estimator \eqref{eq:ipw_trial_eval} are used to construct $\{\mc{C}_n^{(\alpha)}(a)\}_{n \geq 1}$, then Assumptions \ref{assump:unif_cons}--\ref{assump:moment} hold.
    \item Under (A1), if Assumptions \ref{assump:unif_cons}--\ref{assump:moment} hold,
    then 
    $\{\mc{C}_n^{(\alpha)}(a)\}_{n \geq 1}$ is a $(1-\alpha)$-AsympCS for {$\Delta(a)$}. 
\end{itemize}
\end{theorem}

Theorem \ref{thm:main_theorem} only requires the regression estimators $\{\hat{\mu}^{(a)}_{n}\}_{a=1}^K$ converge uniformly to fixed functions $\{\Bar{\mu}^{(a)}\}_{a=1}^K$, which is not necessarily the true functions $\{\mu^{(a)}\}_{a=1}^K$. In practice, it is often difficult to model the efficacy and safety correctly. This flexibility makes our framework model-assisted: the parametric model \eqref{eq:lin_rwd} assists in trial optimization and estimation of effect sizes, but mis-specifying \eqref{eq:lin_rwd} does not void inferential results from AsympCS. Nevertheless, the existence of $\{\Bar{\mu}^{(a)}\}_{a=1}^K$ relies on the choice of estimator, especially when the model is mis-specified and the data are adaptively collected. The inverse propensity weighting in \eqref{eq:ipw_trial_eval} ensures the existence of $\{\Bar{\mu}^{(a)}\}_{a=1}^K$, while without this adjustment, $\{\Bar{\mu}^{(a)}\}_{a=1}^K$ may not exist at all.
Moreover, enough participants have to be allocated to each arm; otherwise, $\hat{\mu}^{(a)}_{n}$ may not be sufficiently convergent to $\Bar{\mu}^{(a)}$ uniformly. Fixing $n_0, \delta$, and $m$ to moderate values improves this uniform convergence with a relatively small sample size. Additional discussions on the burn-in sample size $m$ is deferred to Web Appendix \ref{subsec:burin_in}.

We also note that Theorem 6 in \cite{cook_mishler_ramdas_2024_clear} is not applicable here, as their proof does not incorporate estimation of the efficient influence function (EIF) using observed data.

\section{Simulation Studies} \label{sec:simulation}
The experiment is structured using the ADEMP setup \citep{morris_et_al_2019_stat_medicine}. A total of $n_{sim}=1000$ replications are performed on simulated datasets of size $n_{obs}=200$.

\paragraph{Aims} The best arm is defined as $a^* := \argmax_{a \in [K]/\{1\}} \Delta(a).$ We design simulation experiments to examine whether RiTS can isolate the best arm with confidence, and investigate the finite sample performance of estimates $\{\hat{\Delta}_{n}^{\times}(a)\}_{a=2}^{K}$, and AsympCSs $\{\mc{C}_{n}^{(\alpha)}(a): n \geq 1\}_{a=2}^{K}.$ For simplicity, we focus only on the efficacy endpoint and use ``winner dose'' and ``the most efficacious arm'' interchangeably. Depending on $w$, AsympCS for average utility can also be used instead to identify the winner dose for Phase III. We also assess whether RiTS can allocate better arms to more participants to maximize utility, compared to competitors.

\paragraph{Data-generating Mechanisms} Early-phase clinical trials typically involve a limited cohort of participants. Although many demographic covariates and assays are measured, drug developers generally focus on a small subset of participant-level covariates to assess heterogeneous dose responses. Key covariates are often chosen based on domain knowledge from previous studies. Following standard practice, we consider a single participant-level covariate, denoted by $Z$, which is simulated from the standard normal distribution, $Z \sim \mc{N}(0, 1)$.

We first consider a high signal-to-noise (High-SNR) scenario, where the true average efficacy as a function of $Z$ across each arm is fixed as follows, $\wt{\mu}_0(z, 1) = 2 - 0.01(z+1/2)^2 - 0.01(z-1/2)^2$, $\wt{\mu}_0(z, 2) = 2.7-0.2(z-1/2)^2-0.2(z+1/2)^2$, $\wt{\mu}_0(z, 3) = 2.7 - 0.01(z-1/2)^2 - 0.01(z+1/2)^2$ and $\wt{\mu}_0(z, 4) = 3.2 - 0.2(z-1/2)^2 - 0.2(z+1/2)^2$. True effect sizes are $\mbi{\Delta} := (\Delta(2), \Delta(3), \Delta(4)) = (0.225, 0.475, 0.725)$. The overall effects are calculated by taking expectations of $\wt{\mu}_0(Z, a)$ with respect to $Z\sim \mathcal N(0,1)$. The true average safety across each arm is also functions of one-dimensional variable with $\wt{\nu}_0(z, 1) = 2$, $\wt{\nu}_0(z, 2) = 2 - 0.01 z^2$, $\wt{\nu}_0(z, 3) = 2 - 0.1 z^2$ and $\wt{\nu}_0(z, 4) = 2 - 0.6 z^2$. Define $\mu_0(\mbi{x}, a) := \wt{\mu}_0(z, a)$, where $\mbi{x}=(z, z^2)^T$ is the covariate for a participant. Therefore, the true generating distributions for efficacy and safety are as follows: $R_i = \mu_0(\mbi{X}_i, A_i) + \epsilon_i$, $S_i = \nu_0(\mbi{X}_i, A_i) + \delta_i$, where $\epsilon_i, \delta_i \sim \mathcal N(0,1)$ are independent. Next, a low signal-to-noise (Low-SNR) setup is also considered where $\mu_0(\cdot, \cdot)$ and $\nu_0(\cdot, \cdot)$ are replaced by $\mu_0(\cdot, \cdot)/2$ and $\nu_0(\cdot, \cdot)/2$, respectively. Effect sizes are $\mbi{\Delta}=(0.1125, 0.2375, 0.3625)$ in the Low-SNR setup. In order to fully characterize the operating characteristics of our proposed method, we include a Null-Efficacy setup, where all acive doses have placebo efficacy, i.e., $\mu_0(\cdot, a) = \mu_0(\cdot, 1)$ for all $a \in [K]/\{1\}.$ Safety functions remain unchanged from the High-SNR setup.

In the above setups, Arm 1 represents the placebo, while Arms 2--4 correspond to the drug's low, medium, and high doses, respectively. For simplicity, we define the average utility as a weighted combination of efficacy and safety with equal weights: $\upsilon_0(z, a) = (1/2) \mu_0(z, a) + (1/2) \nu_0(z, a)$. Arm 1 is the safest but has the lowest efficacy, resulting in the lowest utility. Arm 2 consistently exhibits lower efficacy, safety, and utility across all participants, and adaptive trials are expected to allocate very few participants to this arm. Dose-ranging trials often include a low-dose arm to better characterize the dose-response curve. Arm 2 serves this purpose here. Arm 4 provides greater efficacy than Arm 3 for a small group of participants with $z$ values near $0$, while Arm 3 is the overall optimal arm when considering both efficacy and safety. {However, efficacy is the primary criterion for selecting the winner dose for Phase III, and therefore, we expect to choose Arm 4 as the winner dose. Alternative criteria exist for selecting the best dose for Phase III, which consider both efficacy and safety. For simplicity, we consider the winner dose as the most efficacious arm in our simulation experiments.}

\paragraph{Estimand/target of Analysis} The primary target is to estimate $\{\Delta(a)\}_{a=2}^K$ and identify (with confidence) the winner dose $a^*$ as quickly as possible. The secondary target is to minimize cumulative regrets (to be defined later), which monitors the allocation of better arms in efficacy, safety, and utility. 

\paragraph{Methods} Performance is compared among two adaptive randomization methods: (1) \texttt{RiTS}, our proposed method with $w=0.5$, (2) \texttt{TS} that uses one sampler on efficacy endpoint and ignores safety, and a baseline of equal randomization (3) \texttt{Rand}, where participants receive one of the doses with equal probability. For each replication, we first simulate a dataset of size $n_{sim}$. \texttt{TS} and \texttt{RiTS} allocates a randomly selected arm to each one of the first $n_0$ participants. Since endpoints are expected to be observed with some delay, we consider a \texttt{delay} parameter and fix it to $10$. We utilize observations from the first $n_0-\texttt{delay}$ participant and utilize one of \texttt{Rand}, \texttt{TS} and \texttt{RiTS} to allocate an arm to the $(n_0+1)$-th participant onward. Such delay is continued for all future allocations, i.e., to allocate an arm to the $n$-th participant, we only utilize observations from the first $(n-\texttt{delay})$ participants. We choose $n_0=24$, $m=80$ and $\lambda=10$ in all experiments. Additional choices of $n_0 \in \{40, 56\}$ are considered in Web Appendix \ref{subsec:clip_sim} to assess the approximate time-uniform validity of AsympCS. Effect size $\Delta(a)$ are estimated using $\hat{\Delta}^{\times}_{a}(a)$. Estimates and confidence sequences constructed using AsympCS are referred as \texttt{RiTS-AIPW}, \texttt{TS-AIPW} and \texttt{Rand-AIPW}, based on the method of arm allocation. As a baseline, \texttt{Rand-OF} is used, where $\Delta(a)$ is estimated using difference-in-means for Arms $a$ and $1.$ Confidence sequences are obtained using the O'Brien-Fleming (OF) repeated confidence interval (CI) \citep{jennison_turnbull_1999_book}. Depending on the method of allocation, AsympCSs or OF intervals are used with a stopping criterion, where a trial is stopped under two scenarios: (1) once one of $K-1$ the confidence sequences is above $0.1$, which is denoted as the minimum clinically significant effect size, or (2) once all $K-1$ confidence sequences are below $0.1.$ Note that, two scenarios are mutually exclusive. The winner dose is chosen based on the highest estimated effect size in the first scenario. We however do not declare a winner dose in the second scenario of stopping criterion.

\paragraph{Performance Measures} To assess the quality of \texttt{AsympCS}, we monitor bias, root mean squared error (RMSE), width, and cumulative miscoverage rate. The cumulative miscoverage rate for $\Delta(a)$ with burin-in sample size $m$ is defined as, $\lim_{m \to \infty} \P(\exists n \geq m: \Delta(a) \notin \mc{C}_{n}^{(\alpha)}(a)).$ For a finite sample of size $N$, we approximate it as, $\P(\exists n \in [m, N]: \Delta(a) \notin \mc{C}_{n}^{(\alpha)}(a)).$ We also include metrics such as the proportion of replications where Arm 4 has the highest estimated effect size (`Winner (Arm 4)' or \texttt{WA4}, hereafter), and where the stopping criterion is met (`Stopping Criteria' or \texttt{SC}, hereafter). Note that \texttt{WA4} only uses point estimates, whereas \texttt{SC} only uses bounds of \texttt{AsympCS}. For our secondary target, cumulative regrets are monitored, defined as follows. For the $i$-th participant with covariate $\mbi{x}_i$, define the optimal arm based on utility, efficacy, and safety, respectively, as $a_i^* := \argmax_{a \in \mc{A}} \mbi{x}_i^T [w \mbi{\beta}_a^* + (1-w) \mbi{\gamma}_a^*]$, $a_{i}^{(e)*} := \argmax_{a \in \mc{A}} \mbi{x}_i^T \mbi{\beta}_a^*$ and $a_{i}^{(s)*} := \argmax_{a \in \mc{A}} \mbi{x}_i^T \mbi{\gamma}_a^*$. Corresponding cumulative regrets are, $R(n) := \sum_{i=1}^{n} \big(\theta_{a_{i}^*, i} - \theta_{a_{i}, i}\big)$, $R^{(e)}(n) := \sum_{i=1}^{n} \big(\theta_{a_{i}^{(e)*}, i}^{(e)} - \theta_{a_{i}, i}^{(e)} \big)$ and $R^{(s)}(n) := \sum_{i=1}^{n} \big(\theta_{a_{i}^{(s)*}, i}^{(s)} - \theta_{a_{i}, i}^{(s)}\big)$, where $\theta_{a, i} := \mbi{x}_i^T [w \mbi{\beta}_a^* + (1-w) \mbi{\gamma}_a^*]$, $\theta_{a, i}^{(e)} := \mbi{x}_i^T \mbi{\beta}_a^*$ and $\theta_{a, i}^{(s)} := \mbi{x}_i^T \mbi{\gamma}_a^*$.
\begin{figure}
    \centering
    \includegraphics[width=\linewidth]{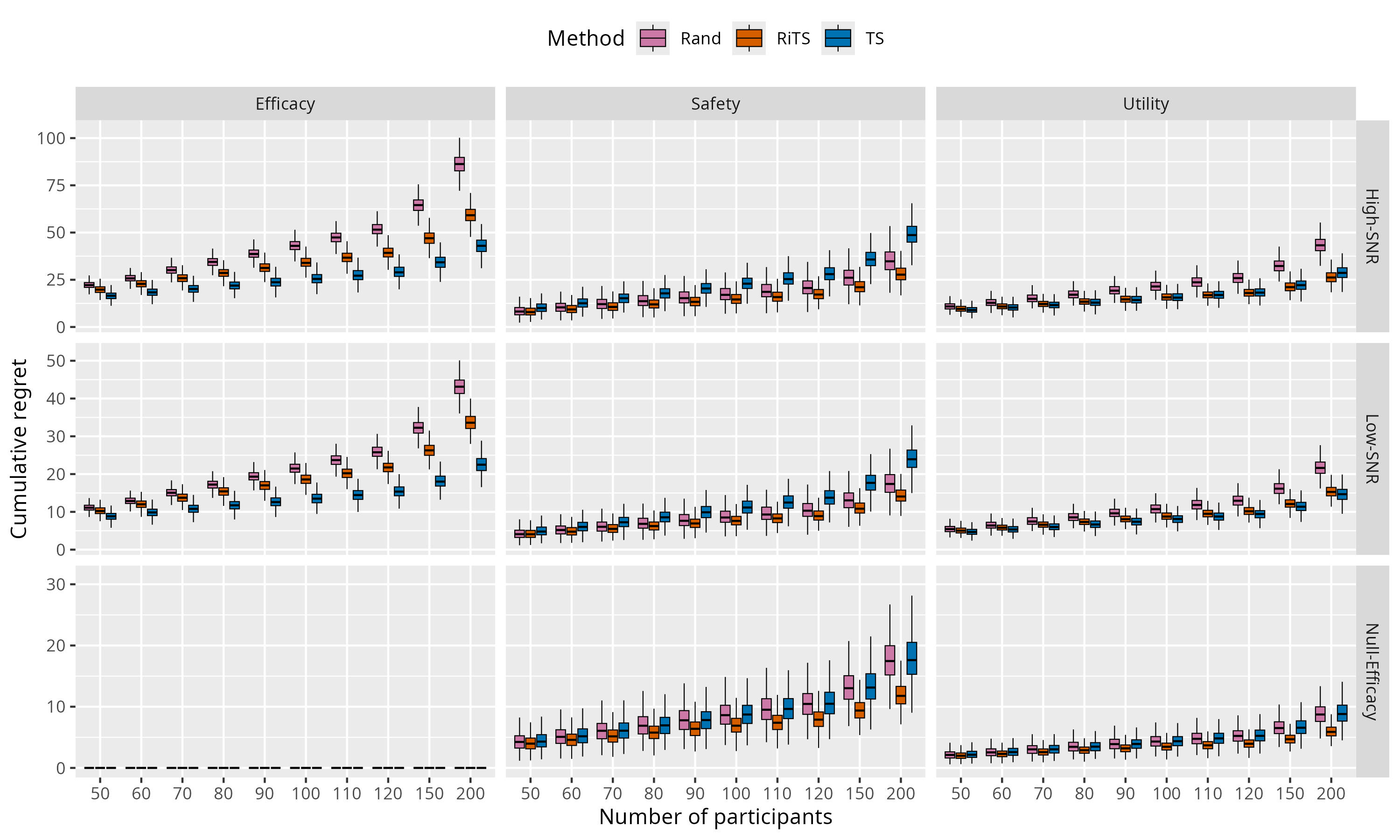}
    \caption{Box plots of cumulative regret based on three different criteria: utility, efficacy, and safety, at different stages of trials across $n_{sim}$ replications. Panels are arranged in columns based on three criteria, and two different rows correspond to the High-SNR and Low-SNR setups. The y-limits are set at different ranges across two different rows in order to compare cumulative regrets within each data-generating setup. A lower value of regret indicates better performance. }
    \label{fig:regret_sim_bwplot}
\end{figure}

\paragraph{Results} {Figure \ref{fig:regret_sim_bwplot} presents three different cumulative regrets incurred by all methods. Variability of regrets is similar within each data-generating setup, up to a different scaling.} {Lower regret is better as it indicates allocation of better treatments.} {\texttt{Rand} naturally has high cumulative regrets in all scenarios.} \texttt{TS} optimizes arm allocation based solely on efficacy, resulting in the lowest cumulative efficacy regret, but an even higher cumulative safety regret than \texttt{Rand}. \texttt{RiTS} achieves the lowest cumulative safety regret and comparable utility regret as compared to \texttt{TS}. Although \texttt{RiTS} incurs higher efficacy regret than \texttt{TS}. Appendix A.3 also compares methods when functions $\{\mu^{(a)}(\cdot)\}_{a=1}^{K}, \ \{\nu^{(a)}(\cdot)\}_{a=1}^{K}$ in \eqref{eq:lin_rwd} are mis-specified.

\begin{figure}[htpb]
    \centering
    \includegraphics[width=\linewidth]{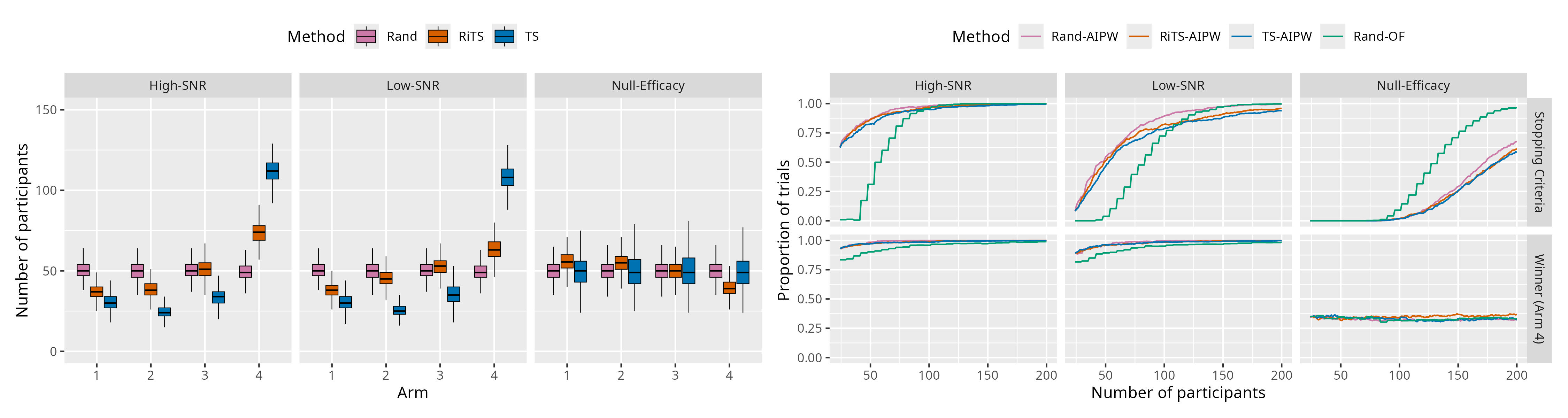}
    \caption{\textit{Left:} Box plot for the frequency of arm allocations by different methods across
    $n_{sim}$ replications. \textit{Right:} Panels in the first row present the proportion of replication where the stopping criteria are met across different stages of the trial. Panels in the second row depict the proportion replications where Arm 4 has the highest estimated effect size.}
    \label{fig:metric_alloc_plot}
\end{figure}

Left panels in Figure \ref{fig:metric_alloc_plot} depict frequency of arms allocated by different methods. \texttt{TS} allocates Arm 4 to more participants in the majority of trials, disregarding its lower safety profile. In contrast, \texttt{RiTS} allocates many participants to Arm 4 but also prioritizes Arm 3 for many due to Arm 3’s higher utility. Right panels of Figure \ref{fig:metric_alloc_plot} summarize performance based on two metrics introduced before. For both \texttt{SC} and \texttt{WA4}, a higher value indicates better performance. \texttt{RiTS-AIPW} meets expectation and performs the best in identifying Arm 4 as The winner dose (i.e., \texttt{WA4} metric) for both SNR setups. Both \texttt{Rand-AIPW} and \texttt{TS-AIPW} demonstrate comparable performance in terms of \texttt{WA4}. \texttt{Rand-OF} performance is slightly worse than rest of the methods at the beginning, whereas it's performance is comparable to other methods at the end stages. During earlier stages, \texttt{RiTS-AIPW} shows comparable performance with \texttt{Rand-AIPW} to stop trials early with some confidence (i.e., \texttt{SC} metric). During later stages, \texttt{Rand-AIPW} leads to the largest number of abruptly stopped trials. \texttt{RiTS-AIPW} fares comparably with \texttt{Rand-AIPW} in High-SNR. For Low-SNR, \texttt{RiTS-AIPW} shows a slight drop in performance. \texttt{TS-AIPW}'s performance is comparable with \texttt{RiTS-AIPW} and \texttt{Rand-AIPW} in High-SNR. However, for Low-SNR, \texttt{TS-AIPW} leads to fewer abruptly stopped trials. \texttt{Rand-OF} leads to the least number of stops at the initial stages, and shows comparable performance with \texttt{Rand-AIPW} at later stages for both High and Low SNR. However, Web Figure 3 suggests that the cumulative miscoverage rate is not under control for \texttt{Rand-OF} at the end of trials. The anti-conservativeness of \texttt{Rand-OF} may be due to the error incurred in the variance estimation of difference-in-means estimator. In order to gain further insights, we investigate performances based on finer metrics such as width, bias, root mean squared error (RMSE), and cumulative miscoverage of \texttt{AsympCS}. Web Appendices A.1 and A.2 include an elaborate discussion on this direction. We summarize our findings: \texttt{Rand-AIPW} leads to the narrowest confidence sequences (CS) with valid time-uniform coverage. \texttt{RiTS-AIPW} yields slightly wider CS compared to \texttt{Rand-AIPW}. CSs from \texttt{TS-AIPW} are the widest, except for $\Delta(4)$, where widths of \texttt{TS-AIPW} are comparable to \texttt{RiTS-AIPW}. Narrower CSs lead to more abruptly stopped trials, which explains our previous discussions on the performance of these methods based on \texttt{SC} metric. In Null-Efficacy, \texttt{Rand-OF} leads to shorter trials, where rest of the methods require longer trials, showing conservative nature of AsympCS under no treatment effect.
\begin{table}[ht]
\centering
\resizebox{\textwidth}{!}{%
\begin{tabular}{c ccccccccccccc}
  \hline
  & & \multicolumn{4}{c}{Arm 2 - Arm 1} & \multicolumn{4}{c}{Arm 3 - Arm 1} & \multicolumn{4}{c}{Arm 4 - Arm 1} \\
  & & \texttt{Rand-OF} & \texttt{Rand-AIPW} & \texttt{TS-AIPW} & \texttt{RiTS-AIPW} & \texttt{Rand-OF} & \texttt{Rand-AIPW} & \texttt{TS-AIPW} & \texttt{RiTS-AIPW} & \texttt{Rand-OF} & \texttt{Rand-AIPW} & \texttt{TS-AIPW} & \texttt{RiTS-AIPW} \\ 
    \hline
  \multirow{6}{*}{High-SNR} & Bias & -0.001 & 0.024 & 0.018 & 0.019 & 0.005 & 0.022 & 0.018 & 0.021 & 0.015 & 0.036 & 0.038 & 0.036 \\
  & RMSE & 0.159 & 0.157 & 0.164 & 0.157 & 0.087 & 0.112 & 0.119 & 0.113 & 0.162 & 0.141 & 0.141 & 0.142 \\
  & Avg. Width & 2.032 & 1.202 & 1.223 & 1.202 & 1.063 & 0.934 & 0.941 & 0.936 & 1.742 & 1.131 & 1.116 & 1.126 \\ 
  & Avg. Stop-time & 59.082 & 29.385 & 30.983 & 29.999 &  - & - & - & - & - & - & - & - \\ 
  & Stop-time SD & 15.778 & 11.842 & 15.675 & 13.920 & - & - & - & - & - & - & - & - \\ 
  & Cum. Miscoverage & $<1\%$ & $<1\%$ & $<1\%$ & $<1\%$ & $<1\%$ & $<1\%$ & $2\%$ & $2\%$ & $<1\%$ & $<1\%$ & $<1\%$ & $<1\%$ \\ 
  \multirow{6}{*}{Low-SNR} & Bias & 0.001 & 0.017 & 0.014 & 0.013 & 0.002 & 0.015 & 0.014 & 0.012 & 0.012 & 0.036 & 0.037 & 0.033 \\
  & RMSE & 0.072 & 0.078 & 0.090 & 0.082 & 0.046 & 0.062 & 0.068 & 0.064 & 0.070 & 0.080 & 0.082 & 0.080 \\ 
  & Avg. Width & 0.676 & 0.634 & 0.678 & 0.627 & 0.432 & 0.534 & 0.557 & 0.529 & 0.597 & 0.585 & 0.584 & 0.585 \\ 
  & Avg. Stop-time & 84.259 & 49.004 & 56.900 & 53.483 &  - & - & - & - & - & - & - & - \\ 
  & Stop-time SD & 25.438 & 25.851 & 35.461 & 32.349 &  - & - & - & - & - & - & - & - \\ 
  & Cum. Miscoverage & $1\%$ & $<1\%$ & $<1\%$ & $<1\%$ & $<1\%$ & $<1\%$ & $2\%$ & $2\%$ & $1\%$ & $<1\%$ & $1\%$ & $1\%$ \\ 
  \multirow{6}{*}{Null-Efficacy} & Bias & -0.004 & -0.002 & -0.003 & -0.004 & -0.004 & -0.003 & -0.004 & -0.005 & -0.005 & -0.003 & -0.004 \\ 
  & RMSE & 0.025 & 0.024 & 0.025 & 0.023 & 0.025 & 0.024 & 0.026 & 0.024 & 0.025 & 0.024 & 0.025 & 0.027 \\ 
  & Avg. Width & 0.171 & 0.190 & 0.193 & 0.183 & 0.170 & 0.189 & 0.192 & 0.185 & 0.170 & 0.189 & 0.192 & 0.196 \\ 
  & Avg. Stop-time & 133.117 & 164.680 & 169.535 & 168.199 &  - & - & - & - & - & - & - & - \\ 
  & Stop-time SD & 27.544 & 31.749 & 31.732 & 32.325 &  - & - & - & - & - & - & - & - \\
  & Cum. Miscoverage & $1\%$ & $<1\%$ & $1\%$ & $1\%$ & $4\%$ & $<1\%$ & $<1\%$ & $<1\%$ & $2\%$ & $<1\%$ & $<1\%$ & $<1\%$ \\ 
   \hline
\end{tabular}}
\caption{Average bias, RMSE, average width, stop-time, standard deviation (SD) of stop-time and cumulative miscoverage rate of estimated effect sizes based on different methods across 1000 replications at stop times. The first second and last six rows correspond to High-SNR, Low-SNR and Null-Efficacy setups. Our proposed method is \texttt{RiTS-AIPW}. Lower bias and RMSE indicate better estimation of effect sizes. Lower average width indicates more efficient sequential inference. Lower stop-time and stop-time SD indicates requirement of shorter trials to identify the winner dose. Stop-times are determined based on the confidence sequences of all active doses, and therefore included only once in the `Arm 2 - Arm 1' columns. Cumulative miscoverage rate is controlled at a nominal level of $5\%.$ }
\label{tab:bias_rmse}
\end{table}

Table \ref{tab:bias_rmse} presents the average bias, RMSE, average width, stop-time, standard deviation (SD) of stop-time and cumulative miscoverage rate of estimated effect sizes at stop-time. Bias is slightly higher for AsympCS based methods compared to \texttt{Rand-OF} in both High and Low SNR. Average stop-time is higher for \texttt{Rand-OF} compared to AsympCS based methods in High and Low SNR, and the opposite in Null-Efficacy. Web Appendix \ref{subsec:clip_sim} further explores the validity of \texttt{AsympCS} empirically, for different choices of hyperparameters. \texttt{Rand-AIPW} successfully controls cumulative miscoverage in all scenarios. For adaptive methods, burn-in sample size $m$, number of initially randomized participants $n_0$,  and clipping $\delta$ play a crucial role in controlling the cumulative miscoverage rate. We recommend fixing $\delta=0.1$ to encourage adaptive methods to explore sub-optimal arms sufficiently. A larger value of $m$ is also crucial to allow \texttt{AsympCS} to enter its \emph{asymptotic regime}. Web Tables 1 and 3 use $m=80$. Larger values of $n_0, m, \ \delta$ allow adaptive methods to allocate enough sample sizes across all arms, so that Assumption \ref{assump:unif_cons} is satisfied. \texttt{RiTS-AIPW} demonstrates nominal cumulative coverage when $\delta$ is set to $0.1$. In other scenarios, \texttt{TS-AIPW} fails to provide nominal cumulative coverage when adaptivity is higher, e.g., in High-SNR. \texttt{RiTS-AIPW} also fails in some of those cases, but its performance is significantly better than \texttt{TS-AIPW} in controlling the cumulative miscoverage.

We now summarize our insights from the experiments and provide our recommendations to implement this framework in practice: (1) \texttt{RiTS-AIPW} is a much safer alternative compared to \texttt{Rand-AIPW}, \texttt{TS-AIPW}, and \texttt{Rand-OF}. (2) Severely mis-specified trial optimizers can potentially impact allocation of better and safer treatments, so domain knowledge from earlier phases has to be incorporated to mitigate mis-specifications. (3) Setting $\delta= 0.05\  \text{or} \ 0.1$, $n_0=24 \ \text{or} \ 40$, and $m=80$ seems sensible in scenarios similar to our settings. (4) A large burn-in sample size $m$ is required for valid time-uniform coverage. During earlier stages, point estimates of \texttt{RiTS-AIPW} can be used safely. With enough burn-in sample size $m$, bounds in \texttt{RiTS-AIPW} can be used safely to stop trials abruptly with confidence. 

\section{Application to Phase IIb Trial for Treating Alopecia Areata} \label{sec:real_data}
Alopecia areata is an autoimmune disorder characterized by sudden hair loss that can affect any hair-bearing area of the body. This condition often has significant psychological and emotional impacts, leading to decreased self-esteem and quality of life. Finding a treatment for alopecia areata is essential, as current treatment options, such as corticosteroids and immunosuppressants, are inadequate. A drug targeting the underlying autoimmune mechanism could potentially restore hair growth,  improving patients' well-being and quality of life. Thus, continued research efforts toward developing effective treatments are crucial to addressing the unmet medical needs of individuals with alopecia areata.

We use a real dataset from a recently conducted Phase IIb trial for a drug to treat alopecia areata. The study is a randomized trial involving five doses of varying levels with a placebo control. We use the percentage change in SALT score from baseline as the efficacy endpoint, which measures the percentage of scalp hair loss.

\begin{figure}[htpb]
    \centering
    \includegraphics[width=\linewidth]{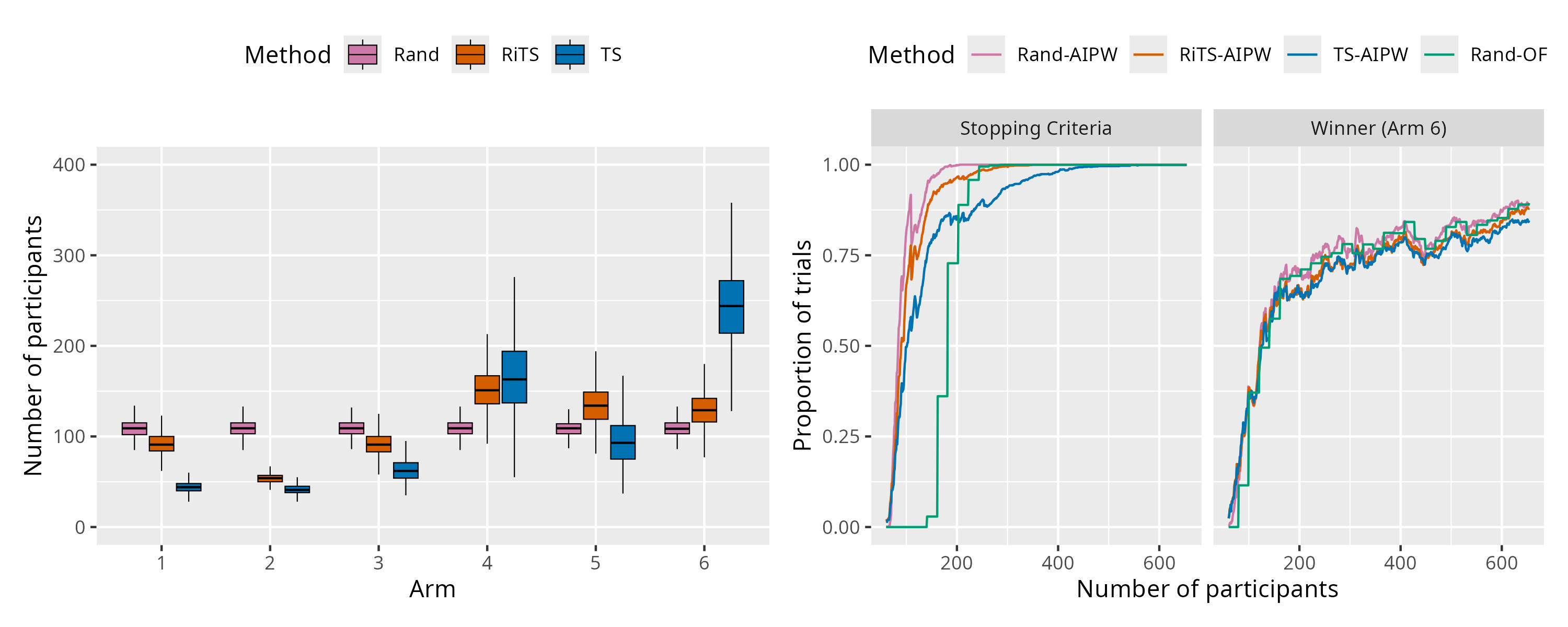}
    \caption{\textbf{Left panel:} Box plot for the frequency of arm allocations by different methods across
    $1000$ replications. \textbf{Right panel:} Left sub-figure presents the proportion of replication where the stopping criteria are met across different stages of the trial, with the minimum clinically significant effect size set to 0.1. The right sub-figure depicts the proportion of replications where Arm 6 has the highest estimated effect size.}
    \label{fig:metric_alloc_plot_real}
\end{figure}

For safety, we use the percentage change in lymphocyte counts in cells per microliter of blood from baseline. We consider the endpoints at Week 24, which is the optimal time for the drug to show its potential effects. Based on related research from the trial, we select baseline SALT score, baseline lymphocyte counts, age, BMI, albumin, and lymphocyte marker as potentially useful covariates to model the efficacy and safety endpoints.
A significant challenge in applying bandit algorithms to historical datasets is the lack of knowledge about counterfactual endpoints for doses not allocated to each participant. Since it is a randomized trial, we use random forests \citep{breiman2001random} to estimate counterfactual endpoints based on a comprehensive set of demographic covariates and baseline lab measurements. The dataset includes 654 participants. We perform 1000 replications, where each replication involves sampling 654 participants with replacement, followed by applying one of \texttt{Rand}, \texttt{TS}, and \texttt{RiTS}. Whenever the allocated arm from the bandit does not match the allocation, we use the counterfactual efficacy and safety; otherwise, we use the observed endpoint values. Initially, arms are randomized to $n_0=60$ participants. The clipping parameter $\delta$ is set to $0.05$. All other tuning parameters are the same as in Section \ref{sec:simulation}. Figure \ref{fig:metric_alloc_plot_real} compares and summarizes the performance of all methods.

Arm 1 is set as the placebo, and Arms 2--6 represent increasing dose levels. Average efficacy of Arms 1--6 are $-0.52$, $-0.63$, $-0.01$,  $0.31$,  $0.22$, and $0.44$, whereas average safety of Arms 1--6 are $0.46$, $0.13$, $-0.06$, $-0.11$, $-0.07$, and $-0.30$, respectively. With $w=0.5$, Arm 4 has the highest utility, and we expect \texttt{RiTS} (with $w=0.5$) to allocate more participants to Arm 4. From the left panel of Figure \ref{fig:metric_alloc_plot_real} we observe the following: \texttt{TS} is more aggressive than \texttt{RiTS}, allocating Arms 1 and 2 to almost no participants after initial exploration. \texttt{TS} prioritizes Arm 6, focusing on higher efficacy while disregarding the safety profile. \texttt{RiTS} continues to allocate Arm 1 or Arm 2 to participants, even at later stages. Notably, \texttt{RiTS} allocates Arm 6 to fewer participants and instead slightly favors Arms 4 and 5 compared to Arm 6 due to the former's better safety profiles. The right panel of Figure \ref{fig:metric_alloc_plot_real} highlights the benefit of \texttt{RiTS} over \texttt{TS} due to former's balanced allocation. \texttt{RiTS-AIPW} recovers Arm 6 as The winner dose more often in earlier stages. It also met the stopping criteria more often compared to \texttt{TS-AIPW}. Naturally, \texttt{Rand-AIPW} performs the best in both aspects due to an almost equal allocation of participants across all arms. Arm 4 was approved for commercial use in the study due to its favorable safety profile, while Arms 5 and 6 were considered potential candidates for approval. While efficacy is commonly prioritized for selecting the optimal dose, safety is also considered during the selection process. Construction of \texttt{AsympCS} can be naturally extended for average utility (for some $w$) instead of average efficacy. We focus on average efficacy for simplicity to illustrate our idea. 
Our empirical results align with the truth that Arm 6 was the most efficacious arm, while Arm 4 has the highest utility and therefore is favored most during adaptive randomization by \texttt{RiTS}.
Good empirical performance provides further evidence that \texttt{RiTS} is a safer and more practical option for applying contextual bandit in clinical trials, a key step towards balancing efficacy and safety.

\section{Discussion} \label{sec:discussion}

We propose a novel method of response adaptive randomization (RAR) called Risk-inclusive Thompson sampling (RiTS) in dose-ranging studies. Our method offers a flexible approach to balance efficacy and safety while leveraging participant-level covariates during arm allocation. Adaptive allocation over allocates superior arms, while under allocates inferior arms. Similar level of efficiency in estimating effect sizes helps to isolate the winner dose. Augmented Inverse Propensity Weighted pseudo-outcomes enable borrowing information across arms. To find the winner dose with confidence, we extend the Asymptotic Confidence Sequences \cite[AsympCS,][]{waudby-smith_et_al_2024+} for data generated using contextual bandits for performing time-uniform inference on effect sizes. Our approach is model-assisted: arm allocation relies on parametric assumptions; however, inference on effect sizes are robust to model mis-specifications. RiTS aids to personalized arm allocation by balancing efficacy and safety, at the cost of longer trials compared to traditional randomized controlled trials (RCTs). Both simulation experiments and application on a real-world dataset provide empirical evidence that while our method is safer compared to traditional randomized trials, it is less efficient compared to RCTs. 

There are several promising directions for future research. First, adaptive designs introduce additional operational challenges. For instance, we have explored a scenario of delayed response in our simulations. In the real data, we have access to endpoints across multiple follow-ups of each participant. It would be interesting to utilize observed endpoints in early follow-ups for adaptive allocation of arms. Moreover, RAR methods often unblind clinical trial staff to results of the trial due to strings of treatment assignments to one arm \citep{proschan_evans_2020}. Adding an independent trial statistician is one mitigation. Second, extending contextual bandits and AsympCS to accommodate different endpoint types, such as binary, count, or time-to-event, is of independent interest; see \cite{dumitrascu_et_al_2018_nips} and references therein. Third, asymptotic approximation in AsympCS requires an adequate burn-in sample size, which may be a key limitation in early-phase trials. Developing alternative anytime valid inferential strategies is of independent interest. Fourth, backfilling has recently proposed as an alternative strategy for dose-finding phase II trials. Existing methods proposes to use Bayesian RAR methods to backfill patients \citep{pin_et_al_2024_cct, zhao_et_al_2025_ct}, by prioritizing only on the efficacy endpoint. Our method can also be used to backfill patients by balancing efficacy and safety. Fifth, \texttt{RiTS} treats efficacy and safety outcome as independent end-points. However, joint modeling the composite outcome can improve treatment allocation \citep{chiaruttini_et_al_2025+}. Development of contextual bandit algorithm for composite outcome that also models the correlation of outcomes is of independent interest. The correlation between composite outcomes may provide more informed guidelines on how to choose the weights in a data-driven manner. Sixth, imputation of missing endpoints improved efficiency in estimation of effect sizes. More research is needed to investigate when to and when not to impute missing endpoints in our setting \citep{tackney_villar_2025_biometrics}.

A fundamental trade-off exists between efficiently estimating effect sizes and ensuring participant safety in clinical trials. ER focuses on estimation by random allocation of subjects to arms with near equal frequencies, but compromises trial safety, whereas nonrandomized (deterministic) procedures exploit the best treatment, while safer, but introduces bias. RAR methods, such as RiTS, aim to balance these extremes. Further research is needed to investigate how to balance these two extremes to conduct safer, shorter, and more efficient trials. Finally, we acknowledge that our method may not be suitable for dose-ranging studies with smaller sample size considered in this paper due to the requirement of burn-in sample size in AsympCS. Note that, while RiTS focuses on finding the winner dose, it also incorporate covariate information and safety end-points. Safety comes at the cost of longer trials. Therefore, we advocate for using classical RAR methods in shorter dose-ranging studies where patient safety is not a priority.




\section*{Acknowledgments}
Part of the research was conducted during Kanrar's 2024 summer internship at Pfizer. Authors thank Professor Philip Dixon for helpful comments on a previous version of this manuscript, and acknowledge the use of generative AI tools to assist in refining the language of the manuscript. Kanrar thanks the Laurence H. Baker Center for Bioinformatics \& Biological Statistics for partly supporting this research. Authors are grateful to the co-editor, associate editor and referees for their insightful comments and suggestions.

\section*{Supporting Information}

Reproducible \texttt{R} code for all experiments is made publicly available at \url{https://github.com/rohitkanrar/RiTS}. Web Appendices A-C are included at the end of this article. 

\vspace*{-8pt}

\newpage

\appendix

{\bf Supplementary Materials for ``Risk-inclusive Contextual Bandits for Early Phase Clinical Trials"
}
         \author{By Rohit Kanrar, Chunlin Li, Zara Ghodsi, and Margaret Gamalo}\\

This document includes supplementary materials to support the content in the main part of the manuscript titled ``Risk-inclusive Contextual
Bandits for Early Phase Clinical Trials.'' In what follows, the number of sections, figures, and tables (include a prefix `Web') with hyperlinks (colored {\color{violet} violet}) are included in this supplementary material. Otherwise, it corresponds to the main part of the manuscript. Web Appendix \ref{sec:add_empiri_res} includes additional empirical results, and Web Appendix \ref{sec:technical_results} outlines all the technical details, including the proof of Theorem 1. Additional discussions for implementations are included in Web Appendix \ref{sec:add_discuss}.

\section{Additional Empirical Results} \label{sec:add_empiri_res}

Web Appendix \ref{subsec:appendix_sim_performance_inference} includes additional discussion on the quality of AsympCS over 1000 replications. Web Appendix \ref{subsec:clip_sim} explores the impact of clipping and other hyperparameters on the cumulative miscoverage of AsympCS. Web Appendix \ref{subsec:sim_performance_mis} investigates the empirical performance of our proposed framework under misspecified models for both efficacy and safety.

\subsection{Quality of AsympCS in Trial Evaluation} \label{subsec:appendix_sim_performance_inference}
\begin{figure}[htpb]
    \centering
    \includegraphics[width=\textwidth]{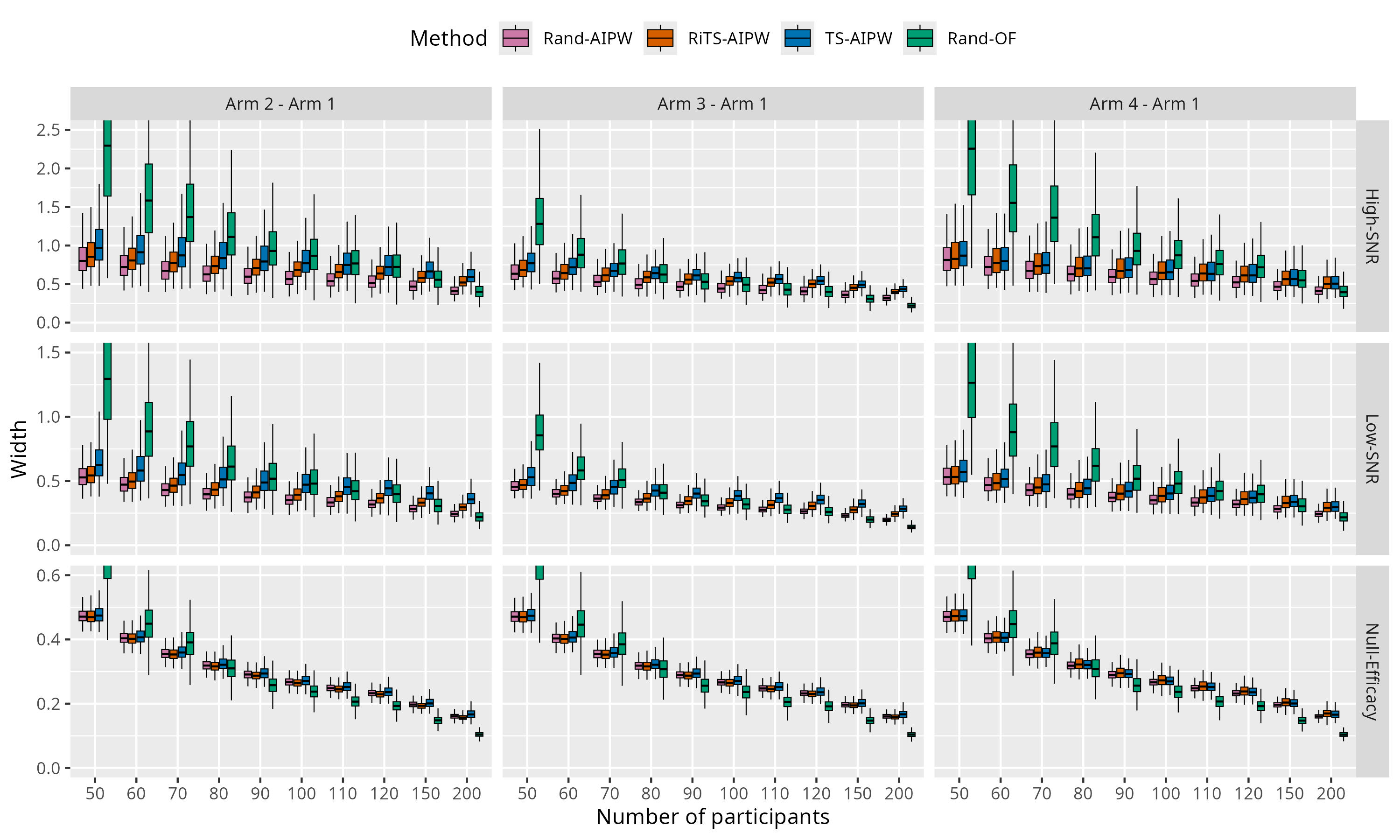}
    \caption{Box-plots of width for confidence intervals at different stages of the trial across 1000 replications. The top and bottom rows correspond to `High-SNR' and `Low-SNR' data-generating mechanisms, respectively. Three different columns of panels correspond to effect sizes of three active doses versus placebo, i.e., $\{\Delta(a)\}_{a=2}^{4}$. }
    \label{fig:bw_width_sim}
\end{figure}

In this section, we continue our investigation from the end of Section 3. Web Figure \ref{fig:bw_width_sim} shows the empirical width of AsympCSs for effect sizes at different stages of the trial across 1000 replications. Widths for \texttt{Rand-IPW}, \texttt{TS-IPW}, and \texttt{RiTS-IPW} are removed as these methods yield much larger confidence intervals. 

In all scenarios, the width decreases in the later stages of trials. Although, \texttt{Rand-T-test} yield the narrowest intervals, it suffers from inflated type-1 errors. See Web Appendix \ref{subsec:clip_sim} for more empirical evidence. Among other methods, \texttt{Rand-AIPW} leads to the narrowest intervals for $\{\Delta(a)\}_{a=2}^{4}$ across all stages of the trial. Note that equal randomization is the gold-standard design, and it is expected to yield the narrowest intervals. Therefore, narrower intervals obtained by \texttt{Rand-AIPW} are expected. \texttt{RiTS-AIPW} yield slightly wider intervals than \texttt{Rand-AIPW}, but narrower than \texttt{TS-AIPW} for $\Delta(2)$ and $\Delta(3)$. For $\Delta(4)$, widths for \texttt{RiTS-AIPW} are comparable to \texttt{TS-AIPW}.

\begin{figure}[htpb]
    \centering
    \includegraphics[width=\textwidth]{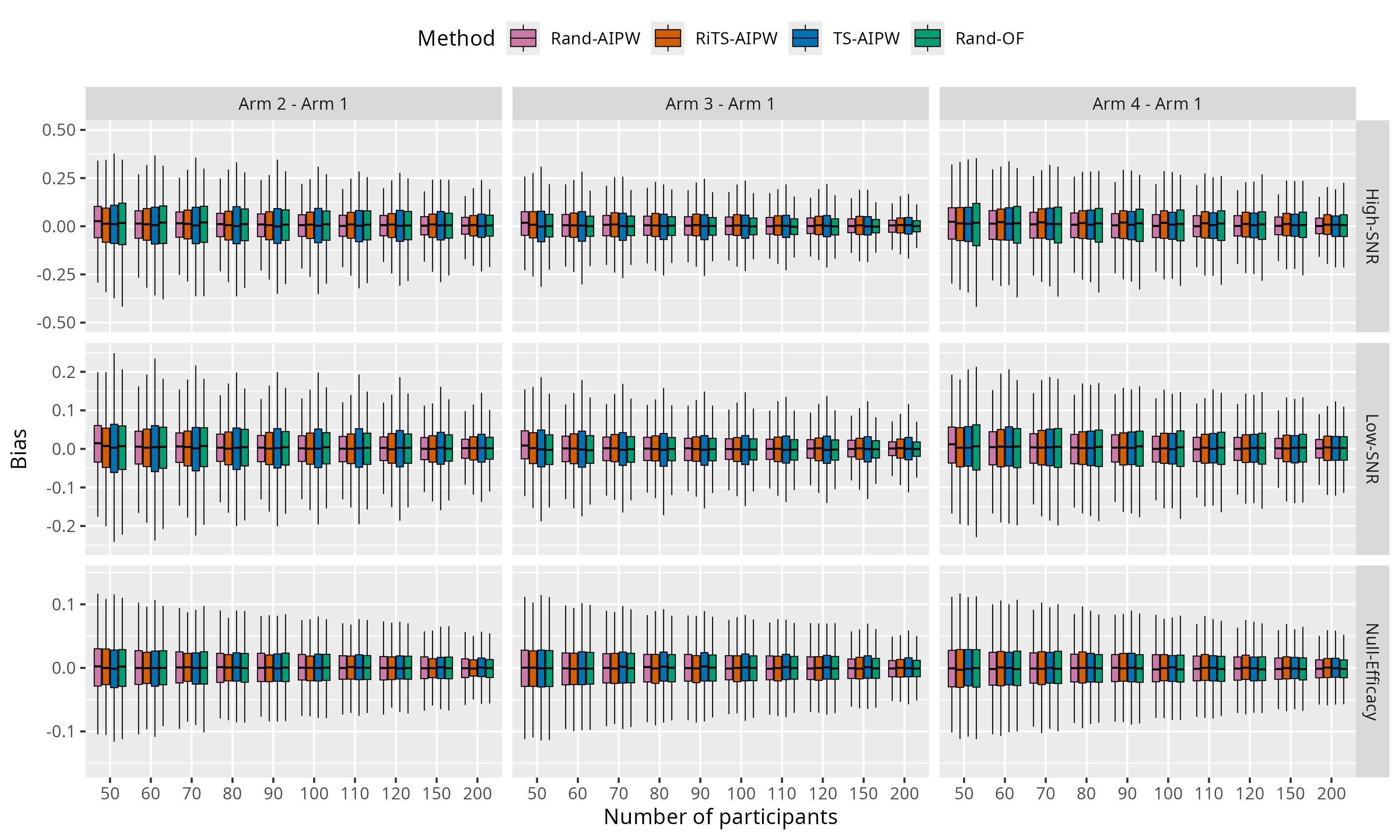}
    \caption{Box-plot depicting the bias of estimated effect sizes across 1000 replications at different stages of the trial. }
    \label{fig:bw_bias_sim}
\end{figure}

Now, we shift our focus to estimation errors for $\{\Delta(a)\}_{a=1}^{4}$. Web Figure \ref{fig:bw_bias_sim} presents box plots for bias in estimating $\{\Delta(a)\}_{a=1}^{4}$ using AsympCSs. Our observation is similar to the discussion after Table 1. \texttt{Rand-AIPW} has the lowest bias across all stages of trials. Performance of \texttt{RiTS-AIPW} is the closest to \texttt{Rand-AIPW}. Unlike previously, \texttt{Rand-T-test} can be used safely here as the point estimates do not suffer from multiple looks. Here \texttt{Rand-T-test} serves as a baseline and shows comparable performance with \texttt{Rand-AIPW} in later stages of trials. Interestingly, \texttt{Rand-AIPW} yields slightly narrower bounds compared to \texttt{Rand-T-test} in early stages. Augmented Inverse Propensity Weighted (AIPW) pseudo-outcomes enable \texttt{Rand-AIPW} to borrow information across arms using regression estimators, which ultimately leads to lower estimation error. This phenomenon further explains the better performance of \texttt{Rand-AIPW} compared to \texttt{Rand-T-test} based on the `Winner (Arm 4)' metric in Figure 4. Performance of \texttt{TS-AIPW} is comparable to \texttt{RiTS-AIPW} for $\Delta(4)$ (i.e., `Arm 4 - Arm 1'). However, \texttt{RiTS-AIPW} performs slightly better than \texttt{TS-AIPW} for $\Delta(2)$ and $\Delta(3)$, which ultimately leads to slightly better performance in terms of \texttt{RiTS-AIPW} compared to \texttt{TS-AIPW} (see Figure 4). Overall, we find that \texttt{Rand-AIPW} yields narrower intervals compared to \texttt{RiTS-AIPW}, however, \texttt{RiTS-AIPW} incorporates safety, which is ignored by \texttt{Rand-AIPW}.

\subsection{Impact of Clipping at Different Choices of Minimum Propensity} \label{subsec:clip_sim}
\begin{figure}[htpb]
    \centering
    \includegraphics[width=\linewidth]{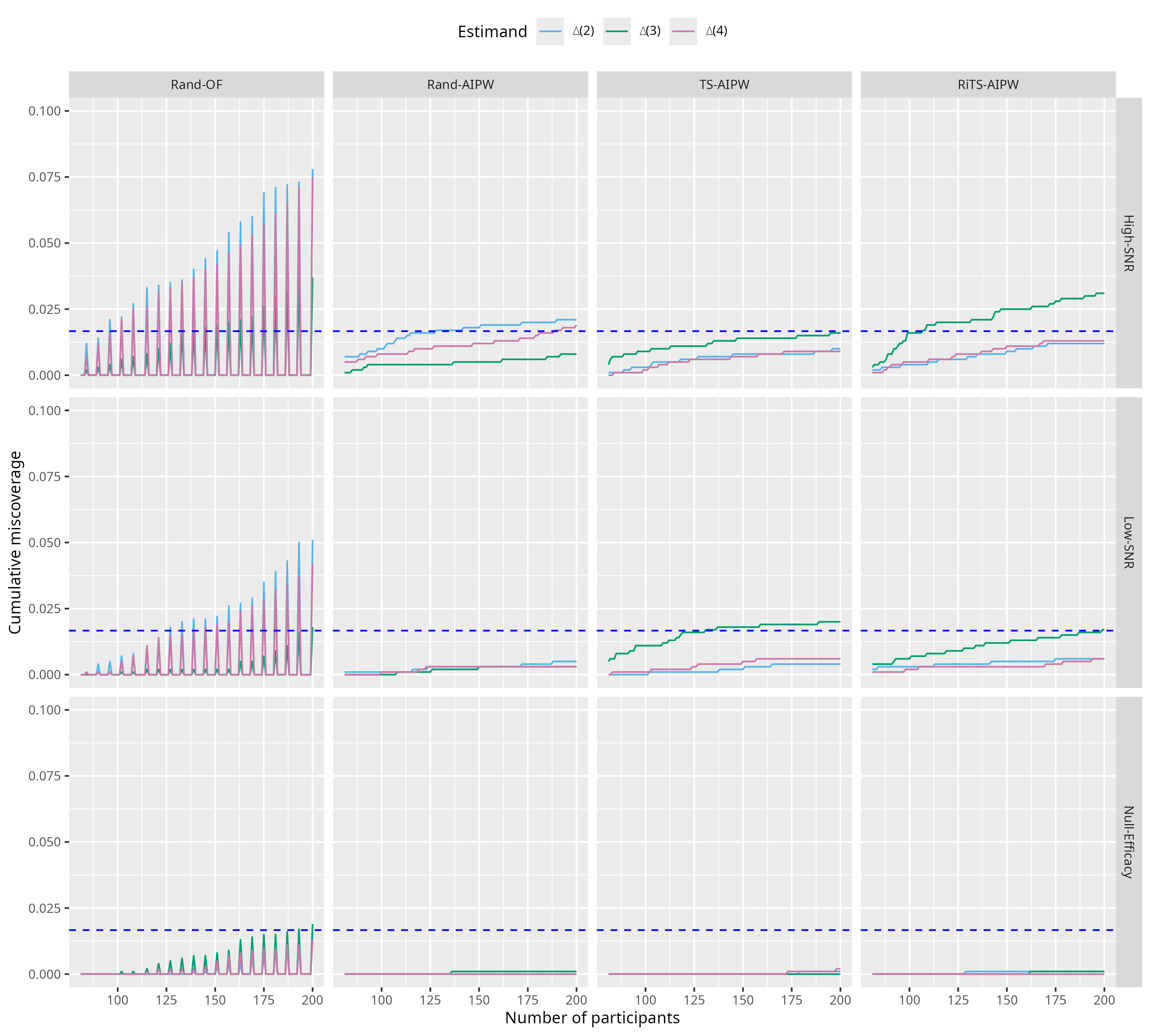}
    \caption{Cumulative miscoverage of \texttt{Rand-OF}, \texttt{Rand-AIPW}, \texttt{TS-AIPW} and \texttt{RiTS-AIPW} for $m=80$, $n_0=24$ and $\delta=0.1$, across different stages of 1000 replicated trials. The blue horizontal line indicates a nominal level of 0.0167. For \texttt{Rand-OF}, 30 equally spaced interim analyses (IA) are used, which leads to spikes if IA is performed.} 
    \label{fig:cum_miscov}
\end{figure}

In order to satisfy Fact \ref{assump:clipping}, it is necessary to restrict propensities within the interval $[\delta, 1-\delta]$. Here, we explore the choice of $\delta \in \{0.005, 0.01, 0.05, 0.1\}$ and its impact on the cumulative miscoverage. Cumulative miscoverage of $\{\mc{C}_{n}^{(\alpha)}(a)\}_{n \geq m}$ for $\Delta(a)$ is defined as the probability of $\mc{C}_{n}^{(\alpha)}(a)$ failing to cover $\Delta(a)$ for any sample size starting and beyond $m$. Empirical cumulative miscoverage approximates the probability using a large number of Monte Carlo replications. Web Figure \ref{fig:cum_miscov} presents the empirical cumulative miscoverage rate (in \%) for different methods across different stages of trials. We observe that \texttt{Rand-OF} fails to control it at a nominal rate of 1.67 \% after Bonferroni correction. \texttt{Rand-AIPW} controls it at the nominal level irrespective of the choices of hyperparameters. Both adaptive methods, i.e., \texttt{RiTS-AIPW} and \texttt{TS-AIPW} require $\delta$, $m$ and $n_0$ to be pre-fixed at large values. Even with such choices, \texttt{RiTS-AIPW} shows better coverage compared to \texttt{TS-AIPW}. This is due to a more balanced allocation of arms using \texttt{RiTS-AIPW}, as depicted in Figure 3, compared to \texttt{TS-AIPW}. Assumption \ref{assump:unif_conv_regression} fails when adaptive methods do not explore some sub-optimal arms sufficiently. Higher values of $\delta$ encourage exploration of sub-optimal arms even at later stages of trials. \texttt{RiTS} or \texttt{TS} with a higher value for $n_0$ loses its advantage of adaptive randomization over equal randomization. Note that a larger $m$ ensures the asymptotic approximation in \texttt{AsympCS}. However, even with a high value of $m$, \texttt{TS-AIPW} and \texttt{RiTS-AIPW} can potentially fail for extremely less explored sub-optimal arms. We propose to avoid such worst-case scenarios by fixing $\delta=0.1$ and $n_0\geq 24$. Web Figure \ref{fig:cum_miscov} presents cumulative miscoverage rate of \texttt{Rand-T-test}, \texttt{Rand-AIPW}, \texttt{TS-AIPW} and \texttt{RiTS-AIPW} for $m=80$, $n_0=24$ and $\delta=0.1$. We observe that \texttt{Rand-T-test} has severely inflated type-1 error. \texttt{Rand-AIPW} controls coverage for both SNRs. Adaptive methods show better coverage for Low-SNR compared to High-SNR, as increased exploration leads to a larger number of allocations in sub-optimal arms. Both \texttt{TS-AIPW} and \texttt{RiTS-AIPW} show comparable coverage with slightly under-coverage for Arm 3. Web Table \ref{tab:cum_miscov_tab_all} further explores the empirical cumulative miscoverage rate with $m=80$ for many choices of hyperparameters. We only include the cumulative miscoverage rate at the end of trials. Values around 1.67\% indicate nominal coverage. Based on Web Table \ref{tab:cum_miscov_tab_all}, we recommend fixing $\delta=0.1$ and $n_0=24$ in practical implementations. 
\begin{table}[htpb]
\centering
\resizebox{\textwidth}{!}{%
    \begin{tabular}{ccc cccc cccc cccc}
    \hline
    \multirow{2}{*}{\textbf{SNR Setup}} & \multirow{2}{*}{\textbf{$\delta$}} & \multirow{2}{*}{\textbf{$m$}} & \multicolumn{4}{c}{\textbf{Arm 2 - Arm 1}} & \multicolumn{4}{c}{\textbf{Arm 3 - Arm 1}} & \multicolumn{4}{c}{\textbf{Arm 4 - Arm 1}} \\
    \cline{4-15}
    & & & \texttt{Rand-OF} & \texttt{Rand-AIPW} & \texttt{TS-AIPW} & \texttt{RiTS-AIPW} & \texttt{Rand-OF} & \texttt{Rand-AIPW} & \texttt{TS-AIPW} & \texttt{RiTS-AIPW} & \texttt{Rand-OF} & \texttt{Rand-AIPW} & \texttt{TS-AIPW} & \texttt{RiTS-AIPW} \\
    \hline
    \multirow{12}{*}{\textbf{Low-SNR}} & \multirow{3}{*}{0.005} & 24 & 5.0 & 1.0 & 33.0 & $<0.1$ & 2.0 & $<0.1$ & 55.0 & 2.0 & 4.0 & $<0.1$ & 49.0 & $<0.1$ \\
    & & 40 & 5.0 & 1.0 & 33.0 & 1.0 & 2.0 & $<0.1$ & 56.0 & 1.0 & 4.0 & $<0.1$ & 47.0 & $<0.1$ \\
    & & 56 & 5.0 & 1.0 & 26.0 & 1.0 & 2.0 & $<0.1$ & 53.0 & 1.0 & 4.0 & $<0.1$ & 35.0 & $<0.1$ \\
    \cline{2-15}
    & \multirow{3}{*}{0.01} & 24 & 5.0 & 1.0 & 22.0 & $<0.1$ & 2.0 & $<0.1$ & 42.0 & 2.0 & 4.0 & $<0.1$ & 36.0 & $<0.1$ \\
    & & 40 & 5.0 & 1.0 & 20.0 & 1.0 & 2.0 & $<0.1$ & 43.0 & 1.0 & 4.0 & $<0.1$ & 35.0 & $<0.1$ \\
    & & 56 & 5.0 & 1.0 & 16.0 & 1.0 & 2.0 & $<0.1$ & 36.0 & 1.0 & 4.0 & $<0.1$ & 26.0 & $<0.1$ \\
    \cline{2-15}
    & \multirow{3}{*}{0.05} & 24 & 5.0 & 1.0 & 2.0 & 1.0 & 2.0 & $<0.1$ & 8.0 & 2.0 & 4.0 & $<0.1$ & 5.0 & $<0.1$ \\
    & & 40 & 5.0 & 1.0 & 2.0 & 1.0 & 2.0 & $<0.1$ & 7.0 & 1.0 & 4.0 & $<0.1$ & 3.0 & $<0.1$ \\
    & & 56 & 5.0 & 1.0 & 1.0 & $<0.1$ & 2.0 & $<0.1$ & 4.0 & 1.0 & 4.0 & $<0.1$ & 1.0 & $<0.1$ \\
    \cline{2-15}
    & \multirow{3}{*}{0.1} & 24 & 5.0 & 1.0 & $<0.1$ & 1.0 & 2.0 & $<0.1$ & 2.0 & 2.0 & 4.0 & $<0.1$ & 1.0 & 1.0 \\
    & & 40 & 5.0 & 1.0 & 1.0 & $<0.1$ & 2.0 & $<0.1$ & 1.0 & 1.0 & 4.0 & $<0.1$ & 1.0 & $<0.1$ \\
    & & 56 & 5.0 & 1.0 & 1.0 & $<0.1$ & 2.0 & $<0.1$ & 1.0 & 1.0 & 4.0 & $<0.1$ & $<0.1$ & 1.0 \\
    \hline
    \multirow{12}{*}{\textbf{High-SNR}} & \multirow{3}{*}{0.005} & 24 & 8.0 & 2.0 & 51.0 & 5.0 & 4.0 & 1.0 & 65.0 & 17.0 & 7.0 & 2.0 & 61.0 & 3.0 \\
    & & 40 & 8.0 & 2.0 & 43.0 & 3.0 & 4.0 & 1.0 & 63.0 & 16.0 & 7.0 & 2.0 & 57.0 & 3.0 \\
    & & 56 & 8.0 & 2.0 & 34.0 & 4.0 & 4.0 & 1.0 & 59.0 & 21.0 & 7.0 & 2.0 & 40.0 & 3.0 \\
    \cline{2-15}
    & \multirow{3}{*}{0.01} & 24 & 8.0 & 2.0 & 31.0 & 5.0 & 4.0 & 1.0 & 49.0 & 17.0 & 7.0 & 2.0 & 44.0 & 3.0 \\
    & & 40 & 8.0 & 2.0 & 26.0 & 3.0 & 4.0 & 1.0 & 48.0 & 15.0 & 7.0 & 2.0 & 39.0 & 3.0 \\
    & & 56 & 8.0 & 2.0 & 19.0 & 4.0 & 4.0 & 1.0 & 42.0 & 20.0 & 7.0 & 2.0 & 27.0 & 3.0 \\
    \cline{2-15}
    & \multirow{3}{*}{0.05} & 24 & 8.0 & 2.0 & 2.0 & 3.0 & 4.0 & 1.0 & 9.0 & 7.0 & 7.0 & 2.0 & 5.0 & 1.0 \\
    & & 40 & 8.0 & 2.0 & 1.0 & 1.0 & 4.0 & 1.0 & 8.0 & 7.0 & 7.0 & 2.0 & 4.0 & 1.0 \\
    & & 56 & 8.0 & 2.0 & 1.0 & 2.0 & 4.0 & 1.0 & 5.0 & 5.0 & 7.0 & 2.0 & 1.0 & 1.0 \\
    \cline{2-15}
    & \multirow{3}{*}{0.1} & 24 & 8.0 & 2.0 & 1.0 & 1.0 & 4.0 & 1.0 & 2.0 & 3.0 & 7.0 & 2.0 & 1.0 & 1.0 \\
    & & 40 & 8.0 & 2.0 & 1.0 & 1.0 & 4.0 & 1.0 & 2.0 & 1.0 & 7.0 & 2.0 & 1.0 & 1.0 \\
    & & 56 & 8.0 & 2.0 & 2.0 & 1.0 & 4.0 & 1.0 & 2.0 & 1.0 & 7.0 & 2.0 & 1.0 & 1.0 \\
    \hline
    \multirow{12}{*}{\textbf{Power}} & \multirow{3}{*}{0.005} & 24 & 100.0 & 95.0 & 91.0 & 96.0 & 100.0 & 100.0 & 100.0 & 100.0 & 100.0 & 100.0 & 100.0 & 100.0 \\
    & & 40 & 100.0 & 95.0 & 94.0 & 96.0 & 100.0 & 100.0 & 100.0 & 100.0 & 100.0 & 100.0 & 100.0 & 100.0 \\
    & & 56 & 100.0 & 95.0 & 92.0 & 96.0 & 100.0 & 100.0 & 99.0 & 100.0 & 100.0 & 100.0 & 99.0 & 100.0 \\
    \cline{2-15}
    & \multirow{3}{*}{0.01} & 24 & 100.0 & 95.0 & 91.0 & 96.0 & 100.0 & 100.0 & 100.0 & 100.0 & 100.0 & 100.0 & 100.0 & 100.0 \\
    & & 40 & 100.0 & 95.0 & 94.0 & 96.0 & 100.0 & 100.0 & 100.0 & 100.0 & 100.0 & 100.0 & 100.0 & 100.0 \\
    & & 56 & 100.0 & 95.0 & 91.0 & 96.0 & 100.0 & 100.0 & 100.0 & 100.0 & 100.0 & 100.0 & 100.0 & 100.0 \\
    \cline{2-15}
    & \multirow{3}{*}{0.05} & 24 & 100.0 & 95.0 & 91.0 & 96.0 & 100.0 & 100.0 & 100.0 & 100.0 & 100.0 & 100.0 & 100.0 & 100.0 \\
    & & 40 & 100.0 & 95.0 & 94.0 & 96.0 & 100.0 & 100.0 & 100.0 & 100.0 & 100.0 & 100.0 & 100.0 & 100.0 \\
    & & 56 & 100.0 & 95.0 & 92.0 & 97.0 & 100.0 & 100.0 & 100.0 & 100.0 & 100.0 & 100.0 & 100.0 & 100.0 \\
    \cline{2-15}
    & \multirow{3}{*}{0.1} & 24 & 100.0 & 95.0 & 92.0 & 96.0 & 100.0 & 100.0 & 100.0 & 100.0 & 100.0 & 100.0 & 100.0 & 100.0 \\
    & & 40 & 100.0 & 95.0 & 94.0 & 97.0 & 100.0 & 100.0 & 100.0 & 100.0 & 100.0 & 100.0 & 100.0 & 100.0 \\
    & & 56 & 100.0 & 95.0 & 93.0 & 97.0 & 100.0 & 100.0 & 100.0 & 100.0 & 100.0 & 100.0 & 100.0 & 100.0 \\
    \hline
    \end{tabular}
    }
    \caption{Cumulative miscoverage rate (in \%) at the end of trials ($n=200$) for \textit{correctly specified} trial optimizer. Each row corresponds to different cases with data generated from either High-SNR or Low-SNR setup, minimum propensity $\delta\in \{0.005, 0.01, 0.05, 0.1\}$ and the number of initial participants randomized $n_0 \in \{24, 40, 56\}$. Number of burn-in samples $m$ is set to 80 for \texttt{AIPW}-based AsympCSs. Cumulative miscoverage rate around $1.67 \ (= 0.05 \cdot 100/3)$ reflects nominal coverage. Last twelve rows depict the power at the nominal significance level.}
    \label{tab:cum_miscov_tab_all}
\end{table}

\subsection{Trial Optimization and Evaluation under a Misspecified Model} \label{subsec:sim_performance_mis}
In practice, the true model for the average efficacy and safety, as outlined in Section 4.1, is unknown. Majority of the existing methods in response adaptive randomization advocate on finding the best arm based on the frequency of subjects allocated to arms \citep{robertson_lee_et_al_2023_stat_science}. The arm which is allocated most frequently in an adaptive trial is likely to be the best dose. However, these method rely on parametric model assumptions, like \texttt{RiTS} and \texttt{TS}. We advocate for finding the best arm using AsympCS. The main purpose of this section is to investigate a scenario where both average efficacy and safety are severely mis-specified. Individual participants are not likely to receive more personalized dose. However, the use of AsympCS still enable us to find the best dose without relying on any parametric model assumptions. 

Let us consider a scenario where the average efficacy and safety are severely misspecified with $\mbi{x}_i =(1, z_i)^T$ (instead of $\mbi{x}_i =(1, z_i, z_i^2)^T$) as follows: \begin{align}
    \begin{split}
        r_i &= \Bar{\mu}_0(\mbi{x}_i, a_i) + \epsilon_i, \quad \text{where}, \Bar{\mu}_0(\mbi{x}, a) := \mbi{x}^T\mbi{\Bar{\beta}}_{a}^{*}, \ \epsilon_i \sim \mc{N}(0, \sigma_0^2), \\
        s_i &= \Bar{\nu}_0(\mbi{x}_i, a_i) + \delta_i, \quad \text{where}, \Bar{\nu}_0(\mbi{x}, a) := \mbi{x}^T\mbi{\Bar{\gamma}}_{a}^{*}, \ \delta_i \sim \mc{N}(0, \sigma_0^2),
    \end{split}
\end{align}  
\paragraph{Cumulative Regrets} Web Figure \ref{fig:regret_mis_sim_all} shows the cumulative regret based on utility, efficacy, and safety under model misspecification. Comparing Web Figure \ref{fig:regret_mis_sim_all} to Figure 2 for correctly specified models, we observe worse regret performance when models are misspecified. { Efficacy regret for \texttt{RiTS} is negatively impacted due to the model misspecification for both the High-SNR and Low-SNR setups. 

\begin{figure}[htpb]
    \centering
    \includegraphics[width=\textwidth]{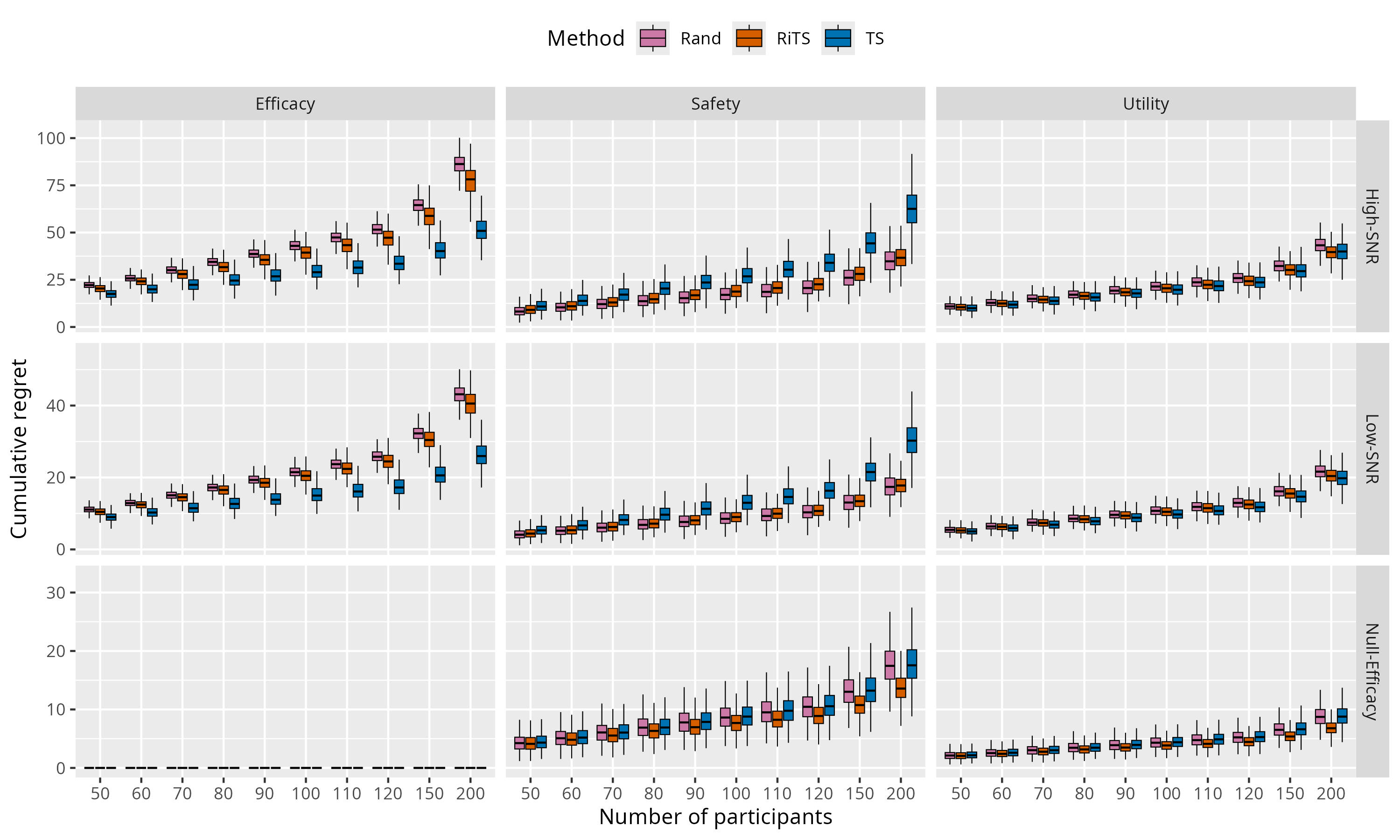}
    \caption{Box-plots of cumulative regret based on utility, efficacy, and safety at different stages of trials across 1000 replications, where both average efficacy and safety are misspecified. }
    \label{fig:regret_mis_sim_all}
\end{figure}

\paragraph{Estimation Error} For a misspecified model, Web Table \ref{tab:bias_rmse_mis} presents average bias and root mean squared error (RMSE) for estimating $\{\Delta(a)\}_{a=1}^{4}$ across 1000 replications at different stages of trials. {Bias is similar to the correctly specified models (see Table 1), with slightly higher RMSE. All other observations remain the same as outlined for the correctly specified model after Table 1 in Section 3.}

\begin{table}[htpb]
\centering
\resizebox{\textwidth}{!}{%
    \begin{tabular}{c c l llllllllll} 
    \hline
        Setup & Arm & Method & 50 & 60 & 70 & 80 & 90 & 100 & 110 & 120 & 150 & 200 \\ 
          \hline
        \multirow{12}{*}{High-SNR} & \multirow{4}{*}{Arm2} & \texttt{Rand-OF} & 0(0.17) & 0(0.15) & 0(0.14) & 0(0.13) & 0(0.12) & 0(0.12) & 0(0.11) & 0(0.11) & 0(0.1) & 0(0.08) \\ 
         &  & \texttt{Rand-AIPW} & 0.01(0.12) & 0(0.11) & 0(0.1) & 0(0.1) & 0(0.09) & 0(0.09) & 0(0.08) & 0(0.08) & 0(0.07) & 0(0.06) \\ 
         &  & \texttt{TS-AIPW} & 0(0.22) & -0.01(0.22) & -0.01(0.2) & -0.01(0.19) & -0.01(0.18) & -0.01(0.18) & -0.01(0.17) & -0.01(0.16) & -0.01(0.15) & -0.01(0.12) \\ 
         &  & \texttt{RiTS-AIPW} & 0(0.19) & 0(0.18) & 0(0.17) & 0(0.16) & 0(0.16) & 0(0.15) & 0(0.14) & 0(0.13) & 0(0.11) & 0(0.1) \\ 
        \cline{3-13}
         & \multirow{4}{*}{Arm3} & \texttt{Rand-OF} & 0(0.09) & 0(0.08) & 0(0.08) & 0(0.07) & 0(0.07) & 0(0.07) & 0(0.06) & 0(0.06) & 0(0.05) & 0(0.05) \\ 
         &  & \texttt{Rand-AIPW} & 0.01(0.09) & 0(0.09) & 0(0.08) & 0(0.08) & 0(0.07) & 0(0.07) & 0(0.06) & 0(0.06) & 0(0.05) & 0(0.05) \\ 
         &  & \texttt{TS-AIPW} & 0(0.15) & 0(0.14) & 0(0.13) & 0(0.12) & 0(0.11) & 0(0.11) & 0(0.1) & 0(0.1) & 0(0.09) & 0(0.07) \\ 
         &  & \texttt{RiTS-AIPW} & 0(0.13) & 0(0.11) & 0(0.1) & 0(0.1) & 0(0.09) & 0(0.09) & 0(0.08) & 0(0.08) & 0(0.07) & 0(0.06) \\ 
        \cline{3-13}
         & \multirow{4}{*}{Arm4} & \texttt{Rand-OF} & -0.01(0.17) & -0.01(0.15) & 0(0.14) & 0(0.13) & 0(0.13) & 0(0.12) & 0(0.12) & 0(0.11) & 0(0.1) & 0(0.09) \\ 
         &  & \texttt{Rand-AIPW} & 0.01(0.12) & 0(0.11) & 0(0.1) & 0(0.09) & 0(0.09) & 0(0.08) & 0(0.08) & 0(0.08) & 0(0.07) & 0(0.06) \\ 
         &  & \texttt{TS-AIPW} & 0.01(0.19) & 0.01(0.18) & 0.01(0.16) & 0(0.15) & 0.01(0.14) & 0.01(0.14) & 0(0.13) & 0(0.12) & 0(0.11) & 0(0.1) \\ 
         &  & \texttt{RiTS-AIPW} & -0.01(0.19) & 0(0.18) & 0(0.17) & 0(0.16) & 0(0.15) & 0(0.15) & 0(0.14) & 0(0.13) & 0(0.12) & 0(0.11) \\ 
        \hline
        \multirow{12}{*}{Low-SNR} & \multirow{4}{*}{Arm2} & \texttt{Rand-OF} & 0(0.09) & 0(0.08) & 0(0.08) & 0(0.07) & 0(0.07) & 0(0.07) & 0(0.06) & 0(0.06) & 0(0.05) & 0(0.04) \\ 
         &  & \texttt{Rand-AIPW} & 0.01(0.07) & 0(0.07) & 0(0.06) & 0(0.06) & 0(0.05) & 0(0.05) & 0(0.05) & 0(0.05) & 0(0.04) & 0(0.04) \\ 
         &  & \texttt{TS-AIPW} & 0(0.12) & 0(0.11) & 0(0.11) & 0(0.1) & 0(0.1) & 0(0.09) & 0(0.09) & 0(0.09) & 0(0.08) & 0(0.07) \\ 
         &  & \texttt{RiTS-AIPW} & 0(0.1) & 0(0.09) & 0(0.09) & 0(0.08) & 0(0.08) & 0(0.07) & 0(0.07) & 0(0.07) & 0(0.06) & 0(0.05) \\ 
        \cline{3-13}
         & \multirow{4}{*}{Arm3} & \texttt{Rand-OF} & 0(0.06) & 0(0.05) & 0(0.05) & 0(0.05) & 0(0.04) & 0(0.04) & 0(0.04) & 0(0.04) & 0(0.03) & 0(0.03) \\ 
         &  & \texttt{Rand-AIPW} & 0.01(0.06) & 0(0.05) & 0(0.05) & 0(0.04) & 0(0.04) & 0(0.04) & 0(0.04) & 0(0.04) & 0(0.03) & 0(0.03) \\ 
         &  & \texttt{TS-AIPW} & 0(0.08) & 0(0.08) & 0.01(0.07) & 0(0.07) & 0(0.07) & 0(0.07) & 0(0.06) & 0(0.06) & 0(0.05) & 0(0.04) \\ 
         &  & \texttt{RiTS-AIPW} & 0(0.07) & 0(0.06) & 0(0.06) & 0(0.06) & 0(0.05) & 0(0.05) & 0(0.05) & 0(0.04) & 0(0.04) & 0(0.03) \\ 
        \cline{3-13}
         & \multirow{4}{*}{Arm4} & \texttt{Rand-OF} & 0(0.09) & 0(0.08) & 0(0.08) & 0(0.07) & 0(0.07) & 0(0.07) & 0(0.06) & 0(0.06) & 0(0.05) & 0(0.05) \\ 
         &  & \texttt{Rand-AIPW} & 0.01(0.07) & 0(0.06) & 0(0.06) & 0(0.05) & 0(0.05) & 0(0.05) & 0(0.05) & 0(0.05) & 0(0.04) & 0(0.04) \\ 
         &  & \texttt{TS-AIPW} & 0.01(0.1) & 0.01(0.09) & 0.01(0.09) & 0.01(0.08) & 0.01(0.08) & 0(0.07) & 0(0.07) & 0(0.06) & 0(0.06) & 0(0.05) \\ 
         &  & \texttt{RiTS-AIPW} & 0(0.11) & 0(0.1) & 0(0.09) & 0(0.09) & 0(0.08) & 0(0.08) & 0(0.08) & 0(0.07) & 0(0.07) & 0(0.06) \\ 
    \hline
    \end{tabular}
    }
    \caption{Average bias (and RMSE) of estimated effect sizes based on different methods across 1000 replications for the \textit{misspecified model}. The first and last twelve rows correspond to High-SNR and Low-SNR setups. Different columns refer to the number of observations used to obtain estimates at different stages of trials. Our proposed method is \texttt{RiTS-AIPW}. }
\label{tab:bias_rmse_mis}
\end{table}

\paragraph{Cumulative Miscoverage} Similar to Web Table \ref{tab:cum_miscov_tab_all}, we include Web Table \ref{tab:cum_miscov_mis_tab_all}, which consists cumulative miscoverage for the misspecified model, at the end of trials across different choices of hyper-parameter, two data generating setup and three different estimands $\{\Delta(a)\}_{a=2}^{4}$. Here also, our findings are almost similar to Web Table \ref{tab:cum_miscov_tab_all}. 
If we have no knowledge about the correctness of model specification in the regression estimators, we recommend using $n_0=24$ with $\delta=0.1$ so that the cumulative miscoverage can be controlled at a nominal level for all active doses.

\begin{table}[htpb]
\centering
\resizebox{\textwidth}{!}{%
    \begin{tabular}{ccc cccc cccc cccc}
    \hline
    \multirow{2}{*}{\textbf{SNR Setup}} & \multirow{2}{*}{\textbf{$\delta$}} & \multirow{2}{*}{\textbf{$m$}} & \multicolumn{4}{c}{\textbf{Arm 2 - Arm 1}} & \multicolumn{4}{c}{\textbf{Arm 3 - Arm 1}} & \multicolumn{4}{c}{\textbf{Arm 4 - Arm 1}} \\
    \cline{4-15}
    & & & \texttt{Rand-OF} & \texttt{Rand-AIPW} & \texttt{TS-AIPW} & \texttt{RiTS-AIPW} & \texttt{Rand-OF} & \texttt{Rand-AIPW} & \texttt{TS-AIPW} & \texttt{RiTS-AIPW} & \texttt{Rand-OF} & \texttt{Rand-AIPW} & \texttt{TS-AIPW} & \texttt{RiTS-AIPW} \\
    \hline
    \multirow{12}{*}{\textbf{Low-SNR}} & \multirow{3}{*}{0.005} & 24 & 19.5 & 0.5 & 35.3 & 0.9 & 17.3 & 0.3 & 44.9 & 0.3 & 20.4 & 0.3 & 35.8 & 3.7 \\
    & & 40 & 19.5 & 0.5 & 30.0 & 1.0 & 17.3 & 0.3 & 41.0 & 0.3 & 20.4 & 0.3 & 27.7 & 3.5 \\
    & & 56 & 19.5 & 0.5 & 24.8 & 0.7 & 17.3 & 0.3 & 38.1 & 0.3 & 20.4 & 0.3 & 17.3 & 2.4 \\
    \cline{2-15}
    & \multirow{3}{*}{0.01} & 24 & 19.5 & 0.5 & 19.4 & 0.9 & 17.3 & 0.3 & 29.9 & 0.3 & 20.4 & 0.3 & 23.3 & 3.4 \\
    & & 40 & 19.5 & 0.5 & 17.9 & 1.0 & 17.3 & 0.3 & 28.2 & 0.3 & 20.4 & 0.3 & 18.3 & 3.1 \\
    & & 56 & 19.5 & 0.5 & 13.9 & 0.6 & 17.3 & 0.3 & 24.7 & 0.2 & 20.4 & 0.3 & 11.0 & 2.2 \\
    \cline{2-15}
    & \multirow{3}{*}{0.05} & 24 & 19.5 & 0.5 & 1.6 & 0.9 & 17.3 & 0.3 & 3.7 & 0.4 & 20.4 & 0.3 & 1.2 & 1.3 \\
    & & 40 & 19.5 & 0.5 & 1.9 & 1.0 & 17.3 & 0.3 & 3.1 & 0.3 & 20.4 & 0.3 & 0.9 & 1.5 \\
    & & 56 & 19.5 & 0.5 & 0.9 & 0.8 & 17.3 & 0.3 & 1.6 & 0.2 & 20.4 & 0.3 & 0.7 & 1.1 \\
    \cline{2-15}
    & \multirow{3}{*}{0.1} & 24 & 19.5 & 0.5 & 1.6 & 1.3 & 17.3 & 0.3 & 0.6 & 0.3 & 20.4 & 0.3 & 0.1 & 0.7 \\
    & & 40 & 19.5 & 0.5 & 1.4 & 1.2 & 17.3 & 0.3 & 0.9 & 0.4 & 20.4 & 0.3 & 0.3 & 0.8 \\
    & & 56 & 19.5 & 0.5 & 0.8 & 0.8 & 17.3 & 0.3 & 0.6 & 0.4 & 20.4 & 0.3 & 0.3 & 1.4 \\
    \hline
    \multirow{12}{*}{\textbf{High-SNR}} & \multirow{3}{*}{0.005} & 24 & 22.5 & 2.1 & 39.1 & 2.8 & 20.1 & 0.8 & 50.2 & 0.3 & 24.1 & 1.9 & 41.7 & 11.2 \\
    & & 40 & 22.5 & 2.1 & 36.7 & 2.1 & 20.1 & 0.8 & 48.9 & 0.6 & 24.1 & 1.9 & 29.8 & 8.2 \\
    & & 56 & 22.5 & 2.1 & 23.5 & 1.5 & 20.1 & 0.8 & 39.0 & $<0.1$ & 24.1 & 1.9 & 9.3 & 6.5 \\
    \cline{2-15}
    & \multirow{3}{*}{0.01} & 24 & 22.5 & 2.1 & 22.5 & 2.4 & 20.1 & 0.8 & 33.6 & 0.3 & 24.1 & 1.9 & 25.2 & 9.4 \\
    & & 40 & 22.5 & 2.1 & 19.9 & 2.0 & 20.1 & 0.8 & 31.3 & 0.7 & 24.1 & 1.9 & 17.1 & 6.7 \\
    & & 56 & 22.5 & 2.1 & 11.8 & 1.6 & 20.1 & 0.8 & 24.3 & $<0.1$ & 24.1 & 1.9 & 6.8 & 5.0 \\
    \cline{2-15}
    & \multirow{3}{*}{0.05} & 24 & 22.5 & 2.1 & 3.6 & 2.1 & 20.1 & 0.8 & 4.2 & 0.4 & 24.1 & 1.9 & 1.3 & 2.5 \\
    & & 40 & 22.5 & 2.1 & 2.1 & 2.0 & 20.1 & 0.8 & 3.9 & 0.4 & 24.1 & 1.9 & 1.1 & 2.4 \\
    & & 56 & 22.5 & 2.1 & 2.1 & 1.8 & 20.1 & 0.8 & 1.5 & $<0.1$ & 24.1 & 1.9 & 0.4 & 2.1 \\
    \cline{2-15}
    & \multirow{3}{*}{0.1} & 24 & 22.5 & 2.1 & 4.0 & 2.1 & 20.1 & 0.8 & 1.0 & 0.8 & 24.1 & 1.9 & 0.6 & 2.6 \\
    & & 40 & 22.5 & 2.1 & 4.0 & 2.9 & 20.1 & 0.8 & 0.8 & 0.3 & 24.1 & 1.9 & 1.3 & 1.9 \\
    & & 56 & 22.5 & 2.1 & 3.1 & 2.5 & 20.1 & 0.8 & 0.6 & 0.2 & 24.1 & 1.9 & 1.2 & 2.8 \\
    \hline
    \multirow{12}{*}{\textbf{Power}} & \multirow{3}{*}{0.005} & 24 & 100.0 & 95.0 & 93.0 & 97.0 & 100.0 & 100.0 & 100.0 & 100.0 & 100.0 & 100.0 & 99.0 & 100.0 \\
    & & 40 & 100.0 & 95.0 & 93.0 & 98.0 & 100.0 & 100.0 & 100.0 & 100.0 & 100.0 & 100.0 & 100.0 & 100.0 \\
    & & 56 & 100.0 & 95.0 & 91.0 & 98.0 & 100.0 & 100.0 & 99.0 & 100.0 & 100.0 & 100.0 & 99.0 & 100.0 \\
    \cline{2-15}
    & \multirow{3}{*}{0.01} & 24 & 100.0 & 95.0 & 92.0 & 97.0 & 100.0 & 100.0 & 100.0 & 100.0 & 100.0 & 100.0 & 100.0 & 100.0 \\
    & & 40 & 100.0 & 95.0 & 93.0 & 97.0 & 100.0 & 100.0 & 100.0 & 100.0 & 100.0 & 100.0 & 100.0 & 100.0 \\
    & & 56 & 100.0 & 95.0 & 91.0 & 97.0 & 100.0 & 100.0 & 99.0 & 100.0 & 100.0 & 100.0 & 100.0 & 100.0 \\
    \cline{2-15}
    & \multirow{3}{*}{0.05} & 24 & 100.0 & 95.0 & 92.0 & 98.0 & 100.0 & 100.0 & 100.0 & 100.0 & 100.0 & 100.0 & 100.0 & 100.0 \\
    & & 40 & 100.0 & 95.0 & 93.0 & 97.0 & 100.0 & 100.0 & 100.0 & 100.0 & 100.0 & 100.0 & 100.0 & 100.0 \\
    & & 56 & 100.0 & 95.0 & 92.0 & 97.0 & 100.0 & 100.0 & 100.0 & 100.0 & 100.0 & 100.0 & 100.0 & 100.0 \\
    \cline{2-15}
    & \multirow{3}{*}{0.1} & 24 & 100.0 & 95.0 & 92.0 & 97.0 & 100.0 & 100.0 & 100.0 & 100.0 & 100.0 & 100.0 & 100.0 & 100.0 \\
    & & 40 & 100.0 & 95.0 & 93.0 & 97.0 & 100.0 & 100.0 & 100.0 & 100.0 & 100.0 & 100.0 & 100.0 & 100.0 \\
    & & 56 & 100.0 & 95.0 & 93.0 & 97.0 & 100.0 & 100.0 & 100.0 & 100.0 & 100.0 & 100.0 & 100.0 & 100.0 \\
    \hline
    \end{tabular}
    }
    \caption{Cumulative miscoverage rate (in \%) at the end of trials ($n=200$) for \textit{mis-specified} trial optimizer. Each row corresponds to different cases with data generated from either High-SNR or Low-SNR setup, minimum propensity $\delta\in \{0.005, 0.01, 0.05, 0.1\}$ and the number of initial participants randomized $n_0 \in \{24, 40, 56\}$. Number of burn-in samples $m$ is set to 24 and 80 for \texttt{IPW}-based and \texttt{AIPW}-based AsympCSs, respectively. Cumulative miscoverage rate around $1.67 \ (= 0.05 \cdot 100/3)$ reflects nominal coverage. Last twelve rows depict the power at the nominal significance level. }
    \label{tab:cum_miscov_mis_tab_all}
\end{table}
}

\section{Technical Results} \label{sec:technical_results}
Web Appendix \ref{subsec:data_gen_setup} defines additional notations and introduces some technical details on the data-generating setup. Web Appendix \ref{subsec:useful_lemmas} introduces some lemmas required to prove Theorem 1. We include the proof of Theorem 1 in Web Appendix \ref{subsec:proof_thm}. Lastly, we include the proofs for lemmas in Web Appendix \ref{subsec:lemma_proofs}.  

\subsection{Additional Details} \label{subsec:data_gen_setup}

Without loss of generality, we only consider Arm $K$ and prove the validity of $\{\mc{C}_{n}^{(\alpha)}(K)\}_{n \geq 1}$ in (3) for parameter $\Delta(K)$. Towards that end, let us simplify notations by removing the specification of arm $K$, starting with $\mc{C}_{n}^{(\alpha)}$ instead of $\mc{C}_{n}^{(\alpha)}(K)$. Define $\mc{F}_{n}^{eval} := \sigma\langle \{Z_{i}\}_{i\in \calI^{eval}_{n-1}} \cup (\mbi{X}_{\floor{n/2}}, A_{\floor{n/2}}) \rangle$ and $\mc{H}_{n}^{eval} := \sigma\langle \{Z_{i}\}_{i \in \calI^{eval}_n} \rangle$. We consider the following Markovian setup: \begin{equation} \label{eq:data_gen_setup}
    p(Z_i|\mc{H}_{i-1}) = p(\mbi{X}_i) \ p(A_i|\mbi{X}_i, \mc{H}_{i-1}) \ p(R_i, S_i|A_i, \mbi{X}_i).
\end{equation} Note that Algorithm 1 satisfies the above assumption. Moreover, we assume that the patient contexts are i.i.d. from a fixed distribution $\mc{P}_X$, i.e., $\{\mbi{X}_n\}_{n \geq 1} \simiid \mc{P}_X$. Hence, we have $\phi =: \mc{H}_{0}^{eval} \subset \mc{F}_{0}^{eval} \subset \ldots \subset \mc{H}_{n-1}^{eval} \subset \mc{F}_{n}^{eval} \subset \mc{H}_{n}^{eval} \subset \ldots$ for all $n \geq 1$. Let $\E_{*}(\cdot)$, $\P_{*}(\cdot)$, $\var_{*}(\cdot)$ be the conditional expectation, probability, and variance on some measurable function of the training (evaluation) data given the evaluation (training) data. For instance, $\E_{*}(R_{i})$ denotes the conditional expectation of $R_{i}$ given $\mc{D}^{trn}_n$. Also denote $\E_{*}^{(n)}(\cdot) := \E_{*}(\cdot|\mc{H}_{n})$. For instance, $\E_{*}^{(n-1)}(R_{n}) = \E(R_{n}|\mc{D}^{trn}, \mc{H}_{n-1}^{eval})$. In a similar spirit, define $\var^{(n)}(\cdot)$ as the conditional variance given $\mc{H}_{n-1}$. We omit subscript $n$ from $\mc{D}^{trn}_n$, $\mc{D}^{eval}_n$, $\mc{I}^{trn}_n$ and $\mc{I}^{eval}_n$ whenever not ambiguous. 

\begin{definition}(Martingale Difference Sequence) \label{def:mds}
    A sequence $(Z_n, \mc{G}_n, n \geq 1)$ defined on $(\Omega, \mc{F}, \P)$ is called a martingale difference sequence if $\mc{G}_n \subset \mc{F}$ are increasing $\sigma$-fields with $Z_n$ being $\mc{G}_n$ measurable and $\E[Z_n|\mc{G}_{n-1}] = 0$ for all $n \geq 2$. 
\end{definition}

\subsection{Some Useful Lemmas} \label{subsec:useful_lemmas}
\begin{lemma}\textbf{[Proof in Section \ref{subsubsec:proof_ubd_eif}]} (Unbiased Efficient Influence Function) \label{lemma:ubd_eif} \\
    For a new random element $Z = (\mbi{X}, A, R, S)$ define \begin{equation*}
        f(Z) =  \firstbigg{ \mu^{(K)}(\mbi{X}) - \mu^{(1)}(\mbi{X}) } + \secondbigg{ \frac{\indic\{A=K\}}{q(K, \mbi{X})} - \frac{\indic\{A=1\}}{q(1, \mbi{X})} } \firstbigg{ R - \mu^{(A)}(\mbi{X}) },
    \end{equation*}
    \begin{equation*} 
        \Bar{f}(Z) =  \firstbigg{ \Bar{\mu}^{(K)}(\mbi{X}) - \Bar{\mu}^{(1)}(\mbi{X}) } + \secondbigg{ \frac{\indic\{A=K\}}{q(K, \mbi{X})} - \frac{\indic\{A=1\}}{q(1, \mbi{X})} } \firstbigg{ R - \Bar{\mu}^{(A)}(\mbi{X}) },
    \end{equation*} and recall \begin{equation*} 
        \hat{f}_{N'}(Z) =  \firstbigg{ \hat{\mu}_{N'}^{(K)}(\mbi{X}) - \hat{\mu}_{N'}^{(1)}(\mbi{X}) } + \secondbigg{ \frac{\indic\{A=K\}}{q(K, \mbi{X})} - \frac{\indic\{A=1\}}{q(1, \mbi{X})} } \firstbigg{ R - \hat{\mu}_{N'}^{(A)}(\mbi{X}) }.
    \end{equation*} Then we have, { $\E\thirdbigg{ \Bar{f}(Z) }$, $\E\thirdbigg{ f(Z) }$ and $\E_{*}\thirdbigg{ \hat{f}(Z) }$ all equal to $\Delta$. }
\end{lemma}

\begin{lemma}\textbf{[Proof in Section \ref{subsubsec:proof_main_decomp}]} (Decomposition of $n \hat{\Delta}_{n}^{\times} - n \Delta$) \label{lemma:main_decomp}\\
    For some $\{\Bar{\mu}^{(a)}(\cdot)\}_{a \in [K]}$ we can write, \begin{equation} \label{eq:decomp_lemma_eq_long}
        n \hat{\Delta}_{n}^{\times} - n \Delta = \wt{S}_{n,eval}^{SA} + \underbrace{ \wt{S}_{n,eval}^{EP} + \wt{S}_{n,trn}^{EP}}_\text{$\wt{S}_{n}^{EP}$} + \underbrace{\wt{S}_{n,trn}^{B} + \wt{S}_{n,eval}^{B}}_\text{$\wt{S}_{n}^{B}$},
    \end{equation} where \begin{align}
        \begin{split}
            \wt{S}_{n}^{SA} &:= \sum_{i=1}^{n} \thirdbigg{ \Bar{f}(Z_i) - \E(\Bar{f}(Z_i)) }, \\
            \wt{S}_{n,eval}^{EP} &:= \sum_{i \in \calI^{eval}_n} \secondbigg{ \thirdbigg{ \hat{f}_{N'}(Z_{i}) - \E_*(\hat{f}_{N'}(Z_{i})) } - \thirdbigg{ \Bar{f}(Z_{i}) - \E(\Bar{f}(Z_{i})) } }, \\
            \wt{S}_{n,trn}^{EP} &:= \sum_{i \in \calI^{trn}_n} \secondbigg{ \thirdbigg{ \hat{f}_{N'}(Z_{i}) - \E_*(\hat{f}_{N'}(Z_{i})) } - \thirdbigg{ \Bar{f}(Z_{i}) - \E(\Bar{f}(Z_{i})) } } \\
            \wt{S}_{n,eval}^{B} &:= \sum_{i \in \calI^{eval}_n} \E_{*}\firstbigg{ \hat{f}_{N'}(Z_{i}) - f(Z_{i}) }, \quad \text{and} \\
            \wt{S}_{n,trn}^{B} &:= \sum_{i \in \calI^{trn}_n} \E_{*}\firstbigg{ \hat{f}_{N}(Z_{i}) - f(Z_{i}) }.
        \end{split}
    \end{align} Furthermore, we have $\wt{S}_{n}^{B}=0$ and \eqref{eq:decomp_lemma_eq_long} reduces to, \begin{equation} \label{eq:decomp_lemma_eq}
        n \hat{\Delta}_{n}^{\times} - n \wt{\Delta}_{n}^{\times} = \wt{S}_{n,eval}^{SA} + \underbrace{ \wt{S}_{n,eval}^{EP} + \wt{S}_{n,trn}^{EP}}_\text{$\wt{S}_{n}^{EP}$}.
    \end{equation}
\end{lemma}

\begin{lemma}\textbf{[Proof in Section \ref{subsubsec:proof_as_behav_ep_term}]} (Almost sure behavior of $\wt{S}_{n}^{EP}$) \label{lemma:as_behav_ep_term} 
    Suppose Fact \ref{assump:clipping} and Assumption \ref{assump:bdd_predictor} hold. Then, \begin{equation}
        \wt{S}_{n}^{EP} = O\firstbigg{ \secondbigg{ \sum_{a=1,K} \norm{\hat{\mu}^{(a)}_{n} - \Bar{\mu}^{(a)} }_{2} } \sqrt{n \log \log n} }.
    \end{equation}
\end{lemma}

\begin{lemma} \textbf{[Proof in Section \ref{subsubsec:proof_cond_l1}]} \label{lemma:cond_l1} (Diverging sum of conditional variances)
    Denote $\sigma_{n}^{2} := \var^{(n-1)}\thirdbigg{ \Bar{f}(Z_n) }$, for $n \geq 2$. Let $\sigma_{0}^{2} := \var(R_i) > 0$. Then, $V_n := \sum_{i=1}^{n} \sigma_{i}^{2} \to \infty$ almost surely. 
\end{lemma}

\begin{lemma} \textbf{[Proof in Section \ref{subsubsec:proof_cond_l2}]} \label{lemma:cond_l2} (Lindeberg-type condition for EIF)
    For $i \geq 2$, let $\sigma_{i}^{2} := \var^{(i-1)}[\Bar{f}(Z_i)]$. Suppose there exists $\delta>0$ such that, $\E\absbigg{R}^{2+\delta} < \infty$ and $\sigma_{0}^{2} > 0$. Then there exists some $0 < \kappa < 1$ such that \begin{equation}
        \sum_{n=2}^{\infty} \frac{1}{V_{n}^{\kappa}} \E^{(n-1)}\thirdbigg{ (\Bar{f}(Z_n) - \Delta)^2 \indic\secondbigg{ ( \Bar{f}(Z_n) - \Delta)^2 > V_{n}^{\kappa} } } < \infty, \quad \text{almost surely}.
    \end{equation}
\end{lemma}

\begin{lemma} \textbf{[Proof in Section \ref{subsubsec:proof_var_est}]} \label{lemma:var_est} (Variance Estimation)
    Let us recall $\hat{\sigma}_{n}^{2}(a)$ as defined in (2.16). Also define $\wt{\sigma}^{2}_{n} = \sum_{i=1}^{n} \sigma_{i}^{2}/n$. Suppose Assumption \ref{assump:unif_conv_regression} holds. Then, $\hat{\sigma}_{n}^{2}(a)/\wt{\sigma}^{2}_{n} \to 1$ almost surely for all $a \in \{2, \ldots, K\}$. 
\end{lemma}

\subsection{Technical Details of Theorem 1} \label{subsec:proof_thm}

\subsubsection{Notations and Technical Assumptions for Theorem 1}

Let $\{Z_n: Z_n = (\mbi{X}_n, A_n, R_n, S_n)\}_{n \geq 1}$ are generated using the same data generating setup outlined in \eqref{eq:data_gen_setup}. Define the pseudo-outcome of $Z=(Z, A, R, S)$ for efficacy endpoint corresponding to arm $a \in [K]$: \begin{equation*}
    g(Z; a) := \mu^{(a)}(\mbi{X}) + \frac{\indic\{A=a\}}{q(a, \mbi{X})} \firstbigg{ R - \mu^{(A)}(\mbi{X}) },
\end{equation*} where $\{q(a, \mbi{X})\}_{a \in [K]}$ are known propensities with $A \sim \text{Multinomial}\thirdbigg{K; \{\mu^{(a)}(\mbi{X})\}_{a \in [K]}}$. Similarly, let us define the pseudo-contrast corresponding to arm $a \in \{2, \ldots, K\}$: \begin{align*}
    \begin{split}
        f(Z; a) &= g(Z; a) - g(Z; 1) \\
        &= \firstbigg{ \mu^{(a)}(\mbi{X}) - \mu^{(1)}(\mbi{X}) } + \secondbigg{ \frac{\indic\{A=K\}}{q(K, \mbi{X})} - \frac{\indic\{A=1\}}{q(1, \mbi{X})} } \firstbigg{ R - \mu^{(A)}(\mbi{X}) }.
    \end{split}
\end{align*} In this article, we assume the propensities are known. However, we estimate the regression functions $\{\mu^{(a)}(\cdot)\}_{a \in [K]}$ using observed data. Denote $\{\hat{\mu}^{(a)}_{N}(\cdot)\}_{a \in [K]}$ and $\{\hat{\mu}^{(a)}_{N'}(\cdot)\}_{a \in [K]}$ as the regression estimator constructed using $\mc{D}^{eval}$ and $\mc{D}^{trn}$ respectively. Following similar notation, we denote estimated pseudo-contrasts respectively as $\{\hat{f}_{N}(\cdot; a)\}_{a \in [K]}$ and $\{\hat{f}_{N'}(\cdot; a)\}_{a \in [K]}$. For some regression functions $\{\Bar{\mu}^{(a)}(\cdot)\}_{a \in [K]}$, let us define the corresponding pseudo-outcomes $\{\Bar{g}(\cdot; a)\}_{a \in [K]}$ and pseudo-contrasts $\{\Bar{f}(\cdot; a)\}_{a \in [K]}$ by replacing $\mu^{(a)}(\cdot)$ with $\Bar{\mu}^{(a)}(\cdot)$. From (2.16), we recall \begin{align} \label{eq:defn_delta_sig_appendix}
    \begin{split}
        &\hat{\Delta}^{\times}_{n}(a) = \frac{1}{n} \secondbigg{ \sum_{i \in \calI^{trn}_n} \hat{f}_{N}(Z_{i}; a) + \sum_{i \in \calI^{eval}_n} \hat{f}_{N'}(Z_{i}; a) }, \\ 
        &\hat{\sigma}_{n}^{2}(a) := \hat{\var}_{n}(\hat{f}; a) := \frac{1}{2} \firstbigg{ \hat{\var}_{N}(\hat{f}_{N'};a) + \hat{\var}_{N'}(\hat{f}_{N};a) },
    \end{split}
\end{align} where $\hat{\var}_{n}(\hat{f}_{N'}; a)$ and $\hat{\var}_{N'}(\hat{f}_{N}; a)$ are the sample variance of the pseudo-outcomes $\{\hat{f}_{N'}(Z_{i}; a): i \in \calI^{eval}_n\}$ and $\{\hat{f}_{N}(Z_{i}; a): i \in \calI^{trn}_n\}$, respectively. Now, we are ready to prove the theorem.

\paragraph{Technical Assumptions} Theorem 1 holds under some assumptions, which we include here:

\begin{assumption} \label{assump:unif_conv_regression} (\textit{Uniform Consistency of Regression Estimator to Some Function})\\
    The following holds true for some functions $\{\Bar{\mu}^{(a)}(\mbi{x}): \Bar{\mu}^{(a)}(\mbi{x}): \reals^{d} \to \reals, \ a \in [K]\}$: \begin{equation*}
        \norm{ \hat{\mu}_{n}^{(a)} - \Bar{\mu}^{(a)} }_{2} = o(1) \ a.s., \quad \text{for each} \ a \in [K].
    \end{equation*}
\end{assumption}

{ 

\begin{assumption} \label{assump:bdd_predictor} (\textit{Higher-order Moment Condition on Regression Estimators}) \\
    There exists some $\eta > 0$, such that, \begin{equation*}
            \max_{a \in [K]} \|\hat{\mu}_{n}^{(a)}\|_{L_{2 +\eta}} \vee \|\Bar{\mu}^{(a)} \|_{L_{2 +\eta}} =O(1) \ a.s
        \end{equation*}
\end{assumption} }

\begin{fact} \label{assump:clipping} (\textit{Clipping Propensity})
    There exists $\delta>0$ such that $q_{i}(a, \mbi{x}_i) \in [\delta, 1-\delta]$.
\end{fact}

Assumption \ref{assump:unif_conv_regression} stipulates that the regression estimators $\{\hat{\mu}_{n}^{(a)}(\cdot)\}_{a=1}^{K}$ converge uniformly to fixed functions $\{\Bar{\mu}^{(a)}(\cdot)\}_{a=1}^{K}$. The construction of AsympCS does not require knowledge of $\{\Bar{\mu}^{(a)}(\cdot)\}_{a=1}^{K}$. Assumption \ref{assump:bdd_predictor} requires that { the higher order moment of approximated function, $\Bar{\mu}^{(a)}(\cdot)$ and the approximation error, $|\hat{\mu}_{N}^{(a)}(\cdot) - \Bar{\mu}^{(a)}(\cdot)|$ are bounded. Assumption \ref{assump:bdd_predictor} is usually satisfied for standard estimators such as linear regression. Fact \ref{assump:clipping} requires clipping the propensities so that all arms have a probability of at least $\delta > 0$ for allocation to each participant. Step 7 in Algorithm 1 ensures Fact \ref{assump:clipping} holds. While this is not desirable from a trial optimization perspective, as suboptimal arms would typically receive fewer allocations, this enforced exploration of suboptimal arms is necessary for valid statistical inference in trial evaluation; see, for example, \cite{shen_et_al_2023_jasa}, where a similar clipping parameter is introduced in a different setting.}
{Theorem 1 includes two statements. The first statement is about the validity of AsympCS, and the second statement is about the validity of Assumption \ref{assump:unif_conv_regression} under a parsimonious and weighted ridge regression estimator. Assumptions \ref{assump:unif_conv_regression}-\ref{assump:bdd_predictor} are sufficient for the first part. Assumption \ref{assump:bdd_cov_reg} is further required to establish the second statement:
\begin{assumption} \label{assump:bdd_cov_reg} (Bounded Higher-order Moment for Covariate and True Regression Function)
    Following statements hold true: $\norm{\mbi{X}}_4 = O(1)$ and $\norm{\mu^{(a)}}_4 = O(1)$ for each $a \in [K]$.
\end{assumption} Note that Assumption \ref{assump:bdd_cov_reg} is mild, although not testable with finite samples, but holds true for sub-Gaussian type variables. 
}

{ The main novelty in the proof of the first statement lies in developing a series of Lemmas in Web Appendix \ref{subsec:useful_lemmas}. Our argument to establish the first part follows exactly the same lines as the Proof of Theorem 3.3 in \cite{waudby-smith_et_al_2024+}. For the sake of completeness, we have included the steps below. }

\subsubsection{Proof of Theorem 1}

\paragraph{Part 1: Validity of AsympCS}

    Due to Lemma \ref{lemma:main_decomp}, we have: \begin{equation*}
    n \hat{\Delta}_{n}^{\times} - n \Delta = \wt{S}_{n}^{SA} + \underbrace{ \wt{S}_{n,eval}^{EP} + \wt{S}_{n,trn}^{EP} }_\text{$:= \wt{S}_{n}^{EP}$}.
    \end{equation*} From Lemma \ref{lemma:as_behav_ep_term}, we have \begin{equation*}
        \wt{S}_{n}^{EP} = O\firstbigg{ \secondbigg{ \sum_{a=1, K} \norm{ \hat{\mu}_{n}^{(a)} - \Bar{\mu}^{(a)} }_{2} } \sqrt{n \log\log n} }
    \end{equation*} By Assumption \ref{assump:unif_conv_regression}, we have \begin{equation*}
        \sum_{a=1, K} \norm{ \hat{\mu}_{n}^{(a)} - \Bar{\mu}^{(a)} }_{2} = o(1).
    \end{equation*} Hence \begin{equation*}
        n \hat{\Delta}_{n}^{\times} - n \Delta = \wt{S}_{n}^{SA} + o(\sqrt{n \log\log n}).
    \end{equation*} So \begin{equation} \label{eq:mds_approx}
        \hat{\Delta}_{n}^{\times} - \Delta = \frac{1}{n} \sum_{i=1}^{n} \firstbigg{ \Bar{f}(Z_i) - \Delta_i } + o(\sqrt{\log\log n / n}).
    \end{equation} Denote $W_i = \Bar{f}(Z_i) - \Delta_{i}$. We note that, $\{W_i\}_{i=1}^{n}$ is a Martingale Difference Sequence (MDS) with $\var(W_i|\{W_j\}_{j=1}^{i-1}) = \sigma_{i}^{2}$ as $\{W_n\}_{t \geq 1}$ satisfies the conditions of Definition \ref{def:mds}. By Lemma A.2 in \cite{waudby-smith_et_al_2024+}, we know that there exists i.i.d. standard Gaussian random variables $\{G_n\}_{n=1}^{\infty}$ such that \begin{equation} \label{eq:strong_approx}
        \frac{1}{n} \sum_{i=1}^{n} W_i - \frac{1}{n} \sum_{i=1}^{n} \sigma_i G_i = o\firstbigg{ \frac{V_{n}^{3/8}\log V_n}{n} },
    \end{equation} where $V_n = \sum_{i=1}^{n} \sigma_{i}^{2}$. Note that, \eqref{eq:strong_approx} requires $V_n \to \infty$, which holds due to Lemma \ref{lemma:cond_l1}. Now, by Step 2 in the proof of Proposition 2.5 \citep{waudby-smith_et_al_2024+}, the following holds with probability at least $(1-\alpha)$, \begin{equation} \label{eq:non_asymp_weighted_wiener}
        \forall n \geq 1, \ \absbigg{\frac{1}{n} \sum_{i=1}^{n} \sigma_i G_i } < \underbrace{\sqrt{\frac{2(V_n \rho^2 + 1)}{n^2 \rho^2} \log\firstbigg{ \frac{\sqrt{V_n \rho^2 + 1}}{\alpha} }} }_\text{$O\firstbigg{ \frac{\sqrt{V_n \log V_n}}{n} }$}.
    \end{equation} Now, the approximation error in \eqref{eq:strong_approx} is faster than the rate at which the width of non-asymptotic CS in \eqref{eq:non_asymp_weighted_wiener} converges to zero. Since $V_n \asymp n$, the approximation error in \eqref{eq:mds_approx} is faster than in \eqref{eq:strong_approx}. Therefore, we combine \eqref{eq:mds_approx}, \eqref{eq:strong_approx}, \eqref{eq:non_asymp_weighted_wiener} and apply Theorem 2.4 in \cite{waudby-smith_et_al_2024+} to have, \begin{equation*}
        [L_n, U_n] := \thirdbigg{ \hat{\Delta}_{n}^{\times} \pm \sqrt{\frac{2(V_n \rho^2 + 1)}{n^2 \rho^2} \log\firstbigg{ \frac{\sqrt{V_n \rho^2 + 1}}{\alpha} }} }, \quad n \geq 1,
    \end{equation*} forms a $(1-\alpha)$-AsympCS for $\Delta$. Now, by Step 4 in the proof of Proposition 2.5 \citep{waudby-smith_et_al_2024+}, the following holds: \begin{equation} \label{eq:bound_approx}
        \sqrt{\frac{2(V_n \rho^2 + 1)}{n^2 \rho^2} \log\firstbigg{ \frac{\sqrt{V_n \rho^2 + 1}}{\alpha} }} \leq \sqrt{\frac{2(n\hat{\sigma}_{n}^{2}\rho^2+1)}{n^2\rho^2} \log\firstbigg{ \frac{\sqrt{1+n\hat{\sigma}_{n}^{2}\rho^2}}{\alpha} }} + o\firstbigg{ \frac{\sqrt{V_n \log V_n}}{n} },
    \end{equation} provided $\hat{\sigma}_{n}^{2} - \wt{\sigma}_{n}^{2} = o(\wt{\sigma}_{n}^{2})$, which holds due to Lemma \ref{lemma:var_est}. Now, let us define, \begin{equation}
        [\hat{L}_n, \hat{U}_n] := \thirdbigg{ \hat{\Delta}_{n}^{\times} \pm \sqrt{\frac{2(n\hat{\sigma}_{n}^{2}\rho^2+1)}{n^2\rho^2} \log\firstbigg{ \frac{\sqrt{1+n\hat{\sigma}_{n}^{2}\rho^2}}{\alpha} }} }
    \end{equation} By \eqref{eq:bound_approx}, we have $\hat{L}_n / L_n \to 1$ and $\hat{U}_n / U_n \to 1$ almost surely since $\hat{\sigma}_{n}^{2} - \wt{\sigma}_{n}^{2} = o(\wt{\sigma}_{n}^{2})$. Therefore, a $(1-\alpha)$-AsympCS for $\Delta$ is \begin{equation*}
        \hat{\Delta}_{n}^{\times} \pm \sqrt{\frac{2(n\hat{\sigma}_{n}^{2}\rho^2+1)}{n^2\rho^2} \log\firstbigg{ \frac{\sqrt{1+n\hat{\sigma}_{n}^{2}\rho^2}}{\alpha} }}, \quad n \geq 1,
    \end{equation*} and we conclude the proof for the first part of Theorem 1.

    {
    \paragraph{Part 2: Validity of Assumption \ref{assump:unif_conv_regression}} Borrowing ideas from Section 3 in \cite{buja2019models}, we can write: \begin{align} \label{eq:popl_ols}
    \begin{split}
        \mu^{(a)}(\mbi{X}) &:= \E[R(a)|\mbi{X}] = \argmin_{f(\mbi{X}) \in L_2} \E[(R(a)-f(\mbi{X}))^2], \quad \text{and}, \\
        \Bar{\mu}^{(a)}(\mbi{X}) &:= \Bar{\beta}_{a,0} + \mbi{X}^T\Bar{\mbi{\beta}}, \ \text{s.t.,} \ \{\Bar\beta_{a,0}\}, \Bar{\mbi{\beta}} =  \argmin_{\beta_{a,0} \in \reals, \ \mbi{\beta} \in \reals^{d}} \sum_{a \in [K]} \E[(R(a)- \beta_{a,0} - \mbi{X}^T \mbi{\beta})^2].
    \end{split}
    \end{align} 
    Note that, we can further write the following: \begin{align} \label{eq:lin_decomp}
        \begin{split}
            R(a) &= \mu^{(a)}(\mbi{X}) + \epsilon, \\
            &= \Bar{\beta}_{a,0} + \mbi{X}^T\Bar{\mbi{\beta}} + \mu^{(a)}(\mbi{X}) - \Bar{\mu}(\mbi{X}) + \epsilon, \\
            &= \Bar{\beta}_{a,0} + \mbi{X}^T\Bar{\mbi{\beta}} + e_a(\mbi{X}),
        \end{split}
    \end{align} where $e_a(\mbi{X}) := \mu^{(a)}(\mbi{X}) - \Bar{\mu}(\mbi{X}) + \epsilon$. Due to \eqref{eq:popl_ols} and the corresponding normal equations, we have: \begin{align} \label{eq:ortho}
        \begin{split}
            &\E[\epsilon \cdot f(\mbi{X})] = 0, \quad \forall \ f \in L_2, \\
            &\E[e_a(\mbi{X})] = 0, \\
            &\E[e_a(\mbi{X}) X_j] = 0, \quad \forall \ j \in [d].
        \end{split}
    \end{align}
    
    To estimate $\Bar{\mu}(\cdot)$ with finite samples, we use $\hat{\mu}_{N'}^{(a)}(\mbi{x}) = \hat{\beta}_{a,0} + \mbi{x}^T\hat{\mbi{\beta}}$, where: \begin{equation} \label{eq:wls}
        \{ \hat{\beta}_{a,0} \}, \ \hat{\mbi{\beta}} := \argmin_{ \{\beta_{a,0}\},\ \mbi{\beta}} \thirdbigg{\sum_{a\in [K]} \sum_{i\in \mathcal I^{trn}_n} \frac{\indic\{A_i = a\}}{q_i(a, \mbi{X}_i)} (R_i - \beta_{a,0} - \mbi{X}_{i}^{T} \mbi{\beta})^2  }. 
    \end{equation} Note that we originally proposed to use a ridge regression estimator in (3). We do not consider the ridge penalty to simplify our exposition here. All of our arguments remain valid for (3) as well. 
    
    Let $\mbf{X} \in \reals^{n \times d}$ denote the standardized design matrix and $\mbi{1}_a \in \reals^n$ denote a vector with entries $\{\indic\{A_i = a\}\}_{i=1}^{n}$. Define $\wt{\mbf{X}} := [\mbi{1}_1, \ldots, \mbi{1}_K, \mbf{X}] \in \reals^{n \times (K+d)}$, $\mbf{W} \in \reals^{n \times n}$, a diagonal matrix with entries $\{1/q_i(A_i, \mbi{X}_i)\}_{i=1}^{n}$, and $\mbi{R} = (R_1, \ldots, R_n)^T$. Then, \eqref{eq:wls} is the weighted least square estimator with the following closed form expression: $(\hat{\beta}_{1,0}, \ldots, \hat{\beta}_{K,0}, \hat{\mbi{\beta}}^T)^T = (\wt{\mbf{X}}^T \mbf{W} \wt{\mbf{X}})^{-1} \wt{\mbf{X}}^T \mbf{W} \mbi{R}$. Note that $\wt{\mbf{X}}^T \mbf{W} \wt{\mbf{X}}$ is a block diagonal matrix with two blocks of non-zero square matrices of size $K$ and $d$ in the diagonal. By \eqref{eq:lin_decomp}, we can write \begin{align*}
        \begin{split}
            \hat{\beta}_{a,0} - \Bar{\beta}_{a,0} &= \firstbigg{\sum_{i\in \calI^{trn}_n} \frac{\indic\{A_i = a\}}{q_i(a, \mbi{X}_i)}}^{-1} \sum_{i\in \calI^{trn}_n} \frac{\indic\{A_i = a\}}{q_i(a, \mbi{X}_i)} e_a(\mbi{X}_i), \quad \forall \ a \in [K], \\
            \hat{\mbi{\beta}} - \Bar{\mbi{\beta}} &= (\mbf{X}^T \mbf{W} \mbf{X})^{-1} \mbf{X}^T \mbf{W} \mbi{e}(\mbf{X})
        \end{split}
    \end{align*} where $\mbi{e}(\mbf{X})^T = (e_{A_1}(\mbi{X}_1), \ldots, e_{A_n}(\mbi{X}_{n}))^T.$ Denote $N'_a := \sum_{i \in \calI_n^{trn}} \indic\{A_i = a\}$. Since $q_i(a, \mbi{X}_i) \in [\delta, 1-\delta]$, we have \begin{equation*}
        \hat{\beta}_{a,0} - \Bar{\beta}_{a,0} \asymp \frac{1}{N'_a} \sum_{i\in \calI^{trn}_n} \frac{\indic\{A_i = a\}}{q_i(a, \mbi{X}_i)} e_a(\mbi{X}_i) 
    \end{equation*} By \eqref{eq:ortho} $\E[e_a(\mbi{X}_i)] = 0$. Therefore $V_n := \sum_{i\in \calI^{trn}_n} \frac{\indic\{A_i = a\}}{q_i(a, \mbi{X}_i)} e_a(\mbi{X}_i) $ is a zero-mean martingale with respect to $\mc{H}_n$. By Assumption \ref{assump:bdd_cov_reg}, we have $\norm{\mu^{(a)}}_4 < \infty, \ \E|\epsilon|^4 < \infty$. Therefore, $S_n$ is a square integrable martingale. Hence, we invoke Theorem 2.15 from \cite{hall_heyde_book} to establish that $V_n / N'_a \to 0$ a.s. provided $N'_a \to \infty$, which is satisfied due to $q_n(a, \mbi{X}_n) > 0$ for $n\geq 1$. Now we establish $\hat{\mbi{\beta}} - \Bar{\mbi{\beta}} = o(1)$ a.s.. 
    
    Let us consider the $(j,k)$-th entry of $\mbf{X}^T \mbf{W} \mbf{X}$ for some $j, k \in [d]$, which can be written as follows: \begin{equation*}
        (\mbf{X}^T \mbf{W} \mbf{X})_{jk} := \sum_{a \in [K]} \sum_{i \in \calI^{eval}_n} \thirdbigg{ \frac{\indic\{A_i = a\}}{q_i(a, \mbi{X}_i)} X_{ij} X_{ik} - \sigma_{jk} } := S_n,
    \end{equation*} where $\sigma_{jk}$ is the $(j,k)$-th entry of $\mbf{\Sigma}_X$, population covariance matrix of $\mbi{X}$. Note that $S_n$ is a zero-mean martingale with respect to $\mc{H}_n$. Moreover, $\E|X_{ij}|^4 < \infty, \ \forall j \in [d]$, and $q_i(a, \mbi{X}_i) \in [\delta, 1-\delta]$ due to Assumption \ref{assump:bdd_cov_reg} and Fact \ref{assump:clipping}, respectively. Therefore, $\{X_{ij} X_{ik}\}_{i \geq 1}$ is a square integrable martingale since, \begin{equation*}
        \sum_{i \in \calI^{eval}_n} \E[(X_{ij}X_{ik})^2 | \mc{H}_{n-1}] \lesssim \firstbigg{\E|X_{ij}|^4 \E|X_{ik}|^4}^2 < \infty,
    \end{equation*} due to the Cauchy-Schwarz inequality and Assumption \ref{assump:bdd_cov_reg}. Therefore, by Theorem 2.15 in \cite{hall_heyde_book}, $S_n / n \to 0$ almost surely. Since $d$ is fixed, we apply the Cramer-Wold device to establish that $\mbf{X}^T \mbf{W} \mbf{X} \to \mbf{\Sigma}_X$ almost surely. 
    
    Now, we consider the $j$-th coordinate of $\mbf{X}^T \mbf{W} \mbi{e}_a(\mbi{X})$: \begin{equation*}
        (\mbf{X}^T \mbf{W} \mbi{e}_a(\mbi{X}))_j := \sum_{a \in [K]} \sum_{i \in \calI^{eval}_n} \thirdbigg{ \frac{\indic\{A_i = a\}}{q_i(a, \mbi{X}_i)} X_{ij} e_a(\mbi{X}_i) } := G_n.
    \end{equation*} By \eqref{eq:ortho}, $\E[X_{ij} e_a(\mbi{X}_i)] = 0$ for all $i \geq 1$, and hence $G_n$ is also a sum zero martingale with respect to $\mc{H}_n$. Since $\norm{\mu^{(a)}}_4 < \infty$, $\E|\epsilon|^4 < \infty$ by Assumption \ref{assump:bdd_cov_reg} and $q_i(a, \mbi{X}_i) \in [\delta, 1-\delta]$ by Fact \ref{assump:clipping}, we have $\E[e_a^4(\mbi{X}_i)] < \infty$. Thus, we again apply Theorem 2.15 in \cite{hall_heyde_book} and the Cramer-Wold device to establish $G_n / n \to 0$ almost surely. By our assumption, $\mbf{\Sigma}_X$ is invertible. Therefore we apply continuous mapping theorem to have $\hat{\mbi{\beta}} - \Bar{\mbi{\beta}} \to \mbi{0}$ almost surely and therefore, $\norm{\hat{\mu}_{N'}^a - \Bar{\mu}}_2 = o(1)$.   
    }

\subsection{Proof of Technical Lemmas} \label{subsec:lemma_proofs}
In this section, we include the proof of technical lemmas outlined in Section \ref{subsec:proof_thm}, which are needed to prove Theorem 1. 

\subsubsection{Proof of Lemma \ref{lemma:ubd_eif}} \label{subsubsec:proof_ubd_eif}

To simplify notations, we omit the specification of `eval' in the rest of this proof. Let us recall, \begin{equation} \label{eq:f_bar_lemma_ubd_eif}
    \Bar{f}(Z_{i}) =  \underbrace{ \firstbigg{ \Bar{\mu}^{(K)}(\mbi{X}_{i}) - \Bar{\mu}^{(1)}(\mbi{X}_{i}) } }_\text{$:= A(\mbi{X}_i)$} + \underbrace{ \secondbigg{ \frac{\indic\{A_{i}=K\}}{q_i(K, \mbi{X}_i)} - \frac{\indic\{A_{i}=1\}}{q_i(1, \mbi{X}_i)} } \firstbigg{ R_{i} - \Bar{\mu}^{(A_{i})}(\mbi{X}_{i}) } }_\text{$:= B(Z_i)$}.
\end{equation}

Note that $\E[\Bar{\mu}^{(a)}(\mbi{X}_i)|\mc{H}_{i-1}] = \E[\Bar{\mu}^{(a)}(\mbi{X}_i)]$ since $\mbi{X}_i \ind \mc{H}_{i-1}$. Therefore, $\E[A(\mbi{X}_i)|\mc{H}_{i-1}^{eval}] = \E[\Bar{\mu}^{(K)}(\mbi{X}_i)] - \E[\Bar{\mu}^{(1)}(\mbi{X}_i)]$. Now, we can obtain the following using repeated application of iterated expectations: \begin{align} \label{eq:lemma1_brain}
    \begin{split}
        &\E^{(i-1)}[B(Z_i)] := \E[B(Z_i)|\mc{H}_{i-1}] \\ 
        &= \E^{(i-1)}\thirdbigg{ \secondbigg{ \frac{\indic\{A_{i}=K\}}{q_i(K, \mbi{X}_i)} - \frac{\indic\{A_{i}=1\}}{q_i(1, \mbi{X}_i)} } \E \thirdbigg{ R_{i} - \Bar{\mu}^{(A_{i})}(\mbi{X}_{i}) | \mc{F}_{i} } } \\
        &= \E^{(i-1)}\thirdbigg{ \secondbigg{ \frac{\indic\{A_{i}=K\}}{q_i(K, \mbi{X}_i)} - \frac{\indic\{A_{i}=1\}}{q_i(1, \mbi{X}_i)} } \firstbigg{ \mu^{(A_i)}(\mbi{X}_i) - \Bar{\mu}^{(A_i)}(\mbi{X}_i) } } \\
        &= \E^{(i-1)}\thirdbigg{ \firstbigg{ \mu^{(K)}(\mbi{X}_i) - \Bar{\mu}^{(K)}(\mbi{X}_i) } \frac{\P(A_i=K|\mbi{X}_i)}{q_i(K, \mbi{X}_i)} - \firstbigg{ \mu^{(1)}(\mbi{X}_i) - \Bar{\mu}^{(1)}(\mbi{X}_i) } \frac{\P(A_i=1|\mbi{X}_i)}{q_i(1, \mbi{X}_i)} } \\
        &= \thirdbigg{ \mu^{(K)}(\mbi{X}_i) - \mu^{(1)}(\mbi{X}_i) } - \thirdbigg{ \Bar{\mu}^{(K)}(\mbi{X}_i) - \Bar{\mu}^{(1)}(\mbi{X}_i) }.
    \end{split}
\end{align} Hence, $\E[B(Z_i)] = \E[\mu^{(K)}(\mbi{X}_i)] - \E[\mu^{(1)}(\mbi{X}_i)] - \E[\Bar{\mu}^{(K)}(\mbi{X}_i)] - \E[\Bar{\mu}^{(1)}(\mbi{X}_i)]$ and therefore $\E[\Bar{f}(Z_{i})|\mc{H}_{i-1}] = \E[\mu^{(K)}(\mbi{X}_i)] - \E[\mu^{(1)}(\mbi{X}_i)] = \Delta$. The proof of $\E[f(Z_{i})|\mc{H}_{i-1}] = \Delta$ follows using the same steps as in \eqref{eq:lemma1_brain}. To show $\E_*[\hat{f}(Z_{i})]=\Delta$, we replace $\E^{(i-1)}(\cdot)$ with $\E^{(i-1)}_{*}(\cdot)$ to obtain $\E_*[\hat{f}(Z_{i})|\mc{H}_{i-1}] = \E_*[\mu^{(K)}(\mbi{X}_i)] - \E_*[\mu^{(1)}(\mbi{X}_i)]$. Since the right-hand side holds conditionally given $\mc{D}^{trn}$, it also holds marginally.

\subsubsection{Proof of Lemma \ref{lemma:main_decomp}} \label{subsubsec:proof_main_decomp}

Note that the decomposition in \eqref{eq:decomp_lemma_eq_long} holds immediately due to Lemma A.8 in \cite{waudby-smith_et_al_2024+}. We only need to show that $\wt{S}_{n}^{B}=0$. To that end, we show $\wt{S}_{n, eval}^{B}=0$. The proof of $\wt{S}_{n, trn}^{B}=0$ holds by following similar steps. We recall \begin{equation*}
    \wt{S}_{n,eval}^{B} := \sum_{i \in \calI^{eval}_n} \E_*\firstbigg{ \hat{f}_{N'}(Z_{i}) - f(Z_{i}) }.
\end{equation*} Due to Lemma \ref{lemma:ubd_eif}, we know $\E_*[\hat{f}_{N'}(Z_{i})] = \E_*[f(Z_{i})] = \E[f(Z_{i})]$. Hence, we have $\wt{S}_{n, eval}^{B} = 0$ and we conclude the proof.

\subsubsection{Proof of Lemma \ref{lemma:as_behav_ep_term}} \label{subsubsec:proof_as_behav_ep_term}
Let us recall, \begin{align*}
        \begin{split}
            \wt{S}_{n, eval}^{EP} &:= \sum_{i \in \calI^{eval}_n} \secondbigg{ \thirdbigg{ \hat{f}_{N'}(Z_{i}) - \E_{*}\firstbigg{ \hat{f}_{N'}(Z_{i}) } } - \thirdbigg{ \Bar{f}(Z_{i}) - \E\firstbigg{ \Bar{f}(Z_{i}) } } } \\
            &= \sum_{i \in \calI^{eval}_n} \secondbigg{ \hat{f}_{N'}(Z_{i}) - \Bar{f}(Z_{i}) },
        \end{split}
    \end{align*} since $\E_*\thirdbigg{ \hat{f}_{N'}(Z_{i}) } = \Delta = \E\thirdbigg{ \Bar{f}(Z_{i}) }$ by Lemma \ref{lemma:ubd_eif}. Define $\mc{S}^{trn} := (\indic(Z_n \in \mc{D}^{trn}))_{n=1}^{\infty}$. Clearly, $\E_*\thirdbigg{ \wt{S}_{n, eval}^{EP} } = 0$ again due to Lemma \ref{lemma:ubd_eif}. The following holds due to Fact \ref{assump:clipping}: \begin{equation} \label{eq:f_hat_diff_ineq}
        \absbigg{\hat{f}_{N'}(Z_{i}) - \Bar{f}(Z_{i})} \leq \firstbigg{\frac{1-\delta}{\delta}} \absbigg{ \sum_{a=1, K} \firstbigg{\hat{\mu}_{N'}^{(a)}(\mbi{x}_{i}^{eval}) - \Bar{\mu}^{(a)}(\mbi{x}_{i}^{eval})} }
    \end{equation} For some positive constants $\{K_i\}_{i \geq 1}$ (to be specified later), we have the following for some $\eta > 0$: \begin{align*}
        \begin{split}
            &\P\firstbigg{ \absbigg{\hat{f}_{N'}(Z_{i}) - \Bar{f}(Z_{i})} > K_i \sqrt{\frac{i}{\log\log i}} } \\
            & \leq \P\firstbigg{ \absbigg{ \sum_{a=1, K} \firstbigg{\hat{\mu}_{N'}^{(a)}(\mbi{x}_{i}^{eval}) - \Bar{\mu}^{(a)}(\mbi{x}_{i}^{eval})} } > \firstbigg{\frac{\delta}{1-\delta}} K_i \sqrt{\frac{i}{\log\log i}} } \\
            & \leqsim \E\absbigg{ \sum_{a=1, K} \firstbigg{\hat{\mu}_{N'}^{(a)}(\mbi{x}_{i}^{eval}) - \Bar{\mu}^{(a)}(\mbi{x}_{i}^{eval})} }^{2+\eta}  \firstbigg{\frac{\log\log i}{i}}^{1+\eta/2}\frac{1}{K_i^{2+\eta}} \\
            & < \thirdbigg{ \sum_{a=1, K} \norm{\hat{\mu}_{N'}^{(a)} - \Bar{\mu}^{(a)}}_{2+\eta} }^{2+\eta} \frac{1}{K_i^{2+\eta}} \firstbigg{\frac{\log\log i}{i}}^{1+\eta/2}.
        \end{split}
    \end{align*} We apply Markov's and Minkowski's inequality to the second, last, and last line above. By Assumption \ref{assump:bdd_predictor}, we have $\max_{a \in [K]} \norm{\hat{\mu}^{(a)}_{T} - \Bar{\mu}^{(a)}}_{2+\eta} = O(1)$ for some $\eta>0$. Therefore, we fix $K_i = \sqrt{\log \log i}/i^{\eta/4(2+\eta)}$ to have, 
    
    \begin{equation*}
        \P\firstbigg{ \absbigg{\hat{f}_{N'}(Z_{i}) - \Bar{f}(Z_{i})} > K_i \sqrt{\frac{i}{\log\log i}} } \leqsim \frac{1}{i^{1+\eta/4}}.
    \end{equation*} Now, we apply the Borel-cantelli lemma to have the following: \begin{equation*}
        \absbigg{\hat{f}_{N'}(Z_{i}) - \Bar{f}(Z_{i})} \leq K_i \sqrt{\frac{i}{\log\log i}}, \quad \ \text{almost surely}.
    \end{equation*} Since, $K_i \to 0$ as $i \to \infty$, we apply the main theorem from \cite{stout_1970} to have the following: \begin{align} \label{eq:st_second_last}
        \begin{split}
            \wt{S}_{n, eval}^{EP} &= \sum_{i \in \calI^{eval}_n} \secondbigg{ \hat{f}_{N'}(Z_{i}; K) - \Bar{f}(Z_{i}; K) } = O\firstbigg{ \sqrt{n \log\log n} \sum_{a=1, K} \norm{\hat{\mu}^{(a)}_{N'} - \Bar{\mu}^{(a)}}_{2} },
        \end{split}
    \end{align} where the last equality holds due to \eqref{eq:f_hat_diff_ineq} and the Minkowski's inequality. Note that, \eqref{eq:st_second_last} holds given $\mc{D}^{trn}$ and $\mc{S}^{trn}$ with probability one. Therefore, \eqref{eq:st_second_last} also holds marginally with probability one, and we conclude the proof. 

\subsubsection{Proof of Lemma \ref{lemma:cond_l1}} \label{subsubsec:proof_cond_l1}

We first show that \begin{equation} \label{eq:cond_var_x}
    \var^{(i-1)}\thirdbigg{\Bar{f}(Z_i)|\mbi{X}_i} > 0,
\end{equation} provided $\sigma_{0}^{2} > 0$. Now, the lower bound above does not depend on $\mbi{X}_i$ hence, $\sigma_i^2 := \var^{(i-1)}\thirdbigg{\Bar{f}(Z_i)} > 0$. Therefore, $V_n = \sum_{i=1}^{n} \sigma_{i}^{2} \to \infty$ as $n \to \infty$. Therefore, it is enough to show \eqref{eq:cond_var_x}.

\paragraph{Proof of \eqref{eq:cond_var_x}:} Note that, \begin{align*}
        \begin{split}
            &\var^{(i-1)}\thirdbigg{ \Bar{f}(Z_i)|\mbi{X}_i } = \var^{(i-1)}\thirdbigg{\E^{(i-1)}(\Bar{f}(Z_i)|A_i, \mbi{X}_i)|\mbi{X}_i} + \E^{(i-1)}\thirdbigg{\var^{(i-1)}(\Bar{f}(Z_i)|A_i, \mbi{X}_i)|\mbi{X}_i}.
        \end{split}
    \end{align*} Now, \begin{align*}
        \begin{split}
            &\var^{(i-1)}\firstbigg{\Bar{f}(Z_i)|A_i, \mbi{X}_i)} \\
            &= \var^{(i-1)}\firstbigg{ \secondbigg{ \Bar{\mu}^{(K)}(\mbi{X}_i) - \Bar{\mu}^{(1)}(\mbi{X}_i) } + \firstbigg{ \frac{\indic\{A_i = K\}}{q_i(K, \mbi{X}_i)} - \frac{\indic\{A_i = 1\}}{q_i(1, \mbi{X}_i)} } \firstbigg{ R_i - \Bar{\mu}^{(A_i)}(\mbi{X}_i) }|A_i, \mbi{X}_i) } \\
            &= \firstbigg{ \frac{\indic\{A_i = K\}}{q_i(K, \mbi{X}_i)} - \frac{\indic\{A_i = 1\}}{q_i(1, \mbi{X}_i)} }^2 \sigma_{0}^{2},
        \end{split}
    \end{align*} and hence, \begin{align*}
        \begin{split}
            &\E^{(i-1)}\thirdbigg{ \var^{(i-1)}(\Bar{f}(Z_i)|A_i, \mbi{X}_i)| \mbi{X}_i) } \\
            &= \sigma_{0}^{2} \thirdbigg{ \frac{\P^{(i-1)}(A_i = K)}{q_{i}^{2}(K)} + \frac{\P^{(i-1)}(A_i = 1)}{q_{i}^{2}(1)} - \frac{2 \P^{(i-1)}(A_i = K, A_i=1)}{q_{i}(K) q_{i}(1)} } = \sigma_{0}^2 \thirdbigg{ \frac{1}{q_i(1, \mbi{X}_i)} + \frac{1}{q_i(K, \mbi{X}_i)} },
        \end{split}
    \end{align*} the last equality holds since $i$-th patient can only receive one arm in our setup. Also, \begin{align} \label{eq:ev_eif}
        \begin{split}
            &\E^{(i-1)}(\Bar{f}(Z_i)|A_i, \mbi{X}_i) \\
            &= \Bar{\mu}^{(K)}(\mbi{X}_i) - \Bar{\mu}^{(1)}(\mbi{X}_i) + \firstbigg{ \frac{\indic\{A_i = K\}}{q_i(K, \mbi{X}_i)} - \frac{\indic\{A_i = 1\}}{q_i(1, \mbi{X}_i)} } \thirdbigg{ \mu^{(A_i)}(\mbi{X}_i) - \Bar{\mu}^{(A_i)}(\mbi{X}_i) },
        \end{split}
    \end{align} therefore, \begin{align*}
        \begin{split}
            &\var^{(i-1)}\thirdbigg{ \E^{(i-1)}(\Bar{f}(Z_i)|A_i, \mbi{X}_i)|\mbi{X}_i } \\
            &= \firstbigg{ \mu^{(K)}(\mbi{X}_i) - \Bar{\mu}^{(K)}(\mbi{X}_i) }^2\thirdbigg{ \frac{\P^{(i-1)}(A_i=K)}{q_{i}^{2}(K, \mbi{X}_i)} + \frac{\P^{(i-1)}(A_i=1)}{q_{i}^{2}(1, \mbi{X}_i)} + 0} \\
            &= \firstbigg{ \mu^{(K)}(\mbi{X}_i) - \Bar{\mu}^{(K)}(\mbi{X}_i) }^2 \thirdbigg{ \frac{1}{q_{i}(K, \mbi{X}_i)} + \frac{1}{q_{i}(1, \mbi{X}_i)} }
        \end{split}
    \end{align*} Hence, \begin{align*}
        \begin{split}
            \var^{(i-1)}(\Bar{f}(Z_i)|\mbi{X}_i) &= \firstbigg{ \frac{1}{q_i(1, \mbi{X}_i)} + \frac{1}{q_i(K, \mbi{X}_i)} } \firstbigg{\sigma_{0}^2 + \firstbigg{ \mu^{(K)}(\mbi{X}_i) - \Bar{\mu}^{(K)}(\mbi{X}_i) }^2 }  \\
            &\geq \frac{2\sigma_{0}^{2}}{\delta} > 0,
        \end{split}
    \end{align*} and we conclude the proof of \eqref{eq:cond_var_x} and the proof of this lemma.

    \subsubsection{Proof of Lemma \ref{lemma:cond_l2}} \label{subsubsec:proof_cond_l2}
    Denote $W_n := \Bar{f}(Z_n) - \Delta$. Due to Section B.5 in \cite{waudby-smith_et_al_2024+}, it is enough to show the following Lyapunov-type condition for some $\eta > 0$: \begin{equation*}
        \sum_{n=2}^{\infty} \frac{\E^{(n-1)}\thirdbigg{ |W_n|^{2+\eta} }}{V_{n}^{1+\eta/2}} < \infty,
    \end{equation*} where we recall $V_n = \sum_{i=1}^{n} \sigma_{i}^{2}$. For any $a, b \in \reals$, it follows that $|a+b+c|^{2+\eta} \leq \{3\max(|a|, |b|, |c|)\}^{2+\eta} \leq 3^{2+\eta}(|a|^{2+\eta}+|b|^{2+\eta}+|c|^{2+\eta})$ for any $\eta>0$. Therefore, we have \begin{equation*}
        |W_n|^{2+\eta} \leq 3^{2+\eta} \secondbigg{  \absbigg{ \Bar{\mu}^{(K)}(\cdot) - \Bar{\mu}^{(1)}(\cdot) }^{2+\eta} + \absbigg{ \frac{1}{q_i(K, \mbi{X}_i)} - \frac{1}{q_i(1, \mbi{X}_i)} }^{2+\eta} \firstbigg{ |R_i|^{2+\eta} + \absbigg{\Bar{\mu}^{(A_i)}(\cdot) }^{2+\eta} } + |\Delta|^{2+\eta} }.
    \end{equation*} { By Assumption \ref{assump:bdd_predictor}, we have $\max_{a \in [K]} \norm{\Bar{\mu}^{(a)}}_{2+\eta} < \infty$ and therefore $ \norm{\Bar{\mu}^{(K)} - \Bar{\mu}^{(1)} }_{2+\eta} < \infty$ }. Due to Fact \ref{assump:clipping}, we have $q_i(a, \mbi{x}) \in [\delta, 1-\delta]$. Hence, $\E^{(n-1)} |\Bar{f}(Z_n)|^{2+\eta} < \infty$ provided $\E|R_n|^{2+\eta}<\infty$. We also assume $|\Delta| < \infty$. Therefore, we have \begin{equation*}
        \sum_{n=1}^{\infty} \frac{\E^{(n-1)}\firstbigg{ W_{n}^{2+\eta} }}{\sqrt{V_n}^{2+\eta} } \leqsim \sum_{n=1}^{\infty} \frac{1}{\sqrt{V_n}^{2+\eta}} < \infty,
    \end{equation*} since $V_n \asymp n$, provided $\sigma_{0}^{2} > 0$. Therefore, we conclude the proof of this lemma.

    \subsubsection{Proof of Lemma \ref{lemma:var_est}} \label{subsubsec:proof_var_est}
    Let us denote $\P_{N}[f] := \sum_{i \in \calI^{eval}_n} f(Z_i) / N$. We recall, \begin{align*}
        \begin{split}
            \hat{\var}_n(\hat{f}) = \frac{1}{2} \firstbigg{ \hat{\var}_{N}(\hat{f}_{N'}) + \hat{\var}_{N'}(\hat{f}_{N}) },
        \end{split}
    \end{align*} and \begin{align} \label{eq:var_decomp}
        \begin{split}
            \hat{\var}_{N}(\hat{f}_{N'}) &= \frac{1}{N} \sum_{i \in \calI^{eval}_n} \secondbigg{ \hat{f}_{N'}(Z_i) - \frac{1}{N} \sum_{i \in \calI^{eval}_n} \hat{f}_{N'}(Z_i) }^2 \\
            &= \P_{N'}\firstbigg{ \hat{f}_{N'} - \P_N \hat{f}_{N'} }^2 \\
            &= \P_{N'}\firstbigg{ \hat{f}_{N'} - \P \Bar{f} + \P \Bar{f} - \P_N \hat{f}_{N'} }^2 \\
            &= \P_{N'}\firstbigg{ \hat{f}_{N'} - \P \Bar{f} }^2 + \thirdbigg{ \P_N \firstbigg{ \hat{f}_{N'} - \P \Bar{f} } }^2 - 2 \thirdbigg{ \P_N \firstbigg{ \hat{f}_{N'} - \P \Bar{f} } }^2 \\
            &= \underbrace{ \P_{N'}\firstbigg{ \hat{f}_{N'} - \P \Bar{f} }^2 }_\text{(i)} - \underbrace{ \thirdbigg{ \P_N \firstbigg{ \hat{f}_{N'} - \P \Bar{f} } }^2 }_\text{(ii)}.
        \end{split}
    \end{align} Note that, \begin{equation*}
        \P_N \firstbigg{ \hat{f}_{N'} - \P \Bar{f} } = \frac{1}{N} \sum_{i \in \calI^{eval}_n}\firstbigg{ \hat{f}_{N'}(Z_{i}) - \P \Bar{f}(Z_{i}) }
    \end{equation*} and due to Lemma \ref{lemma:ubd_eif}, we recall $\E \firstbigg{ \Bar{f}(Z_{i}) } = \Delta = \E_* \firstbigg{ \hat{f}_{N'}(Z_{i}) }$. Therefore, we invoke the Strong Law of Large Numbers (SLLN) for the Martingale Difference Sequences (MDS) to establish $\P_N \firstbigg{ \hat{f}_{N'} - \P \Bar{f} } = o(1)$. By continuous mapping theorem we have $\thirdbigg{ \P_N \firstbigg{ \hat{f}_{N'} - \P \Bar{f} } }^2 = o(1)$. From \eqref{eq:var_decomp}, it is enough to show that $(i) = \var(\Bar{f}) + o(1)$. To that end, we decompose $(ii)$ as, \begin{align*}
        \begin{split}
            &\P_{N}\firstbigg{ \hat{f}_{N'} - \P \Bar{f} }^2 = \P_{N}\firstbigg{ \hat{f}_{N'} - \Bar{f} + \Bar{f} - \P \Bar{f} }^2 \\
            &= \P_{N}\firstbigg{ \hat{f}_{N'} - \Bar{f} }^2 + \P_{N}\firstbigg{ \Bar{f} - \P\Bar{f} }^2 + 2 \P_{N}\firstbigg{ \hat{f}_{N'} - \Bar{f} } \firstbigg{ \Bar{f} - \P\Bar{f} } \\
            &\leq 2 \thirdbigg{ \P_{N}\firstbigg{ \hat{f}_{N'} - \Bar{f} }^2 + \P_{N}\firstbigg{ \Bar{f} - \P\Bar{f} }^2 },
        \end{split}
    \end{align*} where we invoke Cauchy-Schwarz inequality in the last line. Now, $\P_{N}\firstbigg{ \Bar{f} - \P\Bar{f} }^2 \to \var(\Bar{f})$ almost surely due to the SLLN for MDS. Thus, it remains to show $\P_{N}\firstbigg{ \hat{f}_{N'} - \Bar{f} }^2 \to 0$ almost surely.  To that end, we note \begin{equation*}
        \P_{N}\firstbigg{ \hat{f}_{N'} - \Bar{f} }^2 = \frac{1}{N} \sum_{i \in \calI^{trn}_n} \secondbigg{ \hat{f}_{N'}(Z_{i}) - \Bar{f}(Z_{i}) }^2,
    \end{equation*} and therefore \begin{align*} 
        \begin{split}
            &\E_{*}\thirdbigg{ \P_{N}\firstbigg{ \hat{f}_{N'} - \Bar{f} }^2 } = \E_*[\hat{f}_{N'}(Z_{i}) - \Bar{f}(Z_{i}) ]^2.
        \end{split}
    \end{align*} From \eqref{eq:f_hat_diff_ineq} it follows that, \begin{equation} \label{eq:bound_diff_eif}
        \E_*[\hat{f}_{N'}(Z_{i}) - \Bar{f}(Z_{i}) ]^2 \leqsim \sum_{a=1,K} \norm{ \hat{\mu}_{N'}^{(a)} - \Bar{\mu}^{(a)} }_2^2 ,
    \end{equation} since $\delta > 0$ by Fact \ref{assump:clipping}. Let us consider the following event for any $\epsilon>0$, \begin{equation*}
        A(\epsilon) := \secondbigg{ \omega \in \Omega: \P_{N}\firstbigg{ \hat{f}_{N'} - \Bar{f} }^2 \geq \epsilon }.
    \end{equation*} By Markov's inequality, we have \begin{align*}
        \begin{split}
            \P_*\firstbigg{ \P_{N}\firstbigg{ \hat{f}_{N'} - \Bar{f} }^2 > \epsilon } &< \frac{1}{\epsilon} \E_{*}\thirdbigg{ \P_{N}\firstbigg{ \hat{f}_{N'} - \Bar{f} }^2 } \\ 
            &\leqsim \sum_{a=1,K} \norm{ \hat{\mu}^{(a)}_{N'} - \Bar{\mu} }_{2}^{2} = o(1),
        \end{split}
    \end{align*} where the last inequality holds due to \eqref{eq:bound_diff_eif} and last equality holds due to Assumption \ref{assump:unif_conv_regression}. Therefore, $\P_{N}\firstbigg{ \hat{f}_{N'} - \Bar{f} }^2 = o(1)$ by Borel-cantelli Lemma, and we complete the proof of this lemma.

\section{Practical Considerations for Implementation} \label{sec:add_discuss}

\subsection{Practical Considerations to Choose Weight Parameter $w$} \label{subsec:choice_of_w} 
While adaptive methods offer a pathway to efficient, automated clinical trials, human oversight remains essential given the stakes of human lives. Involvement of $w$ serves as a flexible way to incorporate domain knowledge, allowing practitioners to control safety integration and mitigate severe adverse outcomes. In oncology trials, RiTS can be implemented with a higher $w\in(0,1)$ to tolerate higher toxicity; conversely, in dermatology trials, where severe adverse events are less acceptable, RiTS can be set with a lower $w$ to avoid aggressive arm allocations. Choice of $w$ has to be guided by domain knowledge: a higher $w$ prioritizes efficacy, while a lower $w$ emphasizes safety. In practice, different scales of efficacy and safety may complicate the choice of $w$. One solution is to borrow domain knowledge to transform both into the same scale. For example, if the effective ranges of efficacy and safety are $(a, b)$ and $(c, d)$ respectively, then we recommend transforming the efficacy endpoint $R_i$ to $(R_i - a)/(b-a)$ and the safety endpoint $S_i$ to $(S_i - c)/(d-c)$. Here, $(a, b)$ and $(c, d)$ are effective ranges, where the endpoint is expected to have the majority of observations, and not necessarily the range.

\subsection{Prior Distributions} \label{subsec:prior}
Multivariate normal priors are chosen due to the availability of closed-form posteriors in our setup. More informative priors can be incorporated based on domain knowledge. For intractable posteriors, a variational approximation can be performed. Bayesian regression serves as an online regression algorithm, where domain knowledge from previous trials can be easily incorporated.

\subsection{Exact Form of Posteriors} \label{subsec:posterior}
Let $\mc{N}_d(\mbi{\mu}, \mbf{\Omega}^{-1})$ be the $d$-variate normal distribution with mean $\mbi{\mu}$ and precision matrix $\mbf{\Omega}$. For all $n \geq 1$, let $\pi^{(n)}(\mbi{\beta}_a) := \mc{N}_d\big( \mbi{\mu}_{a}^{(n)}, \{\mbf{\Omega}_{a}^{(n)}\}^{-1} \big)$. Fix $\mbi{\mu}_{a}^{(0)} := \mbi{0}$, $\mbf{\Omega}_{a}^{(0)} := \mbf{I}$. By the conjugacy of normal priors, update parameters by $\mbf{\Omega}_{a}^{(n)} := \mbf{\Omega}_{a}^{(n-1)} + \indic\{A_{n-1} = a\} \frac{1}{\sigma_0^2} \mbi{X}_{n-1} \mbi{X}_{n-1}^T$ and $\mbi{\mu}_{a}^{(n)} := \{ \mbf{\Omega}_{a}^{(n)} \}^{-1}\big[\indic\{A_{n-1} = a\} \frac{1}{\sigma_0^2} \mbi{X}_{n-1} R_{n-1} + \mbf{\Omega}_{a}^{(n-1)} \mbi{\mu}_{a}^{(n-1)}\big]$ for all $n \geq 1, \ a \in [K]$. Similar steps update the posteriors for safety, i.e., $\{\pi^{(n)}(\mbi{\gamma}_a)\}_{a=1}^{K}$. For simplicity, we fix $\sigma_0^2 = 1$, and assume efficacy and safety end-points have same scales after appropriate standardization by borrowing domain knowledge and phase I results.   

\subsection{Hyper-parameters in AsympCS} \label{subsec:hyparam_asympcs}
The asymptotic validity of AsympCS is tied to $n_0$, which has to increase with the increase in the expected total sample size of the trial. Fixing $n_0$ to a smaller number and $\delta$ too close to zero leads to an inadequate number of participants in potentially sub-optimal arms. Without any regularization, individual regression estimators may suffer from over-fitting. In all of our experiments, we use the ridge penalty to avoid such issues. Even with regularization, AsympCS fails to control the cumulative miscoverage after fixing both $n_0$ and $\delta$ to smaller values. Web Appendix \ref{subsec:clip_sim} explores different choices of hyperparameters and their impact on the empirical cumulative miscoverage.

\subsection{Batched Training and Delayed Response for Practical Implementation} \label{subsec:pract_cons}
Delayed response and batched training are also allowed in Algorithm 1. Specifically, Line 6 and Line 10 of Algorithm 1 can be executed with potentially delayed responses $(r_i, s_i)$ that are available in $\mc{D}^{trn}_n$ and $\mc{D}^{eval}_n$. Batched training is also possible by implementing Line 6 of Algorithm 1 for batches of observations, as incorporated in our \texttt{R} code.

\subsection{Discussion on Burn-in Sample Size, $m$} \label{subsec:burin_in}
It is important to ensure that each regression estimator in $\{\hat{\mu}_{n}^{(a)}(\cdot)\}_{a=1}^{K}$ is adequately estimated before constructing AsympCS. The choice of the burn-in sample size, $m$, depends on several trial-specific parameters, including the number of arms $K$, the outcome variability $\sigma^2$, the covariate dimension $d$, and the coefficient of determination, $R^2$, associated with the regression estimators $\{\hat{\mu}_{n}^{(a)}(\cdot)\}_{a=1}^{K}$. A larger $m$ is required for trials characterized by a larger number of arms, a higher-dimensional covariate space, or increased outcome variability. Conversely, a smaller $m$ may suffice when the $R^2$ is large. Since AsympCS is a flexible procedure that permits stopping at a random time, $\tau$, it is crucial to ensure that the cumulative miscoverage rate is controlled below the nominal level up to this stopping time. Our primary objective is to identify the `winner dose' corresponding to the maximal effect size $\max_{a \in [K] \setminus \{1\}} \Delta(a)$. For this objective, it is sufficient to control the cumulative miscoverage rate for all active doses at the stopping time, $\tau$, based on a pre-specified rule. Following the initial equal randomization of the first $n_0$ subjects, increased adaptivity in the allocation scheme tends to skew assignments towards superior doses. While beneficial for patient outcomes, this skewed allocation necessarily compromises the accuracy of the regression estimators corresponding to inferior doses. The selection of the burn-in sample size, $m$, is a crucial aspect of our proposed framework. Providing a concrete, universally applicable recommendation for the optimal choice of $m$ would require extensive simulation experiments, which is beyond the scope of the current manuscript.









\bibliographystyle{apalike}

\bibliography{ref}

@book{hall_heyde_book,
  title={Martingale limit theory and its application},
  author={Hall, Peter and Heyde, Christopher C},
  year={2014},
  publisher={Academic press}
}

@article{buja2019models,
  title={Models as Approximations {I}: Consequences Illustrated with Linear Regression},
  author={Buja, Andreas and Brown, Lawrence and Berk, Richard and George, Edward and Pitkin, Emil and Traskin, Mikhail and Zhang, Kai and Zhao, Linda},
  journal={Statistical Science},
  volume={34},
  number={4},
  pages={523--544},
  year={2019},
  publisher={JSTOR}
}

@article{villar2018covariate,
  title={Covariate-adjusted response-adaptive randomization for multi-arm clinical trials using a modified forward looking Gittins index rule},
  author={Villar, Sof{\'\i}a S and Rosenberger, William F},
  journal={Biometrics},
  volume={74},
  number={1},
  pages={49--57},
  year={2018},
  publisher={Oxford University Press}
}

@inproceedings{agrawal_et_al_2013_icml,
  title={Thompson sampling for contextual bandits with linear payoffs},
  author={Agrawal, Shipra and Goyal, Navin},
  booktitle={International conference on machine learning},
  pages={127--135},
  year={2013},
  organization={PMLR}
}

@article{dumitrascu_et_al_2018_nips,
  title={Pg-ts: Improved thompson sampling for logistic contextual bandits},
  author={Dumitrascu, Bianca and Feng, Karen and Engelhardt, Barbara},
  journal={Advances in neural information processing systems},
  volume={31},
  year={2018}
}

@inproceedings{dimakopoulou_et_al_2019_aaai,
  title={Balanced linear contextual bandits},
  author={Dimakopoulou, Maria and Zhou, Zhengyuan and Athey, Susan and Imbens, Guido},
  booktitle={Proceedings of the AAAI Conference on Artificial Intelligence},
  volume={33},
  pages={3445--3453},
  year={2019}
}

@article{bibaut_et_al_2021_neurips,
  title={Post-contextual-bandit inference},
  author={Bibaut, Aur{\'e}lien and Dimakopoulou, Maria and Kallus, Nathan and Chambaz, Antoine and van Der Laan, Mark},
  journal={Advances in neural information processing systems},
  volume={34},
  pages={28548--28559},
  year={2021}
}

@article{dimakopoulou_et_al_2021_nips,
  title={Online multi-armed bandits with adaptive inference},
  author={Dimakopoulou, Maria and Ren, Zhimei and Zhou, Zhengyuan},
  journal={Advances in Neural Information Processing Systems},
  volume={34},
  pages={1939--1951},
  year={2021}
}

@article{shen_et_al_2023_jasa,
  title={Doubly Robust Interval Estimation for Optimal Policy Evaluation in Online Learning},
  author={Shen, Ye and Cai, Hengrui and Song, Rui},
  journal={Journal of the American Statistical Association},
  pages={1--20},
  year={2023},
  publisher={Taylor \& Francis}
}

@article{hadad_et_al_2021_pnas,
  title={Confidence intervals for policy evaluation in adaptive experiments},
  author={Hadad, Vitor and Hirshberg, David A and Zhan, Ruohan and Wager, Stefan and Athey, Susan},
  journal={Proceedings of the national academy of sciences},
  volume={118},
  number={15},
  pages={e2014602118},
  year={2021},
  publisher={National Acad Sciences}
}

@misc{kim_iyengar_zeevi_2024+,
      title={Learning the Pareto Front Using Bootstrapped Observation Samples}, 
      author={Wonyoung Kim and Garud Iyengar and Assaf Zeevi},
      year={2024},
      eprint={2306.00096},
      archivePrefix={arXiv},
      primaryClass={stat.ML}
}

@article{kim_et_al_2021_nips,
  title={Doubly robust thompson sampling with linear payoffs},
  author={Kim, Wonyoung and Kim, Gi-Soo and Paik, Myunghee Cho},
  journal={Advances in Neural Information Processing Systems},
  volume={34},
  pages={15830--15840},
  year={2021}
}

@article{zhang_et_al_2021_nips,
  title={Statistical inference with m-estimators on adaptively collected data},
  author={Zhang, Kelly and Janson, Lucas and Murphy, Susan},
  journal={Advances in neural information processing systems},
  volume={34},
  pages={7460--7471},
  year={2021}
}

@article{zhang_et_al_2020_nips,
  title={Inference for batched bandits},
  author={Zhang, Kelly and Janson, Lucas and Murphy, Susan},
  journal={Advances in neural information processing systems},
  volume={33},
  pages={9818--9829},
  year={2020}
}

@article{norwood_davidian_laber_2024_biometrics,
    author = {Norwood, Peter and Davidian, Marie and Laber, Eric},
    title = {Adaptive randomization methods for sequential multiple assignment randomized trials (smarts) via thompson sampling},
    journal = {Biometrics},
    volume = {80},
    number = {4},
    pages = {ujae152},
    year = {2024},
    month = {12},
    issn = {0006-341X},
    doi = {10.1093/biomtc/ujae152},
    url = {https://doi.org/10.1093/biomtc/ujae152},
    eprint = {https://academic.oup.com/biometrics/article-pdf/80/4/ujae152/61199703/ujae152.pdf},
}

@article{thompson_1933_biometrika,
  title={On the likelihood that one unknown probability exceeds another in view of the evidence of two samples},
  author={Thompson, William R},
  journal={Biometrika},
  volume={25},
  number={3-4},
  pages={285--294},
  year={1933},
  publisher={Oxford University Press}
}

@article{yang_diao_rosenberger_2024_sbs,
author = {Yang, Li and Diao, Guoqing and Rosenberger, William F. },
title = {A Two-Stage Covariate-Adjusted Response-Adaptive Enrichment Design},
journal = {Statistics in Biopharmaceutical Research},
volume = {0},
number = {0},
pages = {1--11},
year = {2024},
publisher = {Taylor \& Francis}

}

@article{zhu_shen_fu_qu_2024_aoas,
author = {Shuying Zhu and Weining Shen and Haoda Fu and Annie Qu},
title = {{Risk-aware restricted outcome learning for individualized treatment regimes of schizophrenia}},
volume = {18},
journal = {The Annals of Applied Statistics},
number = {2},
publisher = {Institute of Mathematical Statistics},
pages = {1319 -- 1336},
year = {2024}
}

@article{luckett_et_al_2021_jmlr,
  author  = {Daniel J. Luckett and Eric B. Laber and Siyeon Kim and Michael R. Kosorok},
  title   = {Estimation and Optimization of Composite Outcomes},
  journal = {Journal of Machine Learning Research},
  year    = {2021},
  volume  = {22},
  number  = {167},
  pages   = {1--40}
}

@inproceedings{cook_mishler_ramdas_2024_clear,
  title={Semiparametric Efficient Inference in Adaptive Experiments},
  author={Cook, Thomas and Mishler, Alan and Ramdas, Aaditya},
  booktitle={Causal Learning and Reasoning},
  pages={1033--1064},
  year={2024},
  organization={PMLR}
}

@article{waudby-smith_ramdas_2023_jrssb,
    author = {Waudby-Smith, Ian and Ramdas, Aaditya},
    title = "{Estimating means of bounded random variables by betting}",
    journal = {Journal of the Royal Statistical Society Series B: Statistical Methodology},
    volume = {86},
    number = {1},
    pages = {1-27},
    year = {2023},
    month = {02},
}

@article{waudby-smith_et_al_2024+,
  title={Time-uniform central limit theory and asymptotic confidence sequences},
  author={Waudby-Smith, Ian and Arbour, David and Sinha, Ritwik and Kennedy, Edward H and Ramdas, Aaditya},
  journal={The Annals of Statistics},
  volume={52},
  number={6},
  pages={2613--2640},
  year={2024},
  publisher={Institute of Mathematical Statistics}
}

@article{howard_ramdas_mcauliffe_sekhon_2021_aos,
author = {Steven R. Howard and Aaditya Ramdas and Jon McAuliffe and Jasjeet Sekhon},
title = {{Time-uniform, nonparametric, nonasymptotic confidence sequences}},
volume = {49},
journal = {The Annals of Statistics},
number = {2},
publisher = {Institute of Mathematical Statistics},
pages = {1055 -- 1080},
year = {2021}
}

@article{hay_et_al_2014_nature,
  title={Clinical development success rates for investigational drugs},
  author={Hay, Michael and Thomas, David W and Craighead, John L and Economides, Celia and Rosenthal, Jesse},
  journal={Nature biotechnology},
  volume={32},
  number={1},
  pages={40--51},
  year={2014},
  publisher={Nature Publishing Group US New York}
}

@book{friedman_et_al_2020_textbook,
  title={Fundamentals of clinical trials},
  author={Friedman, Lawrence M and Furberg, Curt and DeMets, David L and Reboussin, David M and Granger, Christopher B and others},
  volume={4},
  year={2010},
  publisher={Springer}
}

@article{stout_1970,
  title={A martingale analogue of Kolmogorov's law of the iterated logarithm},
  author={Stout, William F},
  journal={Zeitschrift f{\"u}r Wahrscheinlichkeitstheorie und verwandte Gebiete},
  volume={15},
  number={4},
  pages={279--290},
  year={1970},
  publisher={Springer}
}

@article{robertson_lee_et_al_2023_stat_science,
author = {David S. Robertson and Kim May Lee and Boryana C. L{\'o}pez-Kolkovska and Sof{\'i}a S. Villar},
title = {{Response-Adaptive Randomization in Clinical Trials: From Myths to Practical Considerations}},
volume = {38},
journal = {Statistical Science},
number = {2},
publisher = {Institute of Mathematical Statistics},
pages = {185 -- 208},
year = {2023}
}

@article{pallmann2018adaptive,
  title={Adaptive designs in clinical trials: why use them, and how to run and report them},
  author={Pallmann, Philip and Bedding, Alun W and Choodari-Oskooei, Babak and Dimairo, Munyaradzi and Flight, Laura and Hampson, Lisa V and Holmes, Jane and Mander, Adrian P and Odondi, Lang’o and Sydes, Matthew R and others},
  journal={BMC medicine},
  volume={16},
  pages={1--15},
  year={2018},
  publisher={Springer}
}

@article{burnett2020adding,
  title={Adding flexibility to clinical trial designs: an example-based guide to the practical use of adaptive designs},
  author={Burnett, Thomas and Mozgunov, Pavel and Pallmann, Philip and Villar, Sofia S and Wheeler, Graham M and Jaki, Thomas},
  journal={BMC medicine},
  volume={18},
  pages={1--21},
  year={2020},
  publisher={Springer}
}

@article{guidance2018adaptive,
  title={Adaptive Designs for Clinical Trials of Drugs and Biologics},
  author={FDA},
  journal={Center for Biologics Evaluation and Research (CBER)},
  year={2019},
  url = {https://www.fda.gov/media/78495/download},
  organization = {{Food and Drug Administration}}
}

@article{tournaue_et_al_2009_jnci,
    author = {Le Tourneau, Christophe and Lee, J. Jack and Siu, Lillian L.},
    title = {Dose Escalation Methods in Phase I Cancer Clinical Trials},
    journal = {JNCI: Journal of the National Cancer Institute},
    volume = {101},
    number = {10},
    pages = {708-720},
    year = {2009},
    month = {05}    
}

@article{morris_et_al_2019_stat_medicine,
  title={Using simulation studies to evaluate statistical methods},
  author={Morris, Tim P and White, Ian R and Crowther, Michael J},
  journal={Statistics in medicine},
  volume={38},
  number={11},
  pages={2074--2102},
  year={2019},
  publisher={Wiley Online Library}
}

@article{proschan_evans_2020,
    author = {Proschan, Michael and Evans, Scott},
    title = {Resist the Temptation of Response-Adaptive Randomization},
    journal = {Clinical Infectious Diseases},
    volume = {71},
    number = {11},
    pages = {3002-3004},
    year = {2020},
    month = {05},
    issn = {1058-4838},
    doi = {10.1093/cid/ciaa334},
    url = {https://doi.org/10.1093/cid/ciaa334},
    eprint = {https://academic.oup.com/cid/article-pdf/71/11/3002/36540053/ciaa334.pdf},
}

@article{thall_et_al_2015,
title = {Statistical controversies in clinical research: scientific and ethical problems with adaptive randomization in comparative clinical trials},
journal = {Annals of Oncology},
volume = {26},
number = {8},
pages = {1621-1628},
year = {2015},
issn = {0923-7534},
doi = {https://doi.org/10.1093/annonc/mdv238},
url = {https://www.sciencedirect.com/science/article/pii/S0923753419318629},
author = {Thall, P. and Fox, P. and Wathen, J.}
}

@article{williamsom_villar_2020_biometrics,
author = {Williamson, S. Faye and Villar, Sofía S.},
title = {A response-adaptive randomization procedure for multi-armed clinical trials with normally distributed outcomes},
journal = {Biometrics},
volume = {76},
number = {1},
pages = {197-209},
year = {2020}
}

@article{villar_wason_bowden_2015_biometrics,
  title={Response-adaptive randomization for multi-arm clinical trials using the forward looking Gittins index rule},
  author={Villar, Sof{\'\i}a S and Wason, James and Bowden, Jack},
  journal={Biometrics},
  volume={71},
  number={4},
  pages={969--978},
  year={2015},
  publisher={Oxford University Press}
}

@article{ham_et_al_2022+,
  title={Design-based confidence sequences: A general approach to risk mitigation in online experimentation},
  author={Ham, Dae Woong and Bojinov, Iavor and Lindon, Michael and Tingley, Martin},
  journal={arXiv preprint arXiv:2210.08639},
  year={2022}
}

@article{dalal_et_al_2024+,
  title={Anytime-valid inference for double/debiased machine learning of causal parameters},
  author={Dalal, Abhinandan and Bl{\"o}baum, Patrick and Kasiviswanathan, Shiva and Ramdas, Aaditya},
  journal={arXiv preprint arXiv:2408.09598},
  year={2024}
}

@inproceedings{maharaj_et_al_2023, series={WWW ’23},
   title={Anytime-Valid Confidence Sequences in an Enterprise A/B Testing Platform},
   url={http://dx.doi.org/10.1145/3543873.3584635},
   DOI={10.1145/3543873.3584635},
   booktitle={Companion Proceedings of the ACM Web Conference 2023},
   publisher={ACM},
   author={Maharaj, Akash and Sinha, Ritwik and Arbour, David and Waudby-Smith, Ian and Liu, Simon Z. and Sinha, Moumita and Addanki, Raghavendra and Ramdas, Aaditya and Garg, Manas and Swaminathan, Viswanathan},
   year={2023},
   month=apr, pages={396–400},
   collection={WWW ’23} 
}

@inproceedings{lindon_et_al_2022,
  title={Rapid regression detection in software deployments through sequential testing},
  author={Lindon, Michael and Sanden, Chris and Shirikian, Vach{\'e}},
  booktitle={Proceedings of the 28th ACM SIGKDD Conference on Knowledge Discovery and Data Mining},
  pages={3336--3346},
  year={2022}
}

@article{agrawal_goyal_2017_acm,
  title={Near-optimal regret bounds for thompson sampling},
  author={Agrawal, Shipra and Goyal, Navin},
  journal={Journal of the ACM (JACM)},
  volume={64},
  number={5},
  pages={1--24},
  year={2017},
  publisher={ACM New York, NY, USA}
}

@article{breiman2001random,
  title={Random forests},
  author={Breiman, Leo},
  journal={Machine learning},
  volume={45},
  number={1},
  pages={5--32},
  year={2001},
  publisher={Springer}
}

@book{jennison_turnbull_1999_book,
  title={Group sequential methods with applications to clinical trials},
  author={Jennison, Christopher and Turnbull, Bruce W},
  year={1999},
  publisher={CRC Press}
}

@article{pin_et_al_2024_cct,
  title={Implementing and assessing Bayesian response-adaptive randomisation for backfilling in dose-finding trials},
  author={Pin, Lukas and Villar, Sof{\'\i}a S and Dehbi, Hakim-Moulay},
  journal={Contemporary Clinical Trials},
  volume={142},
  pages={107567},
  year={2024},
  publisher={Elsevier}
}

@article{zhao_et_al_2025_ct,
  title={BARD: A seamless two-stage dose optimization design integrating backfill and adaptive randomization},
  author={Zhao, Yixuan and Liu, Rachael and Lin, Jianchang and Yuan, Ying},
  journal={Clinical Trials},
  volume={22},
  number={4},
  pages={393--404},
  year={2025},
  publisher={SAGE Publications Sage UK: London, England}
}

@article{chiaruttini_et_al_2025+,
  title={Safety-Driven Response Adaptive Randomisation: An Application in Non-inferiority Oncology Trials},
  author={Chiaruttini, Maria Vittoria and Pin, Lukas and Villar, Sofia S},
  journal={arXiv preprint arXiv:2506.08864},
  year={2025}
}

@article{tackney_villar_2025_biometrics,
author = {Mia S Tackney and Sofía S Villar},
title ={Implementing response-adaptive designs when responses are missing: Impute or ignore?},
journal = {Statistical Methods in Medical Research},
volume = {0},
number = {0},
pages = {09622802251366843},
year = {2025}
}
\end{document}